\documentclass[aps,10pt,notitlepage,nofootinbib]{revtex4-1}
\usepackage[utf8]{inputenc}
\usepackage{amsmath,amssymb,amsfonts}
\usepackage{newtxtext,newtxmath}

\usepackage{bbold}
\usepackage{bm}
\usepackage{graphicx}
\usepackage[usenames,dvipsnames]{xcolor}
\usepackage{color}
\usepackage[colorlinks=true,linkcolor=blue,urlcolor=blue,citecolor=blue]{hyperref}

\usepackage{slashed}
\usepackage[english]{babel}
\usepackage{dcolumn}
\usepackage{pifont}
\usepackage{dsfont,mathrsfs}
\usepackage{cancel}
\usepackage{bigints}
\usepackage{accents}
\usepackage{soul}
\usepackage{multirow}

\usepackage{lineno}
\newcommand{\nn}{\nonumber}

\newcommand{\ensembleaverage}[1]{\left\langle#1\right\rangle}

\newcommand{\MB}[1]{\left|#1\right|}
\newcommand{\mb}[1]{|#1|}
\newcommand{\FB}[1]{\left(#1\right)}
\newcommand{\fb}[1]{(#1)}
\newcommand{\SB}[1]{\left\{#1\right\}}
\newcommand{\TB}[1]{\left[#1\right]}
\newcommand{\AB}[1]{\left<#1\right>}

\newcommand{\scrL}{\mathscr{L}}

\newcommand{\munu}{{\mu\nu}}

\newcommand{\psibar}{\overline{\psi}}

\newcommand{\del}{\partial}

\newcommand{\fsl}{\slashed}

\newcommand{\intzinf}{\int_{-\infty}^{\infty}}

\newcommand{\half}{\dfrac{1}{2}}

\newcommand{\kzint}[1]{\int_{-\infty}^\infty \dfrac{d{#1}_z}{2\pi}}

\newcommand{\potU}{\mathcal{U}\FB{ \Phi,\bar{\Phi} ;T}}
\newcommand{\fnppbT}{\FB{\Phi, \bar{\Phi}, T } }
\newcommand{\F}{\Phi}
\newcommand{\Fb}{\bar{\Phi}}
\newcommand{\paroneder}[2]{\frac{\partial {#1}}{\partial {#2}} }

\newcommand{\derpartone}[2]{\frac{d {#1} }{d {#2}}}
\newcommand{\derparttwo}[2]{\frac{d^2 {#1} }{d {#2}^2}}
\newcommand{\Np}{N^\prime}

\begin{document}
\title{Insignificance of the anomalous magnetic moment of the quarks in presence of chiral imbalance}

\author{Nilanjan Chaudhuri$^{a,e}$}
\email{sovon.nilanjan@gmail.com}
\email{n.chaudhri@vecc.gov.in}
\author{Arghya Mukherjee$^{b,e}$}
\email{arbp.phy@gmail.com}
\author{Snigdha Ghosh$^{c}$}
\email{snigdha.physics@gmail.com}
\author{Sourav Sarkar$^{a,e}$}
\email{sourav@vecc.gov.in}
\author{ Pradip Roy$^{d,e}$}
\email{pradipk.roy@saha.ac.in}

\affiliation{$^a$Variable Energy Cyclotron Centre, 1/AF Bidhannagar, Kolkata 700 064, India}
\affiliation{$^b$School of Physical Sciences, National Institute of Science Education and Research, HBNI, Jatni, Khurda 752050, India}
\affiliation{$^c$Government General Degree College at Kharagpur-II, Madpur, Paschim Medinipur - 721149, West Bengal, India}
\affiliation{$^d$Saha Institute of Nuclear Physics, 1/AF Bidhannagar, Kolkata - 700064, India}
\affiliation{$^e$Homi Bhabha National Institute, Training School Complex, Anushaktinagar, Mumbai - 400085, India}


\begin{abstract}

 We incorporate  the  anomalous magnetic moment (AMM) of quarks in the framework of PNJL model to study hot and dense magnetised matter with chiral imbalance. For this purpose, the  eigen energy solution of the  Dirac equation is obtained in presence of  constant  background magnetic field and chiral chemical potential (CCP) along with the minimal anomalous magnetic moment interaction of the fermion. Although there is a marginal enhancement in the  IMC behaviour of the quark condensate due to  the  combined effects of AMM and CCP, we find that  the overall behaviour of the  Polyakov loop and the chiral charge density is  dominated by the chiral chemical potential. It is further shown that  the AMM effects in presence of CCP  remains insignificant even after consideration of thermo-magnetically modified moments.   

\end{abstract}

\maketitle
%

\section{INTRODUCTION}\label{sec.intro}

Quantum chromodynamics (QCD) is the established theory of strong interaction between quarks and gluons. The influence of the background magnetic field on various microscopic as well as bulk properties of strongly interacting matter at finite temperature and/or baryon density, has seen intensive research activities for the last couple of decades (see Refs.~\cite{Kharzeev:2013jha,Miransky:2015ava,Andersen:2014xxa} for review).  Current studies~\cite{Kharzeev:2007jp,Skokov:2009qp} indicate that, in a non-central heavy ion collision(HIC) experiment, very strong transient magnetic fields of the order $ \sim 10^{18} $ Gauss or larger might  be produced. The presence of sizable electrical conductivity of the hot and dense magnetized medium results in  substantial delay in the decay process of these time dependent fields~\cite{Tuchin:2013apa,Tuchin:2015oka,Tuchin:2013ie,Gursoy:2014aka}.  Since, the strength of the magnetic field reaches up to the typical QCD energy scale ($eB\sim \Lambda_\text{QCD}^2$), the properties of the QCD matter could be significantly modified~\cite{Kharzeev:2013jha}.  Apart from that, a large  magnetic field can exist in several other physical situations. For example, in the interior of  magnetars~\cite{Duncan:1992hi,Thompson:1993hn}, the magnetic field strength can be $\sim 10^{15}$ Gauss whereas the  value of the magnetic field  during the electroweak phase transition can be as high as $ \sim 10^{23}$ Gauss~\cite{Vachaspati:1991nm,Campanelli:2013mea}. 

The most fundamental features of QCD  vacuum in the low energy region are the spontaneous breakdown of chiral symmetry, axial anomaly and color confinement. At vanishing baryon density, most of our current understanding of these non-perturbative aspects are derived from the Lattice QCD simulations~\cite{deForcrand:2006pv,Aoki:2006br,Aoki:2009sc,Bazavov:2009zn,Cheng:2007jq,Muroya:2003qs}. However, a straightforward Monte Carlo simulation in three-color QCD on Lattice cannot be performed at finite baryon density owing to the (in)famous sign problem~\cite{Muroya:2003qs,Splittorff:2006fu,Fukushima:2006uv}, although, using indirect methods results are available at baryon densities as large as $  \mu_B/T\sim 2.5 $~\cite{Bazavov:2017dus,Sharma:2019wiv}. An alternative approach is to work with effective model description such as the Nambu$ - $Jona-Lasinio (NJL) model~\cite{Nambu1,Nambu2} which  respects the global symmetries of QCD, most importantly the chiral symmetry. Being an effective model,  within its domain of applicability it provides a useful scheme to study some of the important non-perturbative properties of the QCD vacuum ~\cite{Klevansky,Vogl,Buballa,Volkov:2005kw}. As the gluonic degrees of freedom have been integrated out in this effective description,  point-like interaction among the quarks  appears   ~\cite{Klevansky} which makes the NJL  model non-renormalizable. Therefore, one has to choose a proper regularization scheme  to  deal with the divergent integrals and subsequently, fix the model parameters by  reproducing  a set of   phenomenological quantities, for example the pion-decay constant, quark condensate \textit{etc}. 
It should be noted here that the usual NJL model does not incorporate the  confinement feature of  QCD. 
On the other hand, as the de-confinement transition can be associated with the spontaneous  breaking of the center symmetry, one can consider the Polyakov loop as an  approximate order parameter for the 	de-confinement phase transition~\cite{McLerran:1981pb,Cheng:2007jq}. Thus, in order to obtain a simultaneous description of the confinement and the chiral symmetry breaking within the effective model approach, an extension of the NJL model has been proposed where a temporal, static and homogeneous gluon-like field~\cite{Ratti:2005jh,Ratti:2006wg} is introduced. This particular  extension is known as the Polyakov loop extended Nambu$ - $Jona-Lasinio (PNJL) model and it will be used here as the basic framework for the subsequent study.

The breakdown of chiral and axial symmetries of quarks are intrinsically related to the topological properties of the QCD vacuum which are determined by the gluonic sector of the theory~\cite{Shifman:1988zk}. It is well known that, at small values of temperature, the existence of certain gluon configurations (i.e., instantons) leads to the assignment of an integer-valued topological winding number to the QCD vacuum~\cite{Belavin:1975fg,tHooft:1976rip,tHooft:1976snw}. The infrared instanton structure can provide a mechanism for the chiral phase transition~\cite{Schafer:1996wv}.
However, at high temperatures, an abundant production of the QCD sphalerons, another kind of topological gluon configurations, is predicted~\cite{Manton:1983nd,Klinkhamer:1984di,Kuzmin:1985mm,Arnold:1987mh,Khlebnikov:1988sr,Arnold:1987zg}. This topologically nontrivial sphaleron transitions can generate chiral imbalance in the hot QCD matter and thus lead to the breaking of the parity ($ P $) and charge-parity ($ CP $) symmetry via the axial anomaly of QCD. Since there is no direct observation  of $P$ and $CP$ violations in QCD, there can only be local domains with non-zero chirality which vanishes globally\cite{McLerran:1990de,Moore:2010jd}.
 Thus, at finite temperature 
there might be local domains with non-zero topological charge vis-a-vis 
chiral  imbalance, i.e. unequal number of left handed and right handed fermions which can be conveniently characterised by the introduction of a chiral chemical potential (CCP). Local $P$ and $CP$ violations and the presence of strong magnetic field in peripheral heavy ion collisions lead to several exotic phenomena such as Chiral magnetic effect (CME),
Charge Separation Effect and so on~\cite{Fukushima:2008xe,Kharzeev:2007jp,Kharzeev:2009pj,Bali:2011qj}. 
Thus, heavy ion collisions could provide an excellent environment for
the observation of local $P$ and $CP$ violations. Apart from this, the
chiral imbalance has also important effects
on microscopic transport phenomena~\cite{Vilenkin:1979ui,Vilenkin:1980fu,Fukushima:2008xe,Son:2009tf}, the collective modes 
propagating in the
medium~\cite{Akamatsu:2013pjd,Carignano:2018thu,Carignano:2021mrn}, fermion damping rate~\cite{Carignano:2019ivp}
and collisional energy loss of a fermion~\cite{Carignano:2021mrn}. Moreover, 
chiral matter characterized by different densities of right- and left- 
handed massless fermions has applications in Weyl and Dirac  semimetals
in condensed matter physics as well as in cosmological and astrophysical
context~\cite{Kharzeev:2013ffa,Kharzeev:2015znc,Huang:2015oca,Landsteiner:2016led,Gorbar:2017lnp,Joyce:1997uy,Tashiro:2012mf}.

In addition to the above applications of QCD anomaly leading to local chirally 
imbalanced matter, there are also 
important effects of chiral imbalance on the phase structure of the
strongly interacting matter. Some studies show that, in heavy ion collisions,
chiral charge density reaches equilibrium shortly after the collision
and it remains in equilibrium for comparatively longer period of 
time~\cite{Ruggieri:2016lrn,Ruggieri:2016asg,Ruggieri:2020qtq}. Various studies on the effects of chiral imbalance
on the phase structure in presence of magnetic field have been 
done~\cite{Fukushima:2010fe,Yu:2014xoa,Yu:2015hym,Chao:2013qpa} using both NJL and PNJL models with 
different interactions among the fermions. In all these studies the 
effect of  chiral imbalance is to de-catalize the chiral condensate
implying that the critical temperature of the phase transition decreases with
the magnetic field which is known as inverse maganetic catalysis (IMC).

We  also note here that both in the NJL and PNJL models, the presence of the 
magnetic field without chiral imbalance and at zero baryon chemical potential leads
to magnetic catalysis (MC) with constant coupling, i. e. the critical 
temperature increases with the magnetic field~\cite{Gusynin:1994re,Gusynin:1995nb,Fayazbakhsh:2012vr,Kharzeev:2013jha}. 
This result is independent of the method of regularization and is
inconsistent with the lattice results with same environment. However, using
Pauli-Villiars regularization and magnetic field dependent coupling it
has been shown in Ref.~\cite{Mao:2016fha} that the critical temperature
decreases with the magnetic field leading to IMC which is consistent with the lattice
results. Another approach to obtain results in both
NJL and PNJL models (with constant coupling) consistent with the
lattice calculation is the inclusion of anomalous magnetic moment (AMM) of the
quarks. Recently, there has been a lot of activity in this direction
involving both NJL and PNJL models to study the phase structure as well as the mesonic
excititations ($\sigma$ and $\pi$)~\cite{Fayazbakhsh:2014mca,Farias:2021fci,Chaudhuri:2019lbw,Chaudhuri:2020lga,Ghosh:2021dlo}. 
The values of the AMM of the quarks can be
taken as constant~\cite{Fayazbakhsh:2014mca,Chaudhuri:2019lbw,Chaudhuri:2020lga} or temperature and magnetic field
dependent~\cite{Ghosh:2021dlo}; in either case the values must reproduce
the vacuum values of the AMM of the proton and neutron. With the inclusion of
the AMM of the quarks both in the NJL and PNJL models with constant coupling,
it has been found that the critical temperature decreases with the 
magnetic field leading to inverse magnetic catalysis  (IMC)~\cite{Fayazbakhsh:2014mca,Chaudhuri:2019lbw,Chaudhuri:2020lga}.
Moreover, the inclusion of AMM leads to interesting consequences in the dilepton production rate in both NJL and PNJL models~\cite{Ghosh:2020xwp,Chaudhuri:2021skc}. It is found that there are contributions both 
from the unitarity cut (UC) and Landau cut (LC) with a
forbidden gap in-between. The contribution in the low invariant
mass region comes due to the appearance of the LC which is present only if
the magnetic field is non-zero. It has also been found that for judicious
choice of parameters (such as $B$, $T$) the gap vanishes resulting in a continuous dilepton spectrum. We also
find an increase in the DPR in three-flavor case due to the presence of 
strange quarks.

From the above discussions it is clear that either the  presence of 
chiral imbalance or the AMM of the quarks have been found to influence the phase structure of
the strongly interacting medium. In the present work, we include both
the parameters ($\mu_5$ and AMM of the quarks)  in the Dirac equation for the first time and explore the effects on the phase structure and mass of the quarks 
in a strongly interacing thermo-magnetic medium described by 
the PNJL model with constant coupling. 
To avoid regularization artifacts, we have used a smooth three-momentum cutoff  as our regularization scheme~\cite{Fukushima:2010fe}.  In this work, an extensive study of temperature ($T$), and, background magnetic field ($eB$) dependence of the constituent quark mass ($ M $) and expectation value of Polyakov loop ($ \Phi $) has been performed for different values of CCP $\mu_5 $.  The variation of chiral charge density ( $ n_5 $) as function of $ \mu_5 $ and $ T $ for different values of background magnetic field for both zero and non-zero values of the AMM of the quarks is also studied in detail. It should be noted that, all the calculations presented in the work have been  performed by considering all the Landau levels without resorting to any approximation on the strength of the magnetic field.

The paper is organized as follows. Sec.~\ref{Eigenvalue_calculation} is devoted for the calculation of the energy eigenvalues of a fermions with non-zero AMM in presence of chiral chemical potential and constant background magnetic field of arbitrary strength. In Secs.~\ref{sec.PNJL} and \ref{sec_CQM_TD} we have briefly introduced PNJL model and outline few important steps for estimation of the gap equations and other thermodynamical quantities respectively. In Sec.~\ref{sec.results} we show our numerical results and finally summarize and conclude in Sec.~\ref{sec.summary}. Details of the calculations are provided in the Appendix.

%
\section{ENERGY SPECTRUM OF FERMIONS WITH NON-ZERO AMM AT FINITE CCP AND EXTERNAL MAGNETIC FIELD}\label{Eigenvalue_calculation}
Let us consider a spin-$\frac{1}{2}$ charged fermion of mass $m$ and AMM $ \kappa $ in a medium with fermion-number chemical potential $ \mu $ and CCP $ \mu_5 $ in the presence of an external electromagnetic field characterized by the four-potential $A_\mu^\text{ext}$. In such case the Dirac Lagrangian (density) is given by~\cite{OConnell:1968spc,Fukushima:2008xe,Sheng:2017lfu}  
\begin{equation}
\scrL = \overline{\psi} \TB{i\gamma^\mu \FB{\partial_\mu + ieA_\mu^\text{ext} } -m + \mu \gamma^0 + \mu_5 \gamma^0\gamma^5 + \dfrac{1}{2} a \sigma^{\munu} F_\munu^\text{ext} }\psi~, 
\label{eq_Lagrangian}
\end{equation}
where, $ e $ is the electronic charge of the Fermion (for electrons $ e<0 $), $ a = \kappa e $ and $F_\munu^\text{ext} = \fb{\del_\mu A_\nu^\text{ext} - \del_\nu A_\mu^\text{ext}}$ is the electromagnetic field strength tensor.
Let us restrict ourselves to the the case of pure magnetic field i.e. $ \vec{E} = 0$, so that the non-zero components of $F_\munu^\text{ext}$ are 
$ F_\text{ext}^{12} = - F_\text{ext}^{21} = -B_z  $, $ F_\text{ext}^{13} = - F_\text{ext}^{21} = B_y  $ and $ F_\text{ext}^{23} = - F_\text{ext}^{32} = -B_x  $. 
We will be using the following Dirac representation of the gamma matrices~\cite{Peskin:1995ev}: 
\begin{equation}
\gamma^0 = \begin{pmatrix}
\mathds{1}& 0 \\
0 & -\mathds{1}
\end{pmatrix} ~~,~~
\gamma^k = \begin{pmatrix}
0 & \sigma^k \\
-\sigma^k & 0
\end{pmatrix} ~~,~~
\gamma^5 = \begin{pmatrix}
0 & \mathds{1} \\
\mathds{1} & 0
\end{pmatrix}~~,
\label{eq_rep_gamma}
\end{equation}
where, $\sigma^i$ are the Pauli matrices. Using Eq.~\eqref{eq_rep_gamma}, the last term within square bracket in Eq.~\eqref{eq_Lagrangian} can be explicitly written as
\begin{eqnarray}
\frac{1}{2}a\sigma^\munu F_\munu^\text{ext} =  a \vec{\Sigma}\cdot\vec{B}~,
\label{eq_sigmaF2}
\end{eqnarray}
where,
\begin{equation}
\Sigma^k = \begin{pmatrix}
\sigma^k & 0 \\
0 & \sigma^k 
\end{pmatrix}~.
\label{eq_Sigma_matrix}
\end{equation}
The Euler-Lagrange equation of motion from the Lagrangian of Eq.~\eqref{eq_Lagrangian} comes out to be
\begin{equation}
\TB{i\gamma^\mu \FB{\partial_\mu + ieA_\mu^\text{ext} } -m + \mu \gamma^0 + \mu_5 \gamma^0\gamma^5 + \dfrac{1}{2} a \sigma^{\munu} F^\text{ext}_\munu }\psi = 0~.
\label{eq_Dirac}
\end{equation}
Substituting  Eq.~\eqref{eq_sigmaF2} into Eq.~\eqref{eq_Dirac}, we get after some simplifications,
\begin{equation}
i\frac{\partial \psi }{\partial t} =  \TB{ \gamma^0 \vec{\gamma} \cdot \vec{\Pi} + m\gamma^0 - \mu - \mu_5 \gamma^5  - a \gamma^0\vec{\Sigma}\cdot\vec{B}   }\psi~,
\label{eq_Dirac2}
\end{equation}
where, $\vec{\Pi} = (-i\vec{\nabla} -e\vec{A}) $.

Let us now consider a constant magnetic field $ \vec{B} = B\hat{z} $ along positive $\hat{z}$ direction. This can be achieved by choosing the cylindrical (or symmetric) gauge $A^\mu \equiv \frac{1}{2}\FB{0,-yB, xB, 0}$. In this gauge, Eq.~\eqref{eq_Dirac2} becomes on using Eqs.~\eqref{eq_rep_gamma} 
\begin{equation}
i\frac{\partial \psi }{\partial t} = \begin{pmatrix}
m-\mu -a B  & 0 &  \mu_5 -i\dfrac{\partial}{\partial z} & \Pi_x - i \Pi_y\\ 
0 & m-\mu +a B  & \Pi_x + i \Pi_y & \mu_5 +i\dfrac{\partial}{\partial z} \\
 \mu_5 -i\dfrac{\partial}{\partial z}  & \Pi_x - i \Pi_y  & -m-\mu +a B & 0 \\
\Pi_x + i \Pi_y & \mu_5 +i\dfrac{\partial}{\partial z} & 0 & -m-\mu -a B 
\end{pmatrix} \psi~.
\label{eq_Dirac_matrix1}
\end{equation}
To solve Eq.~\eqref{eq_Dirac_matrix1}, we choose the cylindrical polar coordinate $(x,y,z)\to (\rho,\phi,z)$ via $x=\rho\cos\phi$ and $y=\rho\sin\phi$. 
We apply the following ansatz for the four component Dirac spinor $\psi$ as
\begin{equation}
\psi  = \begin{pmatrix}
\psi_1 \\ \psi_2 \\ \psi_3 \\ \psi_4
\end{pmatrix}~,
\label{eq_ansatz}
\end{equation}
with 
\begin{equation}
\psi_{1,3} = \dfrac{1}{\sqrt{2\pi}} e^{-iEt+ip_z z} e^{i(l-1)\phi } f_{1,3}(\rho)~~~,~~~
\psi_{2,4} = \dfrac{1}{\sqrt{2\pi}} e^{-iEt+ip_z z} e^{il\phi } f_{2,4}(\rho)~,
\label{eq_psi_comp}
\end{equation}
where, $ f_i(\rho) $'s are the functions of $ \rho $ to be determined. In this work, we do not actually require the explicit form of the functions $ f_i(\rho) $'s, rather we just need to obtain the energy eigenvalues $E$. The calculation of energy eigenvalues is provided in Appendix~\ref{sec.app.1} and we can read off the final result from Eqs.~\eqref{Energy_EV_GS} and \eqref{Energy_EV_ES} as
\begin{eqnarray}
\fb{E_{ns}+ \mu}^2 = \begin{cases}
 (p_z - \mu_5)^2 + (m - \kappa \MB{e B})^2 ~~~~~ \text{if}~~ n=0~, \vspace{0.2 cm} \\
 p_z^2 + m^2 + \mu_5^2 + \fb{\kappa e B}^2  + 2 n \MB{eB} \\
~~~~ - 2 s {~\rm sign} (eB) \TB{  \fb{ m^2 + 2 n  \MB{eB} } \fb{\kappa e B}^2 + \fb{ p_z^2 + 2 n  \MB{eB} } \mu_5^2  -  2 m p_z \mu_5 \kappa e B  }^{1/2} 
 ~~~~~ \text{if}~~ n\ge1~,
\end{cases}
\end{eqnarray}
where, the quantum numbers $s\in\{\pm\}$ and $n\in\{0,\mathds{Z}^+\}$ respectively corresponds to the helicity in the massless case and Landau level. In the limit $ \mu_5\rightarrow 0 $ and $\mu\rightarrow 0$, the expression of the energy eigenvalue reduces to~ \cite{Fayazbakhsh:2014mca}
\begin{eqnarray}
E_{ns} = \begin{cases}
\sqrt{ p_z^2 + (m - \kappa \MB{e B})^2} ~~~~~ \text{if}~~ n=0~, \vspace{0.2 cm} \\
\sqrt{ p_z^2 + \TB{\sqrt{   m^2 + 2 n  \MB{eB} }-s \kappa e B  }^2}
 ~~~~~ \text{if}~~ n\ge1~.
\end{cases}
\end{eqnarray}
Here we note that the  eigen value in the ground state is independent of  ${\rm sgn}(eB)$. This is due to the fact that in the LLL, the only allowed spin orientation  for a positively (negatively) charged particle is along (opposite to) the external  magnetic field direction. To incorporate this fact in an explicit way, one may express  the Landau level index $n$ in terms of the quantum number associated with the orbital motion  (say $l\in\{0,\mathds{Z}^+\}$ ) and spin  ($s$) as \cite{Shovkovy:2012zn}  
\begin{align}
n&=l+\frac{1}{2}-\frac{s}{2}{\rm sgn}(eB)~.
\end{align}  
In that case, the constraint $n,l\geq0$ automatically fixes the spin orientation in LLL and the expression for the  energy eigenvalue reduces to the known result  \cite{Xu:2020yag,Chaudhuri:2019lbw,Chaudhuri:2020lga}
\begin{align}
 E_{ls} & = \sqrt{p_z^2 + \TB{ \sqrt{m^2 + \SB{2l + 1-s~ {\rm sgn}(eB)}\MB{eB} } - s\kappa eB }^2}~.
\end{align}

%
\section{THE PNJL MODEL} \label{sec.PNJL}
The Lagrangian for the two-flavor PNJL model considering the AMM of free quarks in the presence of a constant background magnetic field and CCP is given by~\cite{Chaudhuri:2019lbw,Chaudhuri:2020lga}
\begin{equation}
\scrL = \overline{\Psi}(x)\FB{i\fsl{D}-\hat{m} + \gamma^0 \mu_q +\gamma^0 \gamma^5 \mu_5 +\half \hat{a} \sigma^\munu F^\text{ext}_\munu  }\Psi(x)  
+ G\SB{ \FB{\overline{\Psi}(x) \Psi (x)}^2 + \FB{ \overline{\Psi}(x) i\gamma_5\tau \Psi(x)}^2} - \mathcal{U} \FB{\Phi , \bar{\Phi } ; T}~, \label{PNJL_lagrangian}
\end{equation}
where, the flavour $ f \in\{u,d\} $ and color $ c\in\{r,g,b \}$ indices are omitted from the Dirac field $ \FB{ \Psi^{fc}} $ for convenience. In Eq.~\eqref{PNJL_lagrangian}, $\hat{m} = \texttt{diag} \FB{m_u, m_d}$ is bare quark mass matrix representing the explicit breaking of chiral  symmetry  and  $ \mu_q $ is the chemical potential of the quark . To ensure isospin symmetry of the theory at vanishing magnetic field, we will take $ m_u = m_d = m_0 $ throughout this paper. The constituent quarks interact with the external electromagnetic field $ A_\mu$ and the $ {\rm SU_c(3)} $ gauge field $ \mathcal{A}_\mu^{\rm ext}  $ via the covariant derivative
\begin{eqnarray}
D_\mu = \partial_\mu  -ie\hat{Q} A_\mu^\text{ext} - i \mathcal{A}_\mu^a~.
\end{eqnarray}
The factor $\hat{a}= \hat{Q}\hat{\kappa }$,  where $ \hat{Q} = \texttt{diag}(2/3,-1/3) $ and  $\hat{\kappa}=\texttt{diag}(\kappa_u,\kappa_d) $ are  $2\times 2 $ matrices in the flavour space and $ \sigma^\munu= \frac{i}{2}[\gamma^\mu,\gamma^\nu]$.  All the other details can be found in Ref.~\cite{Chaudhuri:2020lga}. The potential $ \mathcal{U}\FB{\Phi,\bar{\Phi};T} $ in the Lagrangian in Eq.~\eqref{PNJL_lagrangian} governs the dynamics of the traced Polyakov loop and its conjugate and is given by~\cite{Roessner:2006xn}
\begin{eqnarray}
\frac{\mathcal{U}\FB{ \Phi,\bar{\Phi} ;T}}{T^4} = -\frac{a(T) }{2}  \bar{\Phi} \Phi + b(T) \ln \SB{1 - 6 \bar{\Phi} \Phi   + 4 \FB {\bar{\Phi }^3 + \Phi^3   } - 3 \FB{\bar{\Phi} \Phi}^2 }~,
\label{polyakov_potential}
\end{eqnarray} 
where, 
\begin{eqnarray}
a(T) = a_0 + a_1 \FB{\frac{T_0}{T}}+ a_2 \FB{\frac{T_0}{T}}^2 ~~~, ~~~
b(T) = b_3 \FB{\frac{T_0}{T}}^3~.
\end{eqnarray}
Values of the different coefficients~\cite{Roessner:2006xn,Fukushima:2010fe} are shown in Table~\ref{Table_parameters}.
\begin{table}[h]
		\caption{Parameter set for Polyakov potential}
		\begin{tabular} { p{2cm}p{2cm}p{2cm}p{2cm}p{2cm} }
			\hline \hline
			\hspace{0.02IN}$ a_0 $ & \hspace{0.1IN}$ a_1 $ &  \hspace{0.07IN}$ a_2 $   &\hspace{0.082IN} $ b_3 $ & $ T_0 $ (MeV) \\ 
			\hline
			\vspace{0.02IN} $ 3.51 $ \vspace{0.02IN}& \vspace{0.02IN}$ -2.47$ & \vspace{0.02IN} $ 15.2 $ &\vspace{0.02IN}  $ -1.75 $ &\vspace{0.02IN} \hspace{0.082IN} $ 270 $  \\
			\hline
		\end{tabular}
\label{Table_parameters}
\end{table}


\section{Constituent quark mass and Thermodynamics}\label{sec_CQM_TD}
Employing the mean field approximation on the Lagrangian in Eq.~\eqref{PNJL_lagrangian}, one can show that, the thermodynamic potential $(\Omega)$ for a  two-flavor PNJL model can be expressed as
\begin{eqnarray}\label{Omega_PNJL}
\Omega &=&   \frac{(M- m_0)^2}{4 G}+ \potU - 3 \sum_{nfs} \alpha_{ns} \frac{\mb{e_f B}}{2\pi } \kzint{p}\omega_{nfs}  \nn \\&& 
- \frac{1}{\beta} \sum_{nfs} \alpha_{ns} \frac{\mb{e_f B}}{2\pi } \kzint{p}\TB{\ln g^{(+)}\FB{\Phi, \bar{\Phi}, T } + \ln g^{(-)}\FB{\Phi, \bar{\Phi}, T } }~,
\end{eqnarray}
where $ \omega_{nfs} $ are the energy eigenvalues of the quarks as derived in Sec.~\ref{Eigenvalue_calculation} and is given by
\begin{eqnarray} 
\omega_{nfs}^2 =
\begin{cases}
 { (p_z - \mu_5)^2  + (M - \kappa_f \mb{e_f B})^2 }  ~~~~ \text{if}~~ n = 0~, \vspace{0.2 cm} \\ 
p_z^2 + M^2 + \mu_5^2 + \fb{\kappa_f e_f B}^2  + 2 n \mb{e_fB}  \\
~~~~- 2 s {~\rm sign} (e_f) \TB{  \fb{ M^2 + 2 n  \mb{e_fB} } \fb{\kappa_f e_f B}^2 + \fb{ p_z^2 + 2 n  \mb{e_fB} } \mu_5^2  -  2 M p_z \mu_5 \kappa_f e_f B  }^{1/2}, ~~~~\text{if}~~ n > 0~,
\end{cases}
\label{energy}
\end{eqnarray}
and $ \alpha_{ns} = \frac{1}{2}(2 - \delta_{n,0}) $. The quantities $ g^{(+)}\fnppbT $ and $ g^{(-)}\fnppbT $  are defined as
\begin{eqnarray}
g^{(+)}\fnppbT &=& 1 + 3 \FB{  \Phi + \Fb e^ { - \beta ( \omega_{nfs} -\mu_q )} } e^ { - \beta ( \omega_{nfs} -\mu_q )} + e^ { - 3\beta ( \omega_{nfs} -\mu_q )}, \\
g^{(-)}\fnppbT &=& 1 + 3 \FB{  \Fb + \F e^ { - \beta ( \omega_{nfs} + \mu_q )} } e^ { - \beta ( \omega_{nfs} + \mu_q )} + e^ { - 3\beta ( \omega_{nfs}+ \mu_q )} .
\end{eqnarray}
Now from Eq.~\eqref{Omega_PNJL}, one can obtain the expressions for the constituent quark mass ($ M $) and the expectation values of the Polyakov loops $ \F $ and $ \Fb $ using the following stationary conditions
\begin{equation}
\paroneder{\Omega}{M} = 0~~ , ~~~~ \paroneder{\Omega}{\F} = 0 ~~,~~~~ \paroneder{\Omega}{\Fb} =0~,
\end{equation} 
which, in turn, leads to the following set of coupled equations:
\begin{gather}
 M = m_0 + 6G \sum_{nfs} \alpha_{ns} \frac{\mb{e_f B}}{4\pi^2 } \intzinf dp_z \frac{M}{\omega_{nfs} }~ \Xi_{nfs} \SB{ \frac{}{}1-  f^+ \fnppbT  - f^- \fnppbT} \label{Gap_M}, \\
 \SB{ \frac{a ( T )}{2}\Fb + 6b(T)\frac{\Fb  - 2 \F^2 + \FB{ \Fb \F } \Fb}{  1 - 6 \Fb \F + 4 \FB{\F^3 + \Fb^3} + 3\FB{\Fb \F}^2 }   } =  - \frac{3}{T^3} \sum_{nfs} \alpha_{ns} \frac{\mb{e_f B}}{4 \pi^2 } \intzinf dp_z \TB{\frac{e^{- \beta( \omega_{nfs} -\mu_q) }}{g^{(+)}}  + \frac{e^{- 2\beta( \omega_{nfs} +\mu_q) }}{g^{(-)}}        }\label{Gap_p}, \\
 \SB{ \frac{a ( T )}{2}\F + 6b(T)\frac{ \F - 2 \Fb^2 + \FB{ \Fb \F } \F}{  1 - 6 \Fb \F + 4 \FB{\F^3 + \Fb^3} - 3\FB{\Fb \F}^2 }   } = - \frac{3}{T^3} \sum_{nfs} \alpha_{ns} \frac{\mb{e_f B}}{4 \pi^2 } \intzinf dp_z \TB{\frac{e^{- 2\beta( \omega_{nfs} -\mu_q) }}{g^{(+)}}  + \frac{e^{- \beta( \omega_{nfs} +\mu_q) }}{g^{(-)}}        }, \label{Gap_pb}
\end{gather}
where, 
\begin{gather}
f^+ \fnppbT = \frac{	\FB{\F + 2\Fb e^{-\beta( \omega_{nfs}  -\mu_q ) }} e^{-\beta( \omega_{nfs}  -\mu_q ) } + e^{-3\beta( \omega_{nfs}  -\mu_q ) }  	}{ 1 + 3 \FB{  \Phi + \Fb e^ { - \beta ( \omega_{nfs} -\mu_q )} } e^ { - \beta ( \omega_{nfs} -\mu_q )} + e^ { - 3\beta ( \omega_{nfs} -\mu_q )}  }~,\label{eq.fplus} \\
f^- \fnppbT = \frac{	\FB{\Fb + 2\F e^{-\beta( \omega_{nfs}  +\mu_q ) }} e^{-\beta( \omega_{nfs}  +\mu_q ) } + e^{-3\beta( \omega_{nfs} + \mu_q ) }  	}{ 1 + 3 \FB{  \Fb + \F e^ { - \beta ( \omega_{nfs} +\mu_q )} } e^ { - \beta ( \omega_{nfs} +\mu_q )} + e^ { - 3\beta ( \omega_{nfs} +\mu_q )}  }~, \label{eq.fminus} \\
\Xi_{nfs} = \begin{cases}
1 - \dfrac{ \kappa_f \mb{e_f B}}{M} ~~~~~\text{if}~~ n=0, \vspace{0.1 cm} \\
1 - s ~{\rm sgn} (e_f) \dfrac{(\kappa_f e_f B)^2 - p_z \mu_5 \kappa_f e_f B / M}{
	\sqrt{  \FB{ M^2 + 2 n  \mb{e_fB} } \FB{\kappa_f e_f B}^2 + \FB{ p_z^2 + 2 n  \mb{e_fB} } \mu_5^2  -  2 M p_z \mu_5 \kappa_f e_f B  }} 
~~~~~\text{if}~~ n> 0~.
\end{cases}
\end{gather}
We need to solve Eqs.~\eqref{Gap_M} to \eqref{Gap_pb} self-consistently to obtain $ T $, $ \mu_q $, $\mu_5$, and $B$ dependence of $ M, \Phi  $ and $ \bar{\Phi} $. Note that in Eqs.~\eqref{Omega_PNJL} and ~\eqref{Gap_M}, the medium independent integral is ultraviolet (UV) divergent. Since the theory is known to be non-renormalizable owing to the point-like interaction between the quarks, a proper regularization scheme is necessary. It is well-known that sharp momentum cutoff scheme suffers from regularization artifact due to replacement of the continuum momentum by discrete Landau quantized one~\cite{Fukushima:2010fe}. To avoid this, a smooth regularization procedure following~\cite{Fukushima:2010fe} is employed in this work by introducing a multiplicative form factor 
\begin{equation}
f_\Lambda (p) = \sqrt{\frac{\Lambda^{2N}}{\Lambda^{2N} + \mb{\texttt{p}}^{2N}}}~,
\label{FormFactor}
\end{equation}
in the diverging vacuum integrals leaving the convergent medium dependent part unaltered. In Eq.~\eqref{FormFactor} 
\begin{equation}
\texttt{p} = \begin{cases}
\MB{\vec{p}} ~~~~ \text{if}~~ eB = 0~, \vspace{0.1cm} \\
\sqrt{p_z^2 + 2n \mb{e_f B}}  ~~\text{if}~~ eB \ne 0~.
\end{cases}
\label{Defn_p}
\end{equation}


The chiral charge density $ n_5 $ ( i.e. the difference between the densities of right and left handed particles) is defined as
\begin{eqnarray}
n_5 = -\FB{\frac{\del\Omega}{\del \mu_5}} = 3 \sum_{nfs} \alpha_{ns}\frac{\mb{e_f B}}{2\pi} \kzint{p} \frac{\mu_5}{\omega_{nfs}}~ \Xi^{\mu_5}_{nfs} \TB{1 - f^+\fnppbT - f^- \fnppbT}~,
\end{eqnarray}
where, 
\begin{equation}
\Xi^{\mu_5}_{nfs} = \begin{cases}
1  - \dfrac{p_z}{\mu_5} ~~~~\text{if }~~ n = 0~, \\
1 - s ~{\rm sign} (e_f) \dfrac{ \fb{p_z^2 + 2n \mb{e_f B} }- M p_z \kappa_f e_f B / \mu_5 }{
	\sqrt{  \fb{ M^2 + 2 n  \mb{e_fB} } \fb{\kappa_f e_f B}^2 + \fb{ p_z^2 + 2 n  \mb{e_fB} } \mu_5^2  -  2 M p_z \mu_5 \kappa_f e_f B  }} 
~~~~\text{if }~~ n>0~.
\end{cases}
\label{Xi_mu5}
\end{equation}

\section{NUMERICAL RESULTS}\label{sec.results}

In this section, we present numerical results for constituent quark mass, the expectation values of the Polyakov loop and chiral charge density in different physical situations. As already discussed in Sec.~\ref{sec_CQM_TD}, due to the four-fermion contact interaction among the quarks, NJL model is known to be non-renormalizable and we have used a smooth regularization scheme following Ref.~\cite{Fukushima:2010fe} to get rid of the divergent vacuum integrals. We have fixed the parameters such that phenomenological vacuum values of the chiral condensate $\ensembleaverage{\bar{u} u}^{1/3} = \ensembleaverage{\bar{d} d}^{1/3}=-243.5$ MeV, pion decay constant $ f_\pi = 93 $ MeV, pion mass $ m_\pi = 138 $~MeV and the magnetic moment of the nucleons $ \mu_\text{proton} \simeq 2.7928 ~\mu_\text{N}$ and $ \mu_\text{neutron}\simeq -1.9130 ~\mu_\text{N} $ are reproduced. This yields the following values of our model parameters:  $ \Lambda = 568.7 $ MeV, $ G \Lambda^2 = 1.857 $ and $ m_0 = 5.6 $~MeV. The vacuum values of AMM of the quarks are $ \kappa_u = 0.02399,~\kappa_d = 0.09595  $  in units of  $ ~\rm GeV^{-1}$ respectively. It should be noted that, the chosen model parameters ensures that, at $B=T=0$, the relation $(F_2^u \sim F_2^d)\simeq 0.05$ is satisfied which guarantees the isospin symmetry~\cite{Fayazbakhsh:2014mca,Bicudo:1998qb}, where $ F_2^{f} $ represents the magnetic form factor of a quark of flavour $ f $. While showing the numerical results,  we will take quark chemical potential $ \mu_q = 0 $. 
%


\begin{figure}[h]
	\begin{center}
		\includegraphics[angle=-90,scale=0.34]{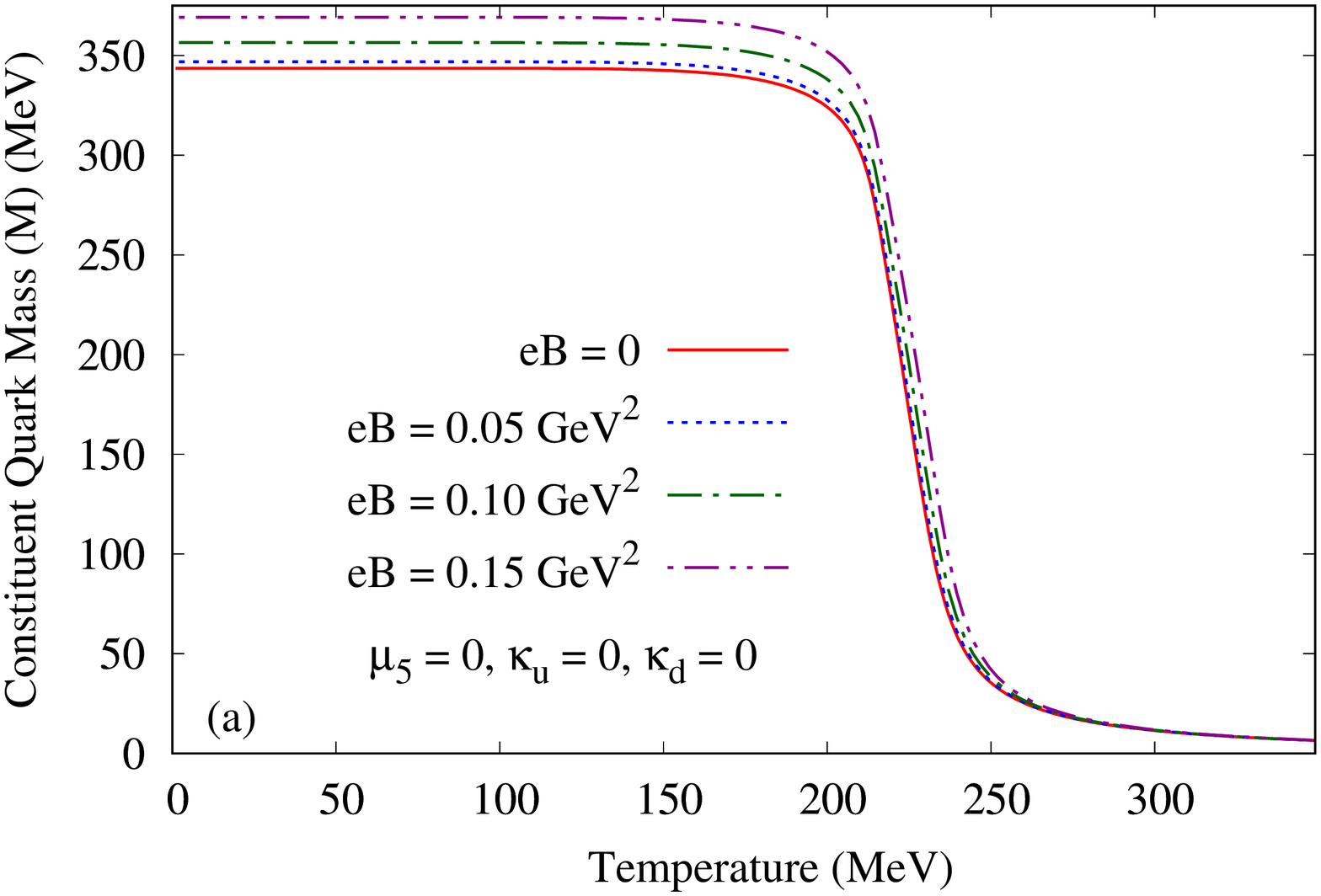}  \includegraphics[angle=-90,scale=0.34]{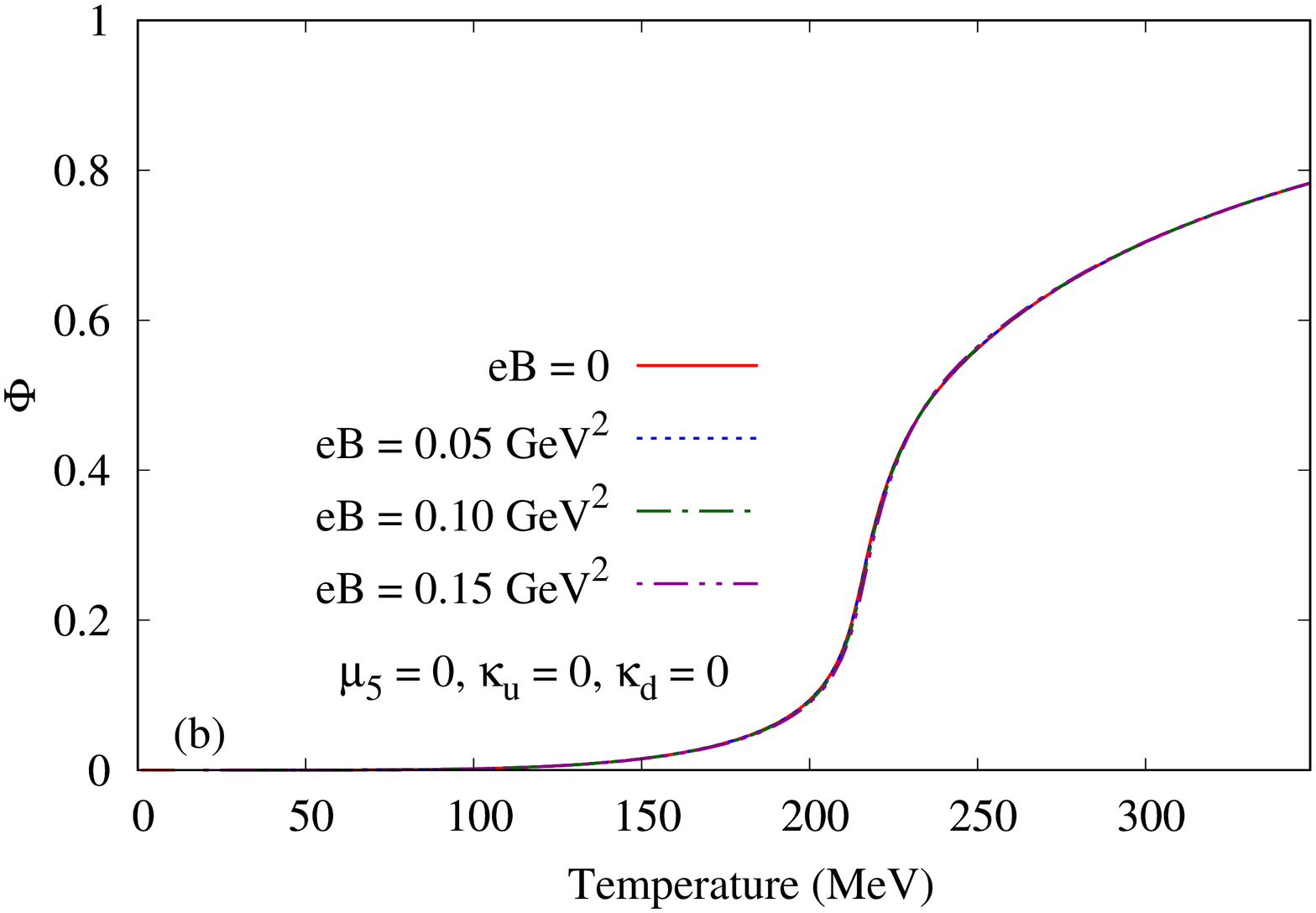} \\
	\end{center}
	\caption{(Color Online) Variation of (a) the constituent quark mass ($ M $) and (b) the expectation value of the Polyakov loop ($ \Phi $) as a function of temperature for background magnetic fields $eB = 0.0, 0.05,0.10  $ and $0.15$~GeV$ ^2 $ respectively, at $ \mu_5 = 0 $, without considering the finite values of AMM of the quarks. }
	\label{fig.MPhi1}
\end{figure}

In Figs.~\ref{fig.MPhi1}(a) and (b) we have shown the temperature dependence of the constituent quark mass ($ M $) and the expectation value of the Polyakov loop ($ \Phi $) for different values of the background magnetic fields ($eB = 0.0, 0.05, 0.10 $ and $0.15$~GeV$ ^2 $ respectively) at zero chiral chemical potential ($ \mu_5 $) without considering the finite values of AMM of the quarks. From Fig.~\ref{fig.MPhi1}(a), it can be observed that, in all the cases, $ M $ almost remains unchanged up to $ T \approx 150 $ MeV, starts decreasing rapidly in a small range of $ T $ and finally goes to the bare mass limit at higher values of $ T $, implying  the transition from chiral symmetry broken (with $ M \ne 0 $) to the restored phase (i.e., $ M \approx m \approx 0 $), is a smooth crossover. Note that, as we have considered non-vanishing current quark mass, $ m_0 =  5.6  $ MeV, the chiral symmetry is only partially restored.  Moreover, for stronger values of the magnetic field, $ M $ increases as $ T \rightarrow 0  $ and the transitions to the symmetry restored phase take place at the larger values of temperature. This is the well-known phenomena of magnetic catalysis (MC)~\cite{Shovkovy:2012zn,Gusynin:1994re,Gusynin:1995nb,Gusynin:1999pq}, which shows that the magnetic field has a strong tendency to enhance (or catalyze) spin-zero fermion-antifermion ($ \AB{\psibar \psi} $) condensates. From Fig.~\ref{fig.MPhi1}(b), it is evident that, the decofinement crossover is marginally affected by the presence of the magnetic field.  For all values of $ eB $, Polyakov loop ($ \Phi $) expectation value remains vanishingly small at  lower values of $ T $ implying a `confined' state, then it starts increasing around  $ T \approx 150 $~MeV and finally reaches $ \simeq 1$ at high $ T $ values indicating `deconfiend' state of matter. This behaviour of $ M  $ and $  \Phi $ mentioned above are consistent with the previous results for NJL-type models as can be seen in Refs.~\cite{Fukushima:2010fe,Chaudhuri:2019lbw,Chaudhuri:2020lga,Ghosh:2020qvg,Andersen:2014xxa,Kharzeev:2012ph,Fayazbakhsh:2012vr,Fayazbakhsh:2014mca,Avancini:2018svs,Gatto:2010qs}.

\begin{figure}[h]
	\begin{center}
		\includegraphics[angle=-90,scale=0.34]{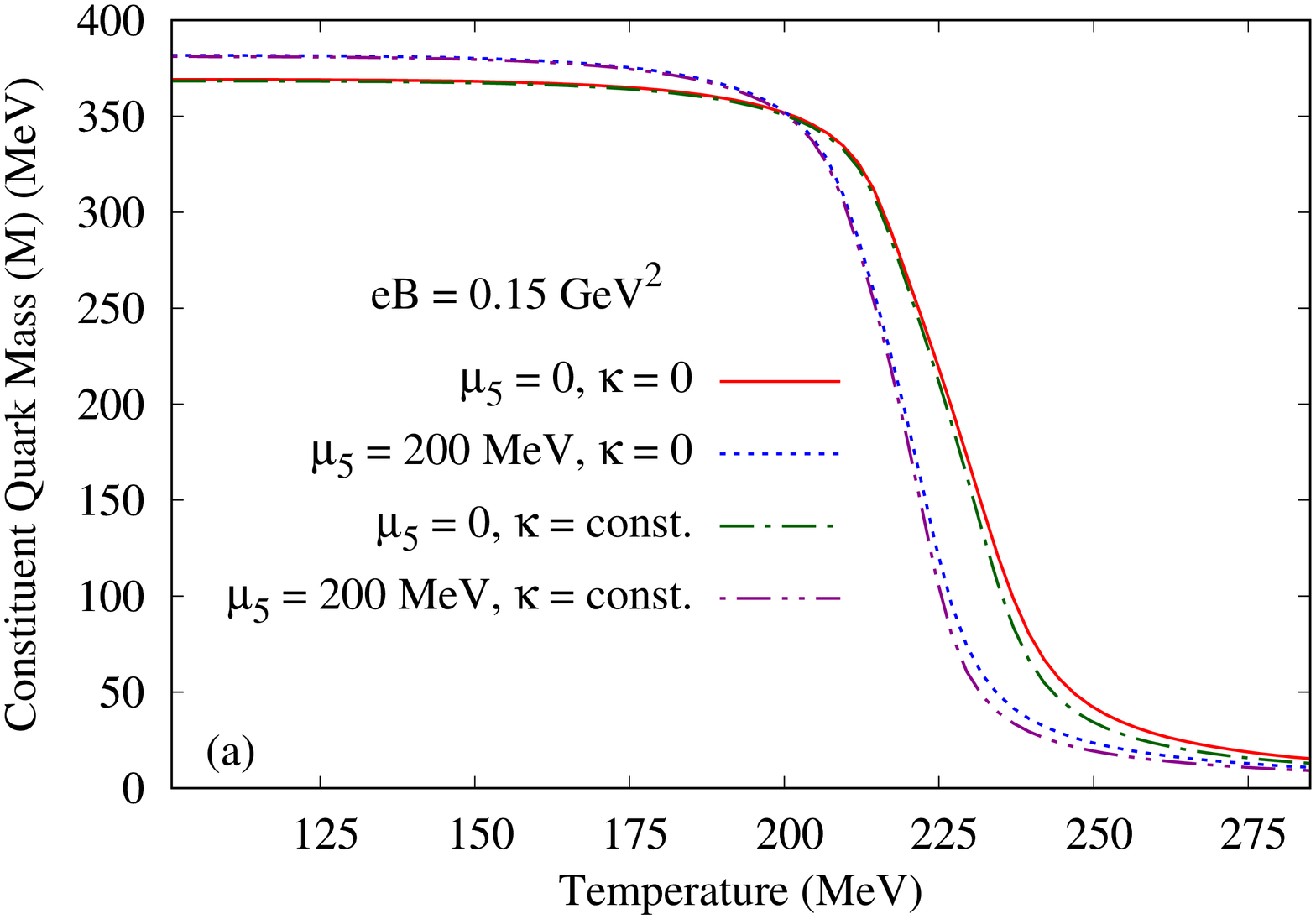}  \includegraphics[angle=-90,scale=0.34]{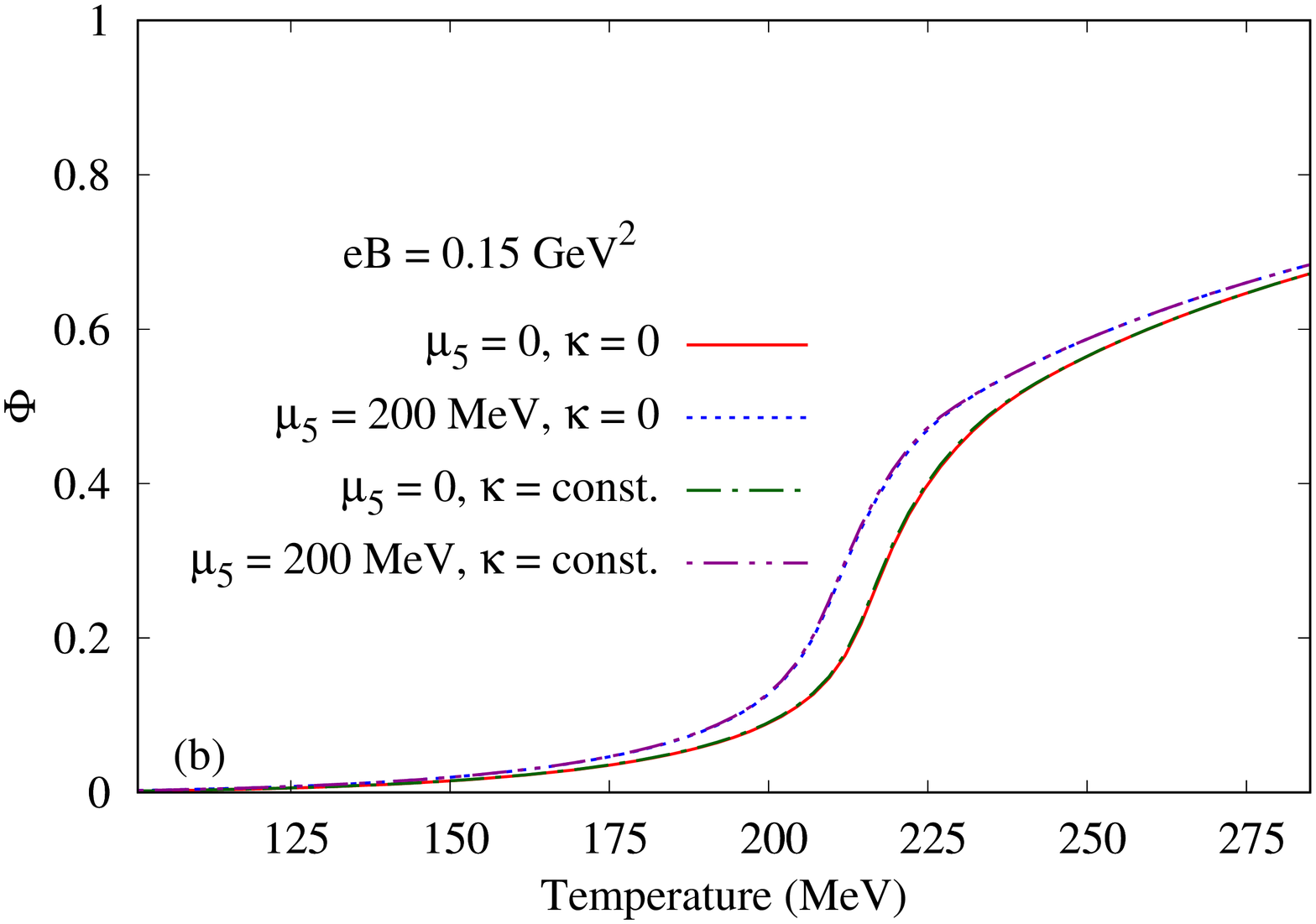} 
	\end{center}
	\caption{(Color Online) Variation of the constituent quark mass ($M$) and polyakov loop ($ \Phi $) as a function of temperature for a fixed value of the external magnetic field using 2-flavour PNJL model. In the cases of  non-vanishing  AMM, constant  $\kappa_u$ and $\kappa_d$  have been considered throughout the temperature range with values 0.024 GeV$^{-1}$ and 0.096 GeV$^{-1}$ respectively.  }
	\label{fig.MPhi2}
\end{figure}

In Figs.~\ref{fig.MPhi2}(a) and (b) the temperature dependence of $ M $ and $ \Phi $ respectively is depicted for non-zero values of $ \mu_5 $ with and without considering the finite values of the AMM of the quarks for $ eB = 0.15 $~GeV$ ^2 $. From Fig.~\ref{fig.MPhi2}(a), it is evident that the inclusion of finite values of $ \mu_5 $, with AMM turned off (blue dotted line), the constituent mass slightly increases compared to $ \mu_5 = 0 $ case (red solid line) at lower values of $ T $ but goes to the current quark mass limit at smaller values of $ T $ (compared to the case when $ \mu_5 = 0 $) indicating inverse magnetic catalysis~\cite{Fukushima:2010fe}.  The similar effect is also observed in deconfinement crossover transition when $ \mu_5 $ is turned on. The green-dash-dot line of Fig.~\ref{fig.MPhi2}(a) represents the $ T $-dependence of  $M $ at $ \mu_5 = 0$ when the finite values of the AMM of the quarks are switched on in presence of the constant background magnetic field $ eB = 0.15  $~GeV$ ^2 $.  It can be seen that, consideration of non-zero values of the AMM of the quarks leads to a marginal decrease in $ M $ at lower values of $ T $ and chiral symmetry restoration occurs at smaller values of $ T $ compared to the zero AMM case. However, inclusion of finite values of the AMM of the quarks at $ \mu_5 = 0 $ does not bring any appreciable change in the $ T $-dependence of $ \Phi $  which can be observed in Fig.~\ref{fig.MPhi2}(b). Consequently, the deconfinement crossover transition remains unaltered. In Fig.~\ref{fig.MPhi2}(a), we have depicted the variation of $ M $ as a function of $ T $ at $ \mu_5 = 200 $ MeV with considering the finite value of the AMM of the quarks (see purple-dash-dot-dot line). The combined effect of non-zero values of  AMM and $ \mu_5 $ results in slight increase in $ M $ at low $ T $ values compared to the scenario when both of them are absent. 
Moreover, the IMC of chiral transition temperature is enhanced further as the restoration of chiral symmetry takes place at even lower values of $ T $ compared to the case when only finite values of  chiral chemical potential are taken into consideration. As has already been discussed switching on finite values of AMM of the quarks results in negligible change in the deconfinement crossover transition. The $ T $-dependence of $ \Phi $, when both AMM and $ \mu_5 $ are turned on, is almost similar to the case when only finite  chiral chemical potential is considered.


\begin{figure}[h]
	\includegraphics[angle= -90, scale = 0.34]{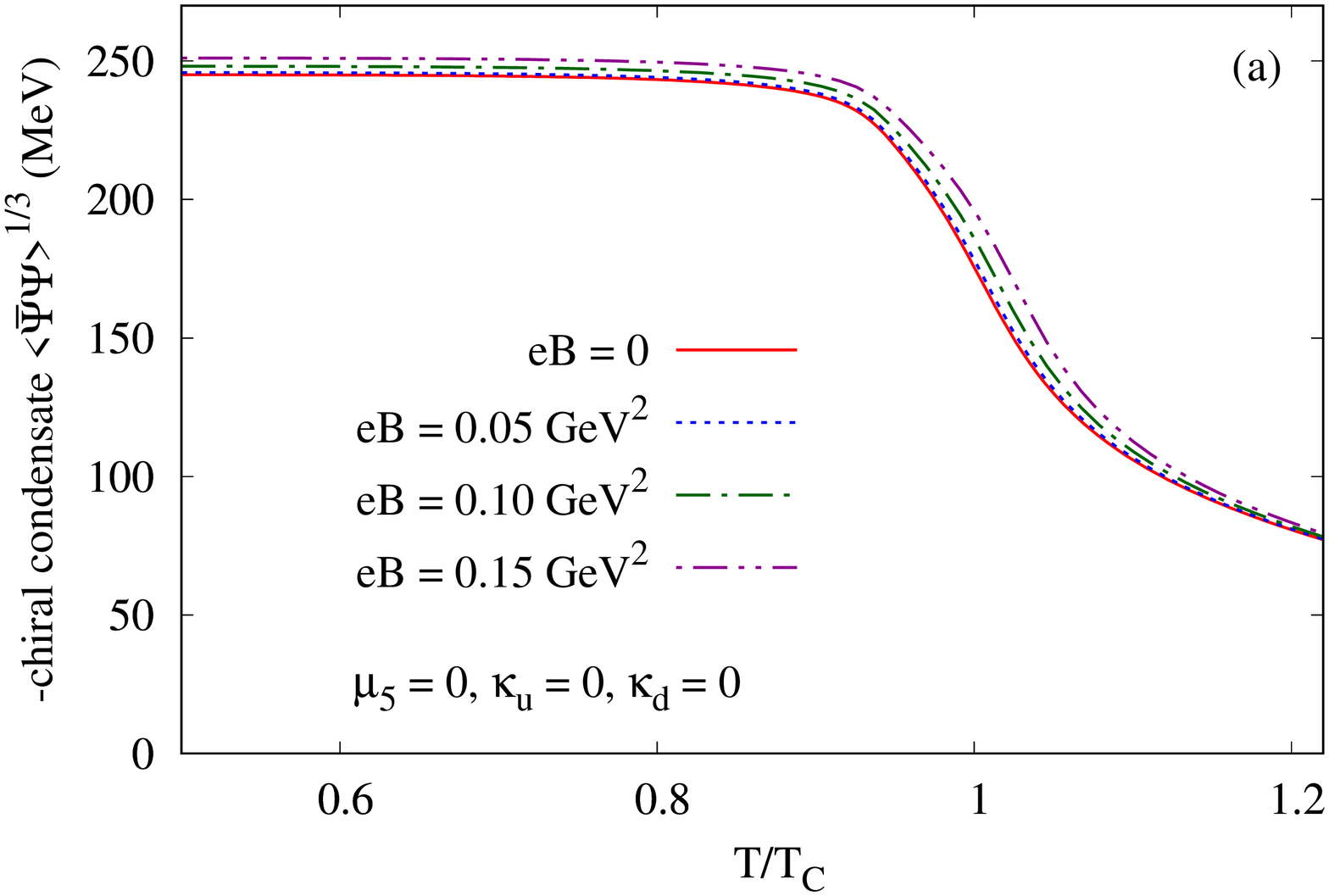}
	~~~\includegraphics[angle= -90, scale = 0.34]{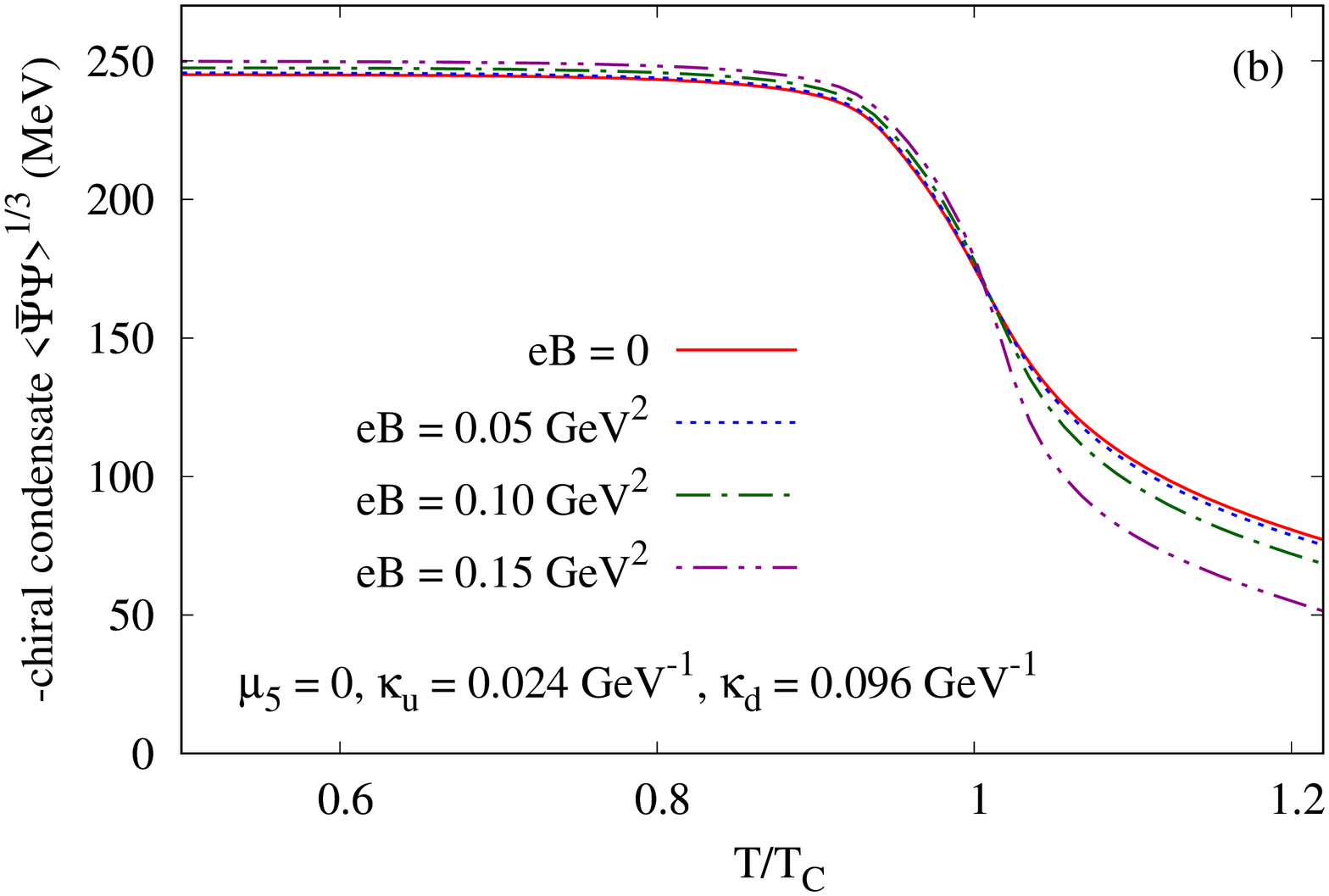}\\
	\includegraphics[angle= -90, scale = 0.34]{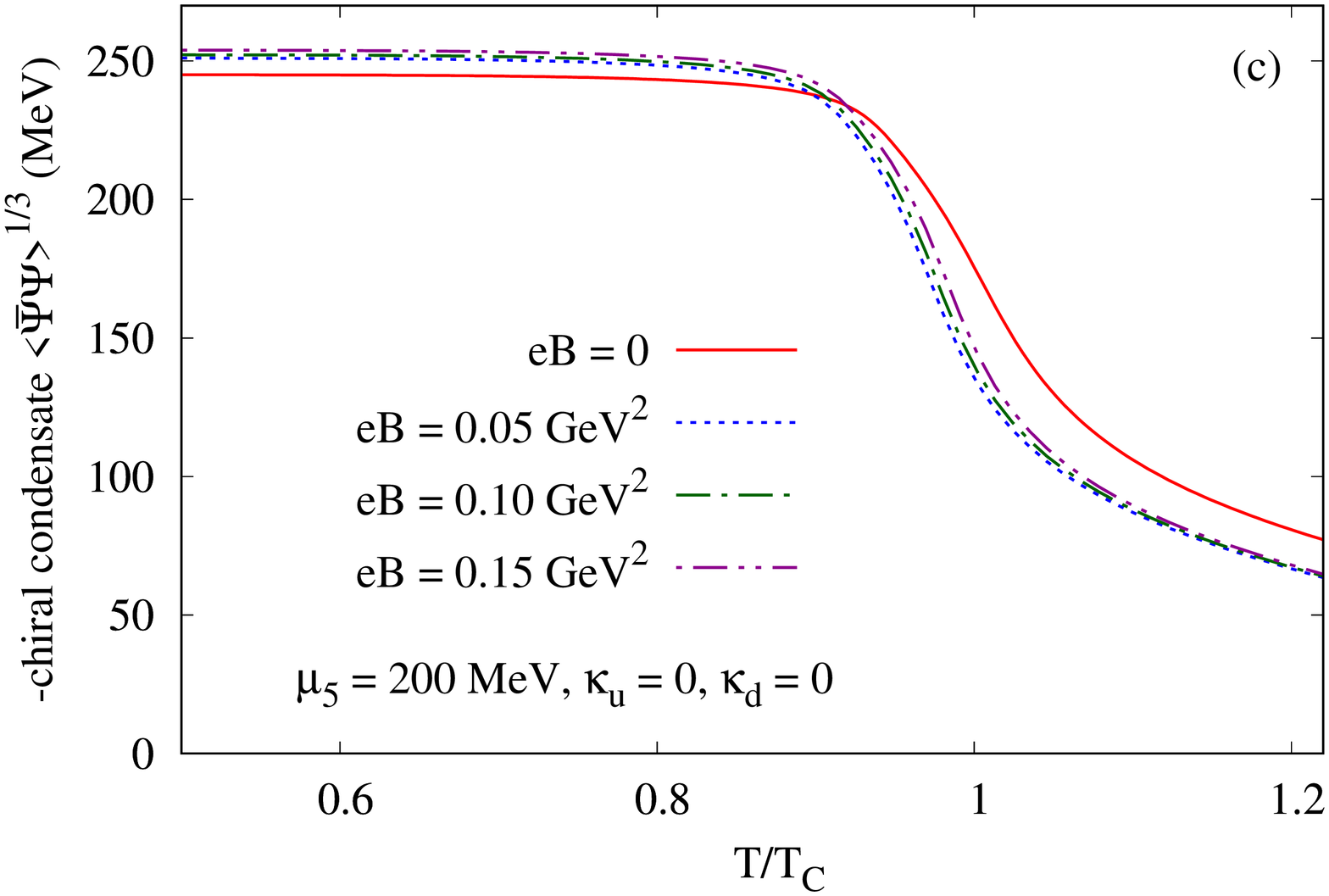}
	~~~\includegraphics[angle= -90, scale = 0.34]{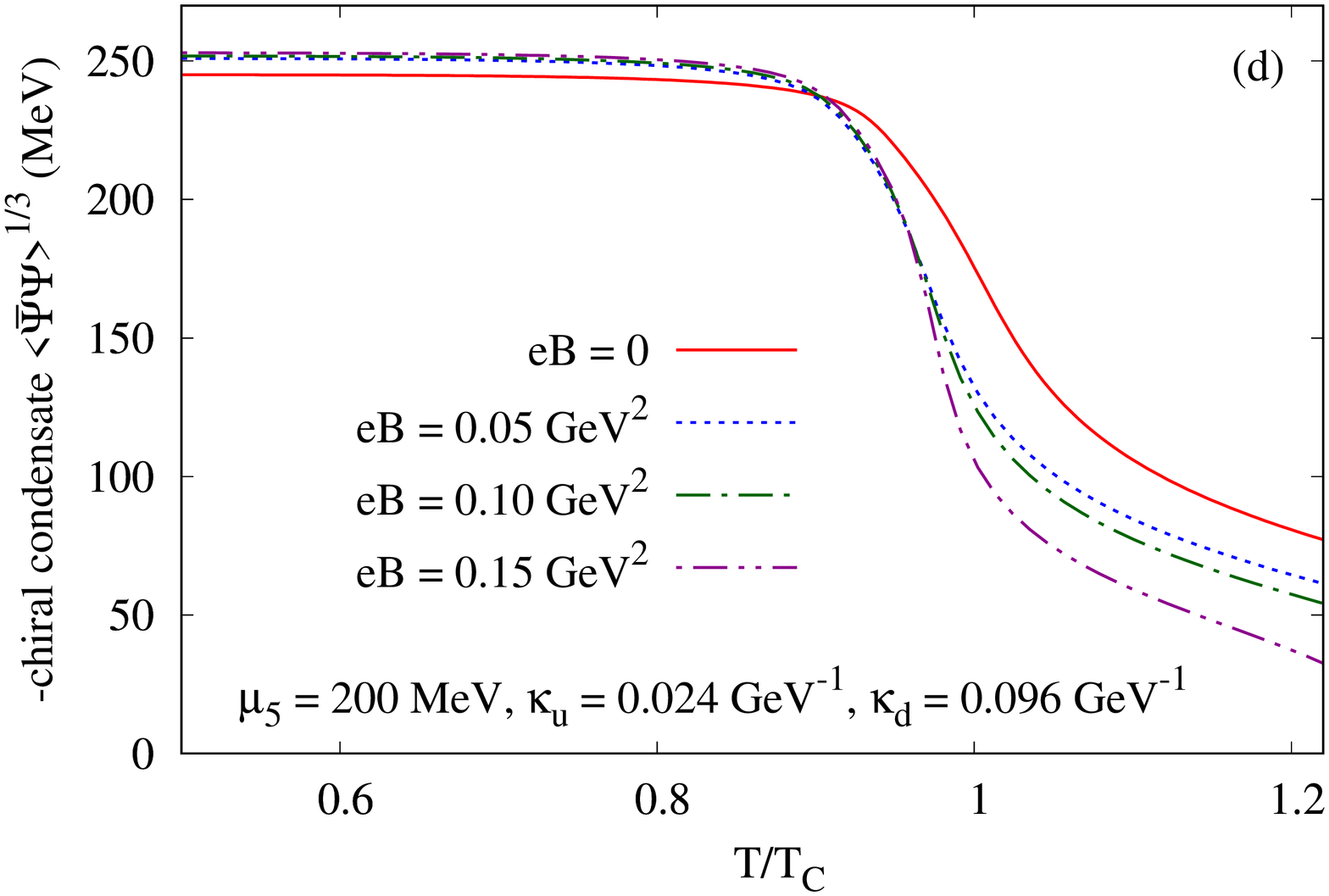}
	\caption{The variation of chiral condensate as function of the ratio $ T/T_C $  for different values of $ eB $, considering zero and finite values of AMM of the quarks as well as CCP. Here $ T_C = 229 $ MeV.}
	\label{Fig_Ref_1}
\end{figure}

\begin{figure}[h]
 	\begin{center}
 		\includegraphics[angle=-90,scale=0.34]{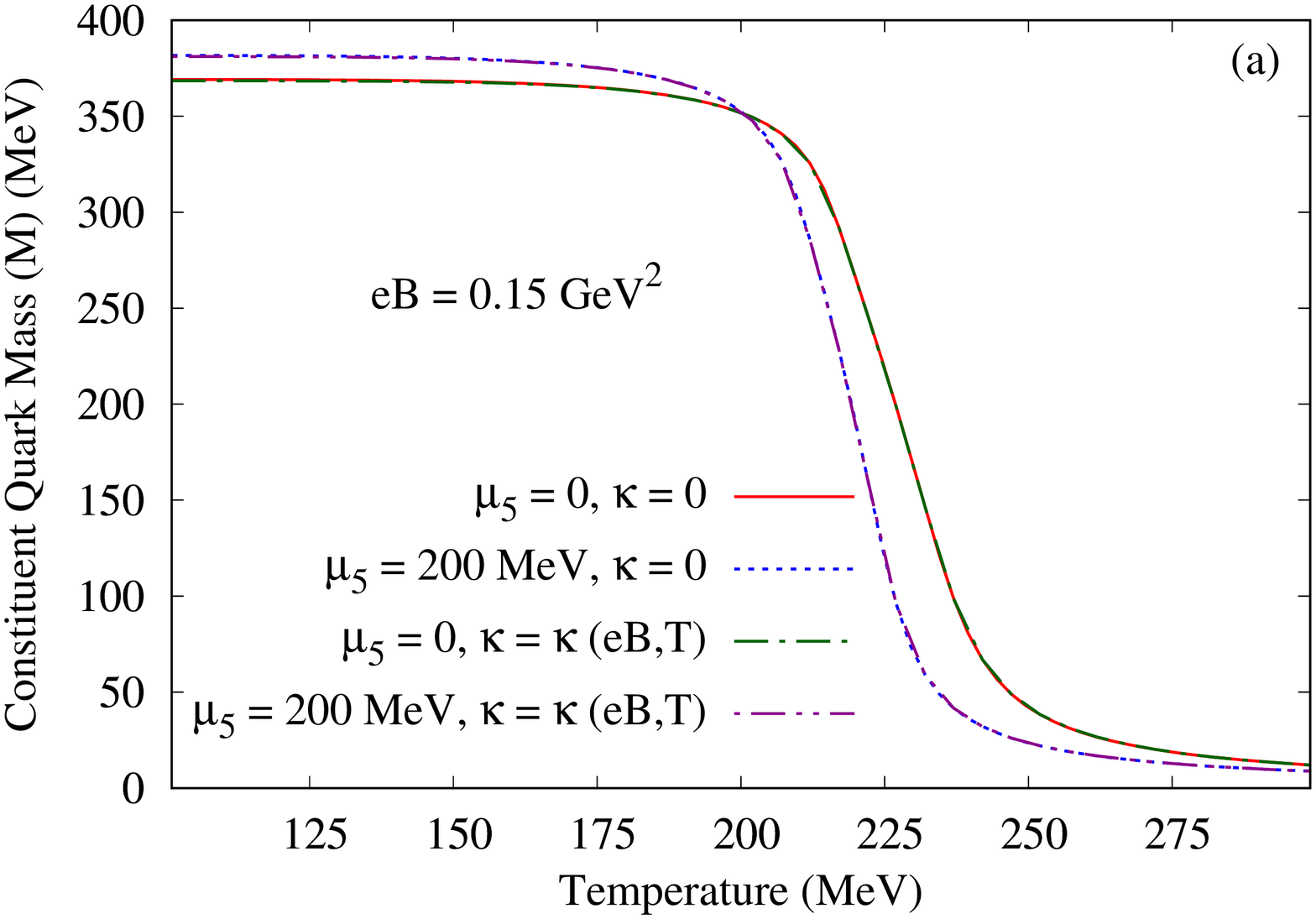}  \includegraphics[angle=-90,scale=0.34]{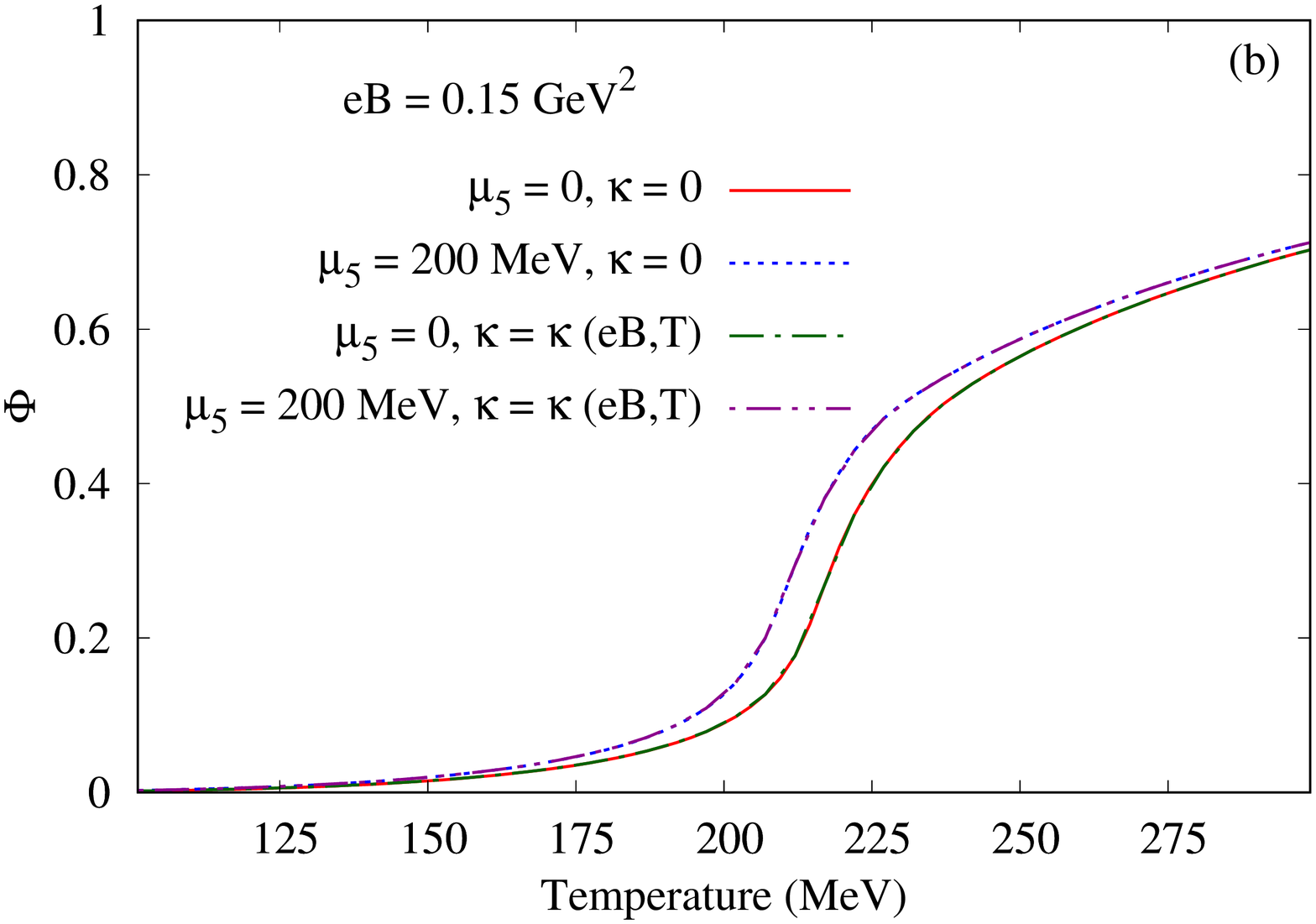} 
 	\end{center}
 	\caption{(Color Online) Variation of the constituent quark mass ($M$) and polyakov loop ($ \Phi $) as a function of temperature for different values of external magnetic field   using  2-flavour PNJL model considering $ T $ and $ eB $ dependent values of the AMM of the quarks following~\cite{Ghosh:2021dlo}. At $eB = 0.15$ GeV$^2$, as we increase the temperature from 100 MeV to 300 MeV, the values of $\kappa_u$ and $\kappa_d$ respectively varies from 0.023 GeV$^{-1}$  and  0.095 GeV$^{-1}$ to 0. }
 	\label{fig.MPhi3}
 \end{figure}

To compare the  individual and the combined effects of $\mu_5$ and AMM systematically on MC and IMC of the chiral transition temperature ($ T_C $), in Fig.~\ref{Fig_Ref_1} we have shown the variation of the chiral condensate as a function of scaled temperature $T/T_C$ for a different values of $eB$.   
In  Fig~\ref{Fig_Ref_1} (a), we first consider the temperature dependence of the chiral condensate without the $ \mu_5 $ and  AMM effects. Here, with increasing $ eB $ values, an enhancement of the condensate is observed throughout the temperature range. The enhancement in the low temperature regime is the  indication of  the well known  MC effect ~\cite{Gusynin:1994re,Gusynin:1999pq,Fayazbakhsh:2012vr,Fayazbakhsh:2014mca,Avancini:2018svs,Chaudhuri:2019lbw}. On the other hand, concentrating on the transition region, one can observe an increasing trend of the transition temperature with $eB$ which is consistent with the earlier works. 
In Fig~\ref{Fig_Ref_1} (b), the AMM of the quarks is taken into consideration in absence of chiral imbalance. It can be noticed that in the low temperature regime, the qualitative behaviour  remains similar to the previous case. However, at higher temperatures, as we increase the $eB$ value, we observe  a decreasing trend of the condensate. The suppression becomes more prominent  above the transition regime and there, the magnitude of the chiral condensate  decreases substantially.  Near $T_C$, the change is  marginal and  the system goes to the symmetry restored phase at slightly smaller values of temperature, leading to IMC~\cite{Fayazbakhsh:2014mca,Chaudhuri:2019lbw,Chaudhuri:2020lga,Bali:2011qj}.
It should be mentioned here that the phenomenological values of the AMM considered in Refs.~\cite{Fayazbakhsh:2014mca,Chaudhuri:2019lbw,Chaudhuri:2020lga} are different than  those of the present article. Here we have followed~\cite{Ghosh:2021dlo}, so that we can incorporate the thermo-magnetic modification of the anomalous magnetic moment of quarks which will be discussed in Figs~\ref{fig.MPhi3} (a) and (b). As a result, the quantitative effects of finite values of AMM  are less prominent compared to what has been observed in Refs.~\cite{Fayazbakhsh:2014mca,Chaudhuri:2019lbw,Chaudhuri:2020lga}. Moreover, to regularize the medium independent integrals, a magnetic field dependent sharp cutoff scheme has been used in Refs.~\cite{Chaudhuri:2019lbw,Chaudhuri:2020lga}, which is known to suffer from  regularization artifact ~\cite{Fukushima:2010fe,Ghosh:2021dlo}. To avoid this issue, a smooth regularization procedure has been implemented in the present article, yielding quantitatively different results than Refs.~\cite{Chaudhuri:2019lbw,Chaudhuri:2020lga}. However, the qualitative nature remains compatible with the previous studies.  
Let us now concentrate in Fig.~\ref{Fig_Ref_1} (c) where we have considered $ \mu_5 = 200 $~MeV ignoring the AMM of the quarks. Again, at lower values of temperature,  the chiral condensate increases (though marginally) with $eB$ which is qualitatively  similar to the previous cases. However, the transition from chiral symmetry broken phase to the restored phase occurs at substantially smaller values of temperature indicating IMC. Finally, in Fig.~\ref{Fig_Ref_1} (d), keeping $ \mu_5$ fixed at $200 $~MeV,  the AMM of the quarks are incorporated. We observe that the combined effect  leads to a mild enhancement in the IMC of chiral transition temperature. Thus the dominance of chiral chemical potential on the IMC of chiral transition temperature over the AMM of the quarks can be inferred.


\begin{figure}[h]
	\begin{center}
		\includegraphics[angle=-90,scale=0.34]{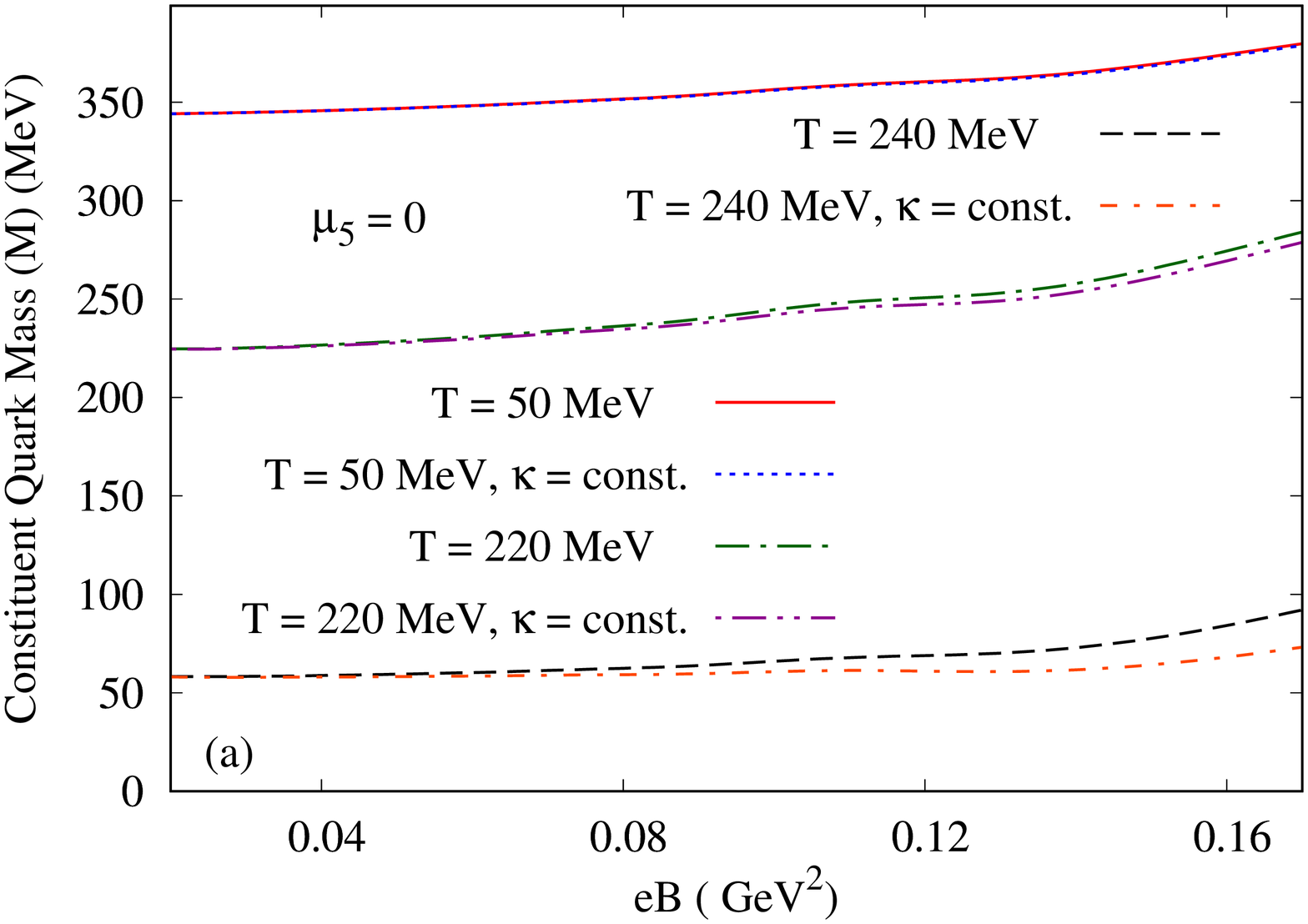}  		\includegraphics[angle=-90,scale=0.34]{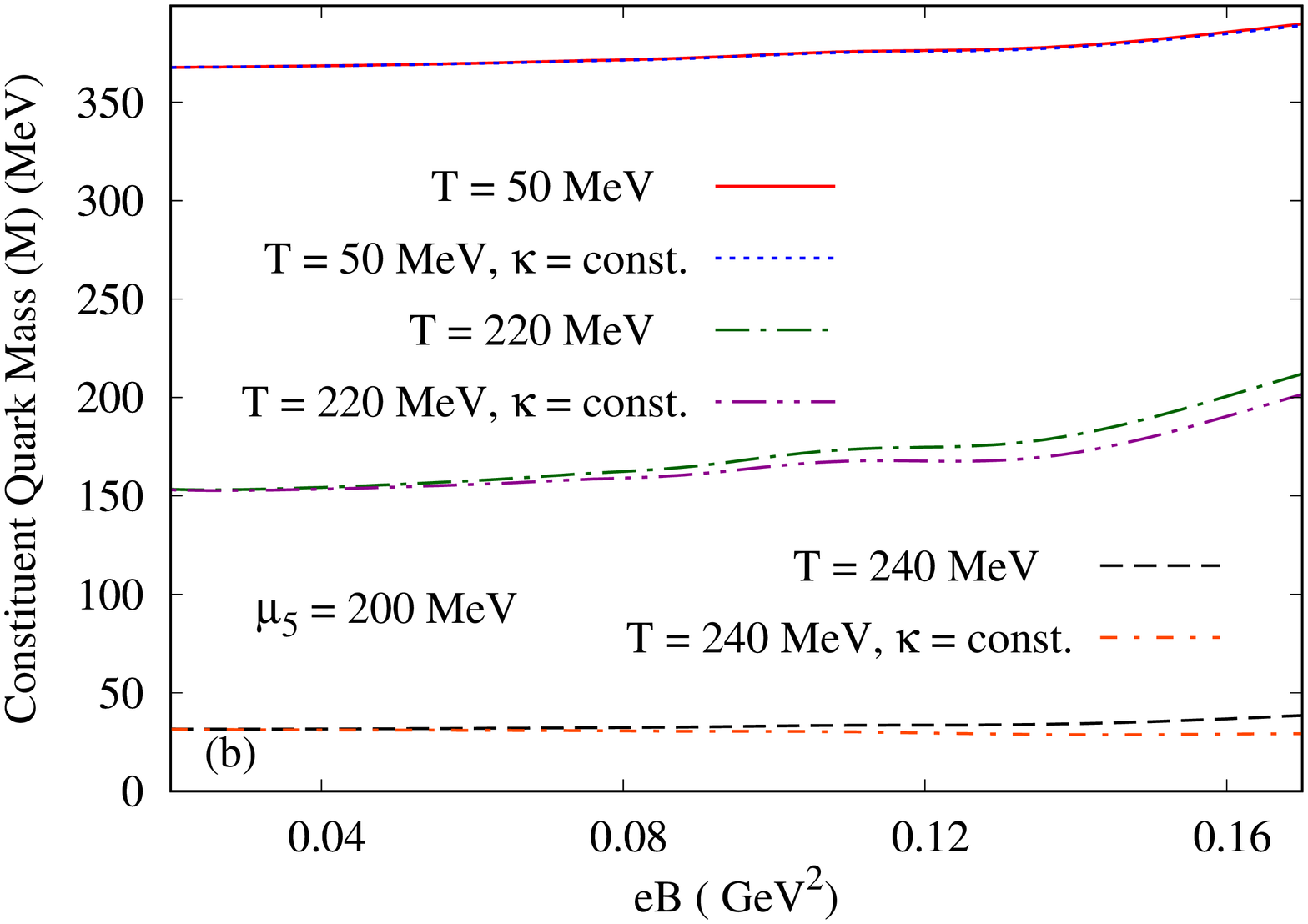}  
		\end{center}
	\caption{(Color Online) Variation of the constituent quark mass ($M$) as a function of $ eB $ for different values of temperature and chiral chemical potential using  2-flavour PNJL model. In the cases of  non-vanishing  AMM, constant  $\kappa_u$ and $\kappa_d$  have been considered  with values 0.024 GeV$^{-1}$ and 0.096 GeV$^{-1}$ respectively.}
	\label{fig.MeB}
\end{figure}

In Figs.~\ref{fig.MeB}(a) and (b) we have plotted $ eB $-dependence of $ M $ at $ \mu_5  = 0$ and  $ 200  $ MeV respectively with and without considering AMM of quarks for three different values of temperatures. At $ T = 50 $ MeV, it is evident from both the figures that, the constituent quark mass increases with the increasing values of $ eB $ for both zero and non-zero values of the AMM of the quarks although the magnitude of $ M $ is marginally small in finite AMM case for both zero and non-zero values of $ \mu_5 $. Also, at $ \mu_5 = 200$ MeV for both zero and finite values of the AMM of the quarks, the magnitude of $ M $ is slightly larger compared to the $ \mu_5 =  0$ case at $ T = 50  $ MeV. This behaviour of $ M $ at $ \mu_5 = 0 $ and $ 200 $ MeV as a function of $ eB $ at $ T = 50 $ MeV is consistent with the observations made while discussing Fig.~\ref{fig.MPhi2}(a). Now as we increase the temperature to $ T = 220   $ MeV,  from Fig.~\ref{fig.MeB}(a) it can be seen that, though there is an overall increase in $ M $ with increasing $ eB $, the inclusion of finite values of AMM of the quarks  leads to slight decrease in the constituent mass of the quarks specifically at high $ eB $ values.  However, from Fig.~\ref{fig.MeB}(b), it is evident that,  for a particular value of $ eB $, when finite chiral chemical potential ($ \mu_5 = 200 $ MeV) is taken into consideration, the overall magnitude of $ M $ decreases, independent of zero and non-zero values of the AMM of the quarks when  compared to the $ \mu_5 = 0 $ case. This is an opposite effect in contrast to what we have seen at $ T = 50 $ MeV which is demonstrating the IMC of chiral transition temperature at finite chiral chemical potential as previously discussed. Moreover, it can be seen from Fig.~\ref{fig.MeB}(b) that at $ T = 220 $ MeV and $ \mu_5 =200 $ MeV, as a consequence of switching on AMM of the quarks, the constituent quark mass decreases further compared to the $ \mu_5 = 0  $ case shown in Fig.~\ref{fig.MeB}(a).  Finally, from Fig.~\ref{fig.MeB}(b) it is evident that, at $ T = 240 $ MeV and $ \mu_5 = 200 $ MeV, the increasing trend of constituent quark mass as a function of $ eB $ diappears completely. It is to be noted that, the small oscillations observed in the $ eB $-dependence of constituent quark mass is well-known de Haas-van Alphen (dHvA) effect~\cite{Landau:1980mil} and had also been observed in Refs.~\cite{Fayazbakhsh:2010bh,Fayazbakhsh:2010gc,Fayazbakhsh:2012vr,Fayazbakhsh:2014mca,Ebert:1998gx,Inagaki:2004ih,Noronha:2007wg,Fukushima:2007fc,Inagaki:2004ih,Chaudhuri:2019lbw,Chaudhuri:2020lga}.  It occurs whenever the Landau levels cross the quark Fermi surface. As can be seen in Figs.~\ref{fig.MeB}(a) and (b), the dHvA oscillations get smoothed out with the increase in magnetic field (as LLL dominates) in agreement with Refs.~\cite{Fayazbakhsh:2014mca,Fayazbakhsh:2012vr,Chaudhuri:2019lbw}.

\begin{figure}[h]
	\begin{center}
 \includegraphics[angle=-90,scale=0.34]{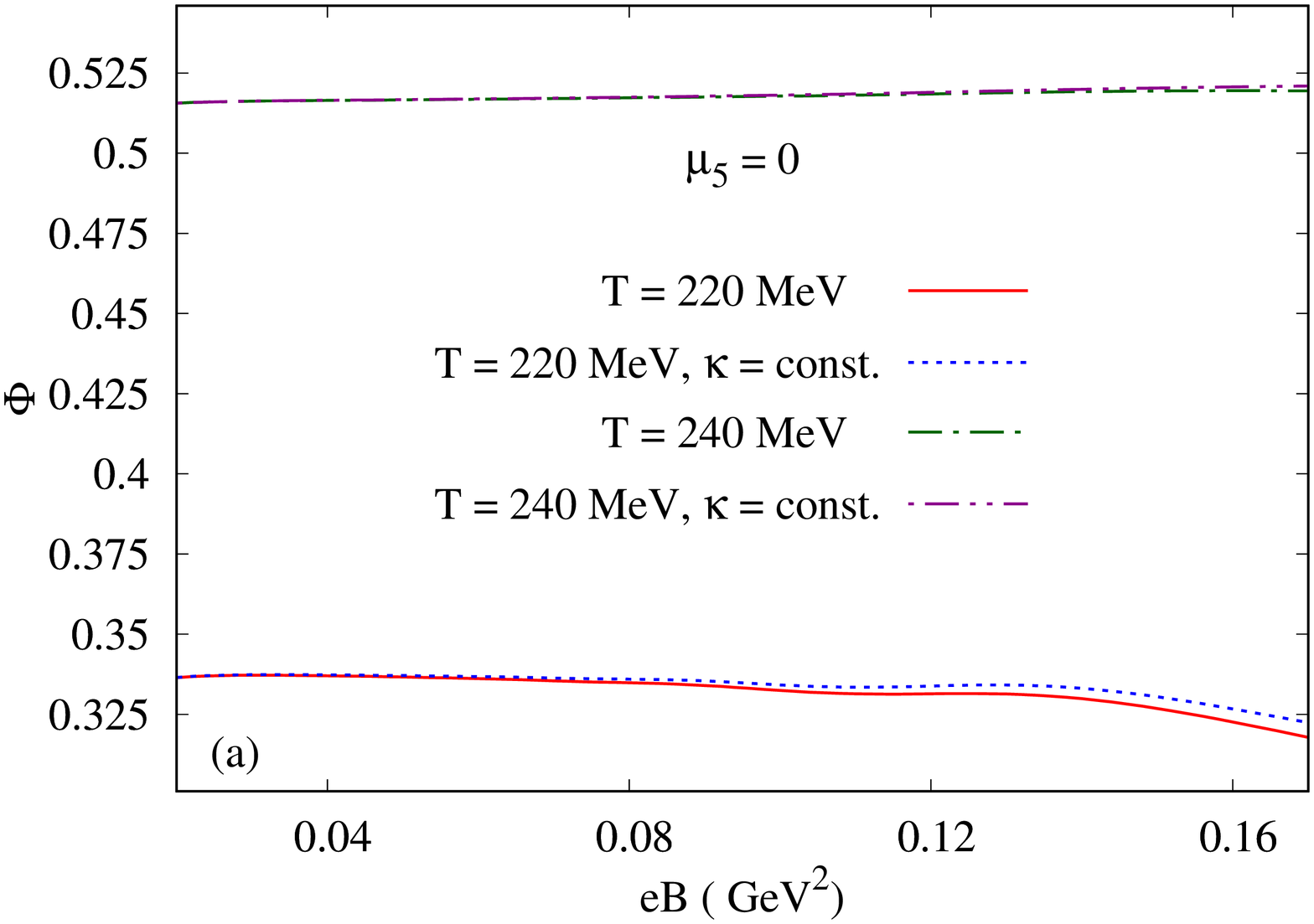} 
 \includegraphics[angle=-90,scale=0.34]{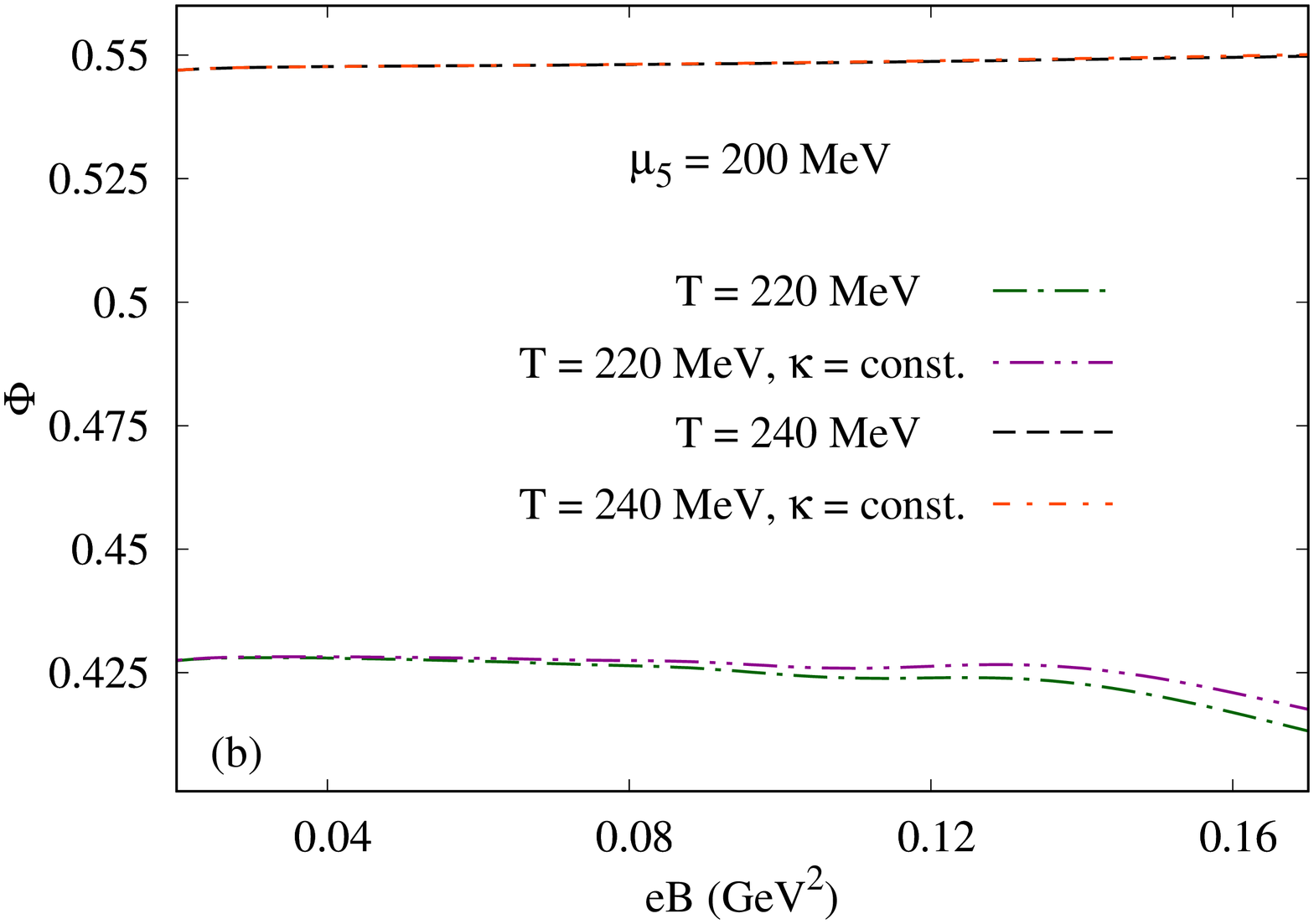} 
	\end{center}
	\caption{(Color Online) Variation of the expectation value of the Polyakov loop ($ \Phi $) as a function of $ eB $ for different values of temperature and chiral chemical potential using  2-flavour PNJL model. In the cases of  non-vanishing  AMM, constant  $\kappa_u$ and $\kappa_d$  have been considered with values 0.024 GeV$^{-1}$ and 0.096 GeV$^{-1}$ respectively.}
	\label{fig.PhieB}
\end{figure}

In Figs.~\ref{fig.PhieB}(a) and (b) we have shown the $ eB $-dependence of $ \Phi $ at $ \mu_5  = 0$ and  $ 200  $ MeV respectively with and without considering AMM of quarks for two different values of temperatures. 
It is evident that, at $T = 220  $ MeV, the magnitude of $ \Phi $ remains almost constant for all $ eB $ values and consideration of finite values of the AMM of the quarks does not lead to any significant modifications in the magnitude of $ \Phi $ for both zero and non-zero values of $ \mu_5 $. 
However at $ T = 240 $ MeV, independent of the values of $ \mu_5 $, when AMM of the quarks is switched on, magnitude of $ \Phi $ decreases at higher values of $ eB $ compared to the scenario when AMM of the quarks are switched off. 
Moreover, unlike $ T  = 220 $ MeV case, $ \Phi $ shows a small oscillatory behaviour at higher values of $ eB $ for both $ \mu_5  = 0$ and  $ 200  $ MeV respectively. 
Comparing Figs.~\ref{fig.PhieB}(a) and (b), it can be seen that, at $ \mu_5 = 200 $~MeV, the magnitude of $ \Phi $, irrespective of the values of $ T $ and AMM of the quarks, are larger compared to $ \mu_5 = 0  $ case which is understandable from the fact that, while discussing Fig.~\ref{fig.MPhi2}(b), we have shown that, finite values of the $ \mu_5  $ results in a decrease in the deconfinement transition temperature .

\begin{figure}[h]
	\begin{center}
		\includegraphics[angle=-90,scale=0.34]{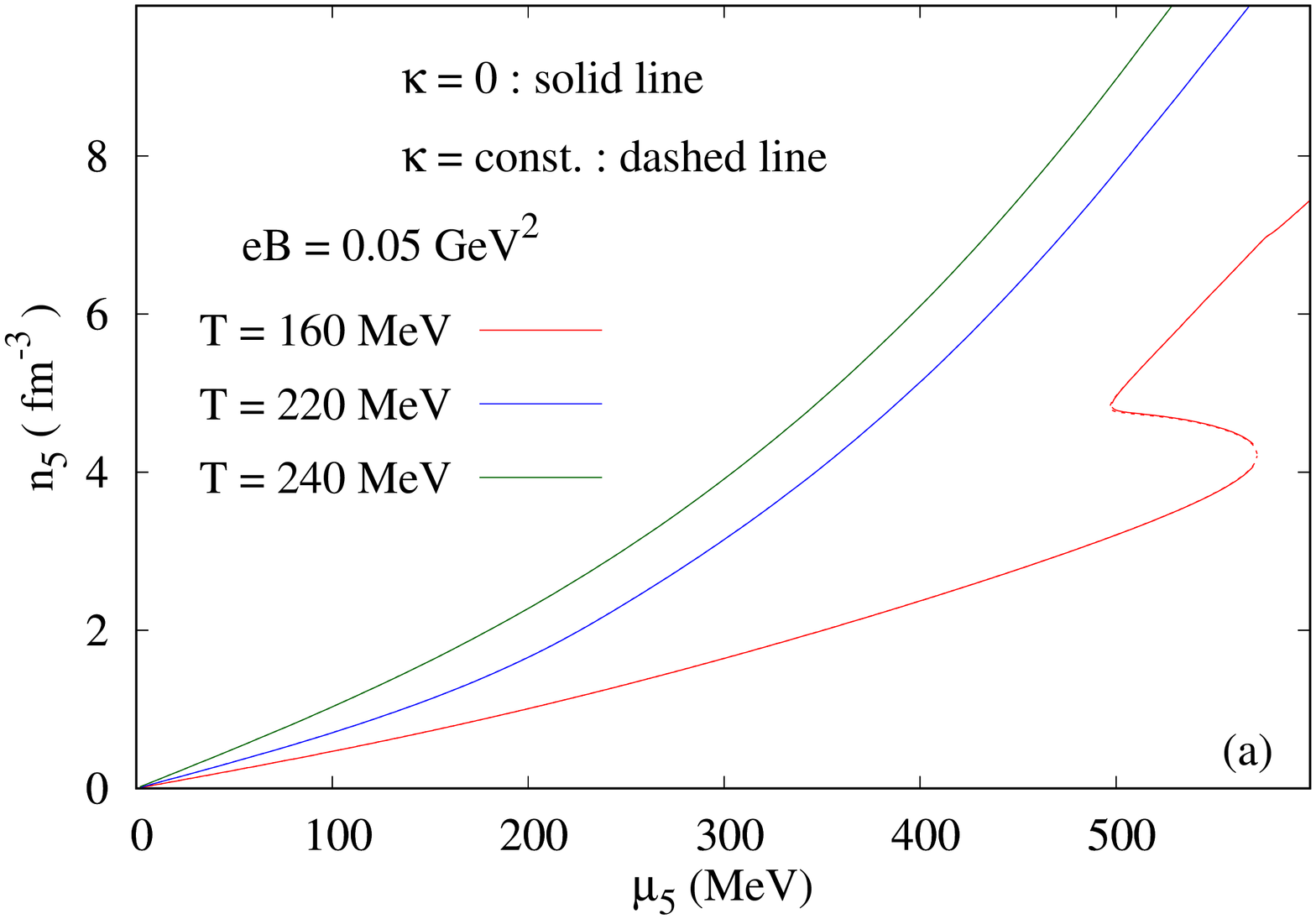} \includegraphics[angle=-90,scale=0.34]{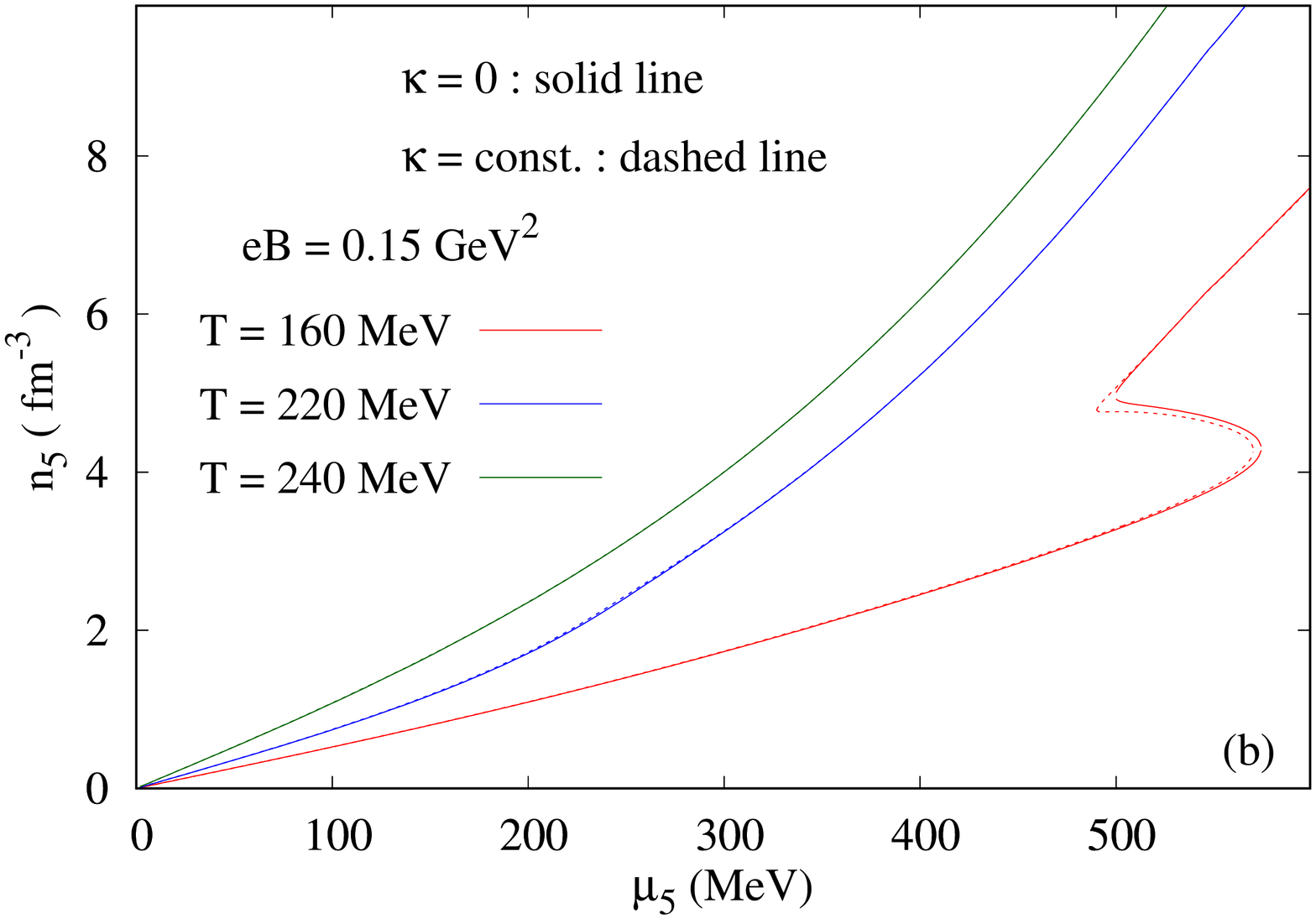} \end{center}
	\caption{(Color Online) Variation of $n_5$ as a function of $ \mu_5 $  using  2-flavour PNJL model. In the cases of  non-vanishing  AMM, constant  $\kappa_u$ and $\kappa_d$  have been considered with values 0.024 GeV$^{-1}$ and 0.096 GeV$^{-1}$ respectively.}
	\label{fig.n5}
\end{figure}

We now focus on the variation of chiral charge density ( $ n_5 $) as function of $ \mu_5 $ and $ T $ which is one of the relevant quantities in studying the chiral magnetic effect. 
In Figs.~\ref{fig.n5}(a) and (b) we have depicted the $ \mu_5 $ dependence of chiral charge density for three different values of $ T $ with and without considering the finite values of the AMM of the quarks for $ eB =  0.05 $ and $ 0.15$~GeV$ ^2 $ respectively.
 It can be observed that as we increase the magnitude of the background magnetic field, the qualitative behaviour of $ n_5 $ as a function of $ \mu_5 $ remains similar for different values of the AMM of the quarks. 
 From both the figures, it is evident that,  at higher values of temperature i.e.  $ T = 220  $ and $ 240  $ MeV,  $ n_5 $ is a monotonically increasing function of $ \mu_5 $ and inclusion of finite values of the  AMM of the quarks does not bring any noticeable difference.
    However, at $ T= 160 $~MeV, at high values of $ \mu_5 $ , chiral charge density becomes multiple valued function of $ \mu_5 $ owing to the several possible solutions of $ M $ and $\Phi $ from gap equations. 
This multivaluedness in the $ \mu_5 $-dependence of $ M $ and $ \Phi  $ is a signature of first order transition and the S-like structure is a typical manifestation of the mixed phase at critical $ \mu_5 $. Moreover, even at $ T = 160 $~MeV, the effects of switching on AMM of the quarks are negligible upto $ \mu_5  \approx 450 $ MeV. Although, for high background magnetic field, i.e. $ eB = 0.15$~GeV$ ^2 $, the enhancement in $ n_5 $ occurs at slightly lower values of $ \mu_5 $. It should be noted that, the above results for zero AMM case are in good agreement with Ref.~\cite{Fukushima:2010fe}.

\begin{figure}[h]
	\begin{center}
		\includegraphics[angle=-90,scale=0.34]{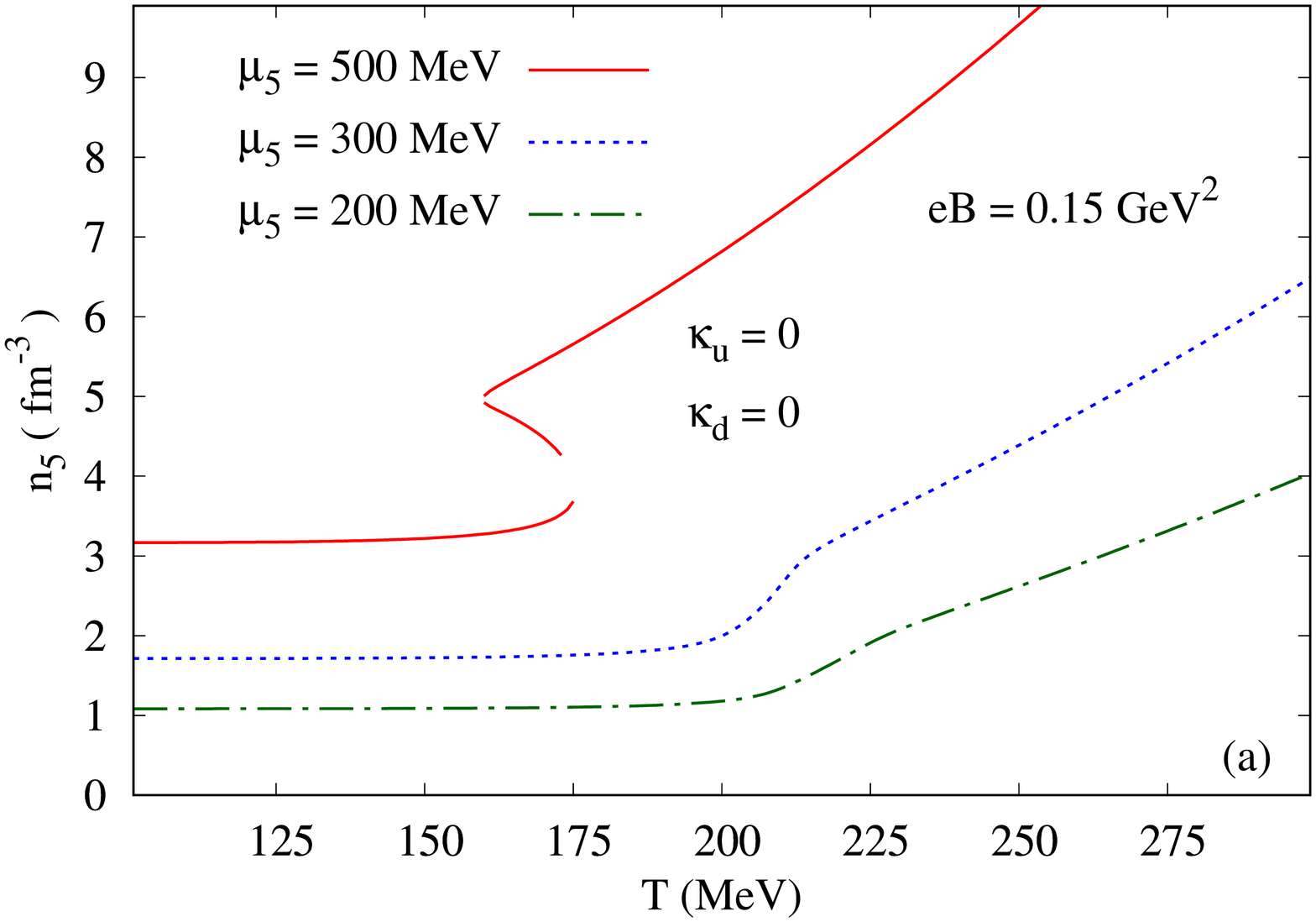}  \includegraphics[angle=-90,scale=0.34]{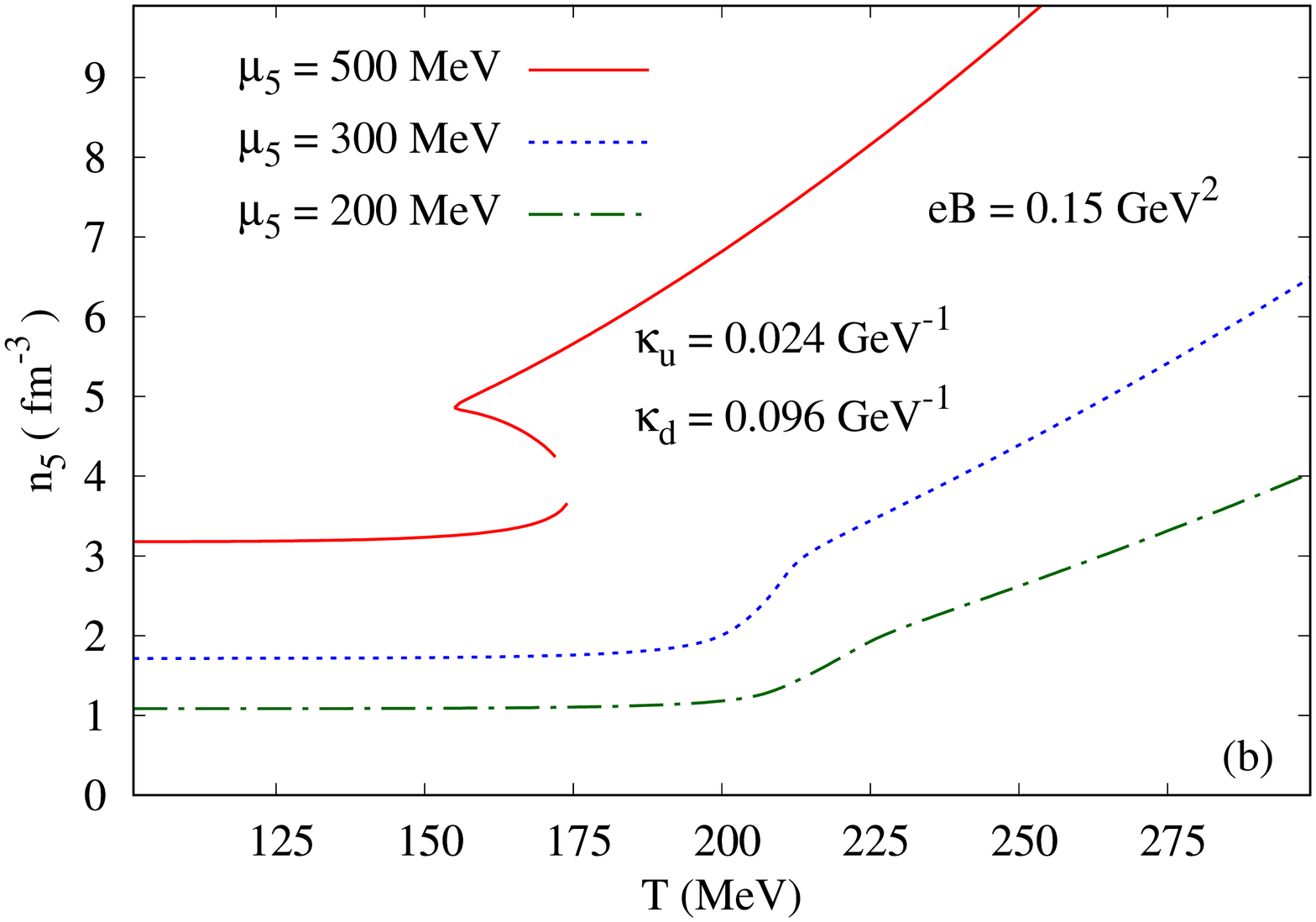}  
	\end{center}
	\caption{(Color Online) Variation of $n_5$ as a function of $ T $  using  2-flavour PNJL model. }
	\label{fig.n51}
\end{figure}

 In Figs.~\ref{fig.n51}(a) and (b) we have depicted the variation of chiral charge density as function of temperature for three different values of $ \mu_5$  for $ eB =  0.15 $~GeV$ ^2 $ with zero and non-zero values of the AMM of the quarks respectively. From Fig.~\ref{fig.n51}(a), it can be seen that, in   the presence of a fixed chemical potential $ \mu_5 $,  chiral charge density is a growing  function of the temperature and is   strongly enhanced near the transition temperature, as the system goes from the chiral symmetry broken to the restored phase. For small values of $ \mu_5 $, i.e.  $ \mu_5 = 200  $ and $ 300  $ MeV, both the chiral as well as the deconfinement transitions are crossover which is manifested by the absence of multiplevalued nature of $ n_5 $. But at high values of $ \mu_5 $, as discussed in the previous paragraph, both $ M $ and $ \Phi  $ show first order transitions, as a consequence the S-like structure is observed, conveying the multiple solutions of $ M $ and $\Phi $ from gap equations. Furthermore, as we switch on non-zero values of AMM of the quarks in Fig.~\ref{fig.n51}(b), no appreciable changes in the qualitative behaviour of  $ n_5 $ is observed for lower values of $ \mu_5 $. However, at $ \mu_5  = 500 $ MeV, the jump in $ n_5 $ is found to shift towards somewhat lower values of $ \mu_5 $.

 		In all the results shown above we have considered constant values of the AMM of the quarks as has been done in most of the literature~\cite{Fayazbakhsh:2014mca,Chaudhuri:2019lbw,Chaudhuri:2020lga,Aguirre:2020tiy,Mei:2020jzn}. However, it should be noted that, in Ref.~\cite{Ghosh:2021dlo}, it is shown that the constituent mass as well as the AMM of the quarks are large in the chiral symmetry broken phase in the low temperature region. Around the pseudo-chiral phase transition, both $ M $ and $ \kappa_{u/d} $ suffer a sudden decrease and at high temperature limit, both of them approach vanishingly small values at the symmetry restored phase. Furthermore, the variation of AMM of the quarks as a function temperature is different for different values of magnetic field. In Figs.~\ref{fig.MPhi3}(a) and (b) we have depicted the variation of $ M $ and $ \Phi $ as a function temperature for non-zero values of $ \mu_5 $ for both zero and $ T $ and $ eB $ dependent values of the AMM of the quarks following~\cite{Ghosh:2021dlo} for $ eB = 0.15 $~GeV$ ^2 $. In Fig.~\ref{fig.MPhi2}(a), we have already seen that, the effect of the (constant) AMM of the quarks  are more prominent at higher values of temperatures. However, since both $ \kappa_u $ and $ \kappa_d $ become vanishingly small at higher values of temperature in~\cite{Ghosh:2021dlo}, the $ T $-dpenedence of the constituent quark mass in Fig.~\ref{fig.MPhi3}(a) remain unaltered when $ T $ and $ eB $ dependent values of the AMM of the quarks are considered. Moreover, variation of $ \Phi $ also remain unchanged for different values of the AMM of the quark. The similar behaviour is also noticed while discussing Fig.~\ref{fig.MPhi2}(b).
 	
\section{SUMMARY \& CONCLUSIONS} \label{sec.summary}
The main motivation of this work is to study the effect  of the  anomalous magnetic moment of quarks on the chiral and de-confinement phase transition in presence of  chiral imbalance and background magnetic field.  For this purpose, the Dirac equation in presence of constant background magnetic field and baryonic chemical potential  is modifed incorporating the minimal AMM interaction as well as the chiral chemical potential $\mu_5$. We find that, the energy eigenvalues obtained from this  modified equation possesses a  more complicated dependence on the quark anomalous magnetic moment and $\mu_5$ compared to the cases where those are considered individually.  The eigen spectrum so obtained is  then  used in the PNJL framework to study the variation of  the constituent quark mass and the Polyakov loop  with the external parameters i.e $T$, $eB$ and $\mu_5$. Note that, in this work we only consider the case of zero baryonic chemical potential. As already mentioned,  if one introduces  chiral chemical potential  in the PNJL model, the critical temperature of the chiral symmetry restoration as well as deconfinement phase transition shows a decreasing trend \cite{Fukushima:2010fe}. With the incorporation of the AMM of the quarks,  we find that the critical temperature of the chiral symmetry restoration further decreases  showing an enhancement in the IMC effect. On the other hand, the behaviour of the Polyakov loop is found to be dominated by the chiral chemical potential and consequently, the  deconfinement transition temperature shows no significant change compared to the case with vanishing $\kappa$. The $ eB $-dependence of $ M $ and $ \Phi  $ are obtained for a given temperature for both zero and non-zero values of $ \mu_5 $ and it is observed that the effect of consideration of finite values of chiral chemical potential is more visible at high values of temperature. The variation of chiral charge density ( $n_5 $ ) as a function of $ \mu_5 $ and $T$ is obtained. We find that  as a function of chiral chemical potential (temperature), $ n_5  $ becomes multiple-valued at lower(higher) values of temperature(chiral chemical potential), indicating a first order transition, which is consistent with previous investigations with this model~\cite{Fukushima:2010fe}. We also observe that, the consideration of finite values of AMM of the quarks results in very small change in the $ T$-dependence of $ n_5 $ at high values of background magnetic field.

A few comments regarding these observations are in order. First of all the enhancement in the IMC effect due to the incorporation of AMM is not surprising, because, even without the chiral imbalance, it is already observed that effect of AMM may alter the trend of chiral symmetry breaking from MC to IMC \cite{Fayazbakhsh:2014mca,Mukherjee:2018ebw,Chaudhuri:2019lbw,Chaudhuri:2020lga}. It is also known that similar decreasing trend of critical temperature  can be observed if one considers the chiral chemical potential 
\cite{Chao:2013qpa,Yu:2014sla,Yu:2014xoa,Yu:2015hym}.  Although the energy dispersion derived for the first time in this work, possesses a non-trivial  dependence on $\mu_5$ and $\kappa$, we find that the overall effect roughly remains additive in nature showing an enhanced IMC behaviour  consistent with the naive expectations. However, one must keep in mind that the decreasing trend of the critical temperature, observed in the effective models in presence of  $\mu_5$,  is  inconsistent with the recent lattice results \cite{Astrakhantsev:2019wnp,Braguta:2015owi,Braguta:2015zta} as well as with the predictions obtained in Dyson-Schwinger approaches \cite{Xu:2015vna,Wang:2015tia,Cui:2016zqp,Shi:2020uyb}. In fact, this contradictory behavior is common with  models with local  interaction kernel \cite{Chernodub:2011fr} (however, see also the discussions in Ref. \cite{Braguta:2016aov} and \cite{Farias:2016let} where catalysis behaviour consistent with lattice results is obtained). Recently, in Ref. \cite{Ruggieri:2020qtq}, it is shown that in the framework of non-local NJL model, enhancement of both
chiral and axial symmetry breaking can be observed where $\mu_5$ is treated as  a new coupling that  requires a renormalization in the ultraviolet domain.   
However, in this work we restrict ourselves to an  effective model with local interactions (i.e PNJL model) to obtain a qualitative idea of the impact of anomalous magnetic moments of quarks on the properties of chiral medium in presence of the background  magnetic field. We conclude that the impact is  insignificant     for the de-confinement transition and almost of similar magnitude for the chiral symmetry restoration.  We expect, with models having catalysing effect with $\mu_5$, there will be a competition between the effects of $\kappa$ and the chiral chemical potential which definitely is an interesting future direction to look for.

\appendix

\section{DETERMINATION OF ENERGY EIGENVALUES} \label{sec.app.1}
In a cylindrical coordinate system, we have
\begin{eqnarray}
\dfrac{\partial}{\partial x} &=& \cos\phi \dfrac{\partial}{\partial \rho} - \sin\phi  \frac{1}{\rho} \dfrac{\partial}{\partial \phi }, \\
\dfrac{\partial}{\partial y} &=& \sin\phi \dfrac{\partial}{\partial \rho} + \cos\phi  \frac{1}{\rho} \dfrac{\partial}{\partial \phi }.
\end{eqnarray}
Also noting that, $ x\pm iy = e^{\pm i \phi} $, we can write 
\begin{eqnarray}
\Pi_x \pm i \Pi_y &=& -i \FB{\dfrac{\partial}{\partial x}\pm i \dfrac{\partial}{\partial y}} - e\FB{A_x \pm i A_y} 
= -i e^{\pm i \phi} \SB{\FB{\dfrac{\partial}{\partial \rho}\pm \frac{i}{\rho} \dfrac{\partial}{\partial \phi}} \pm \rho\frac{1}{2}eB  }. 
\label{eq.Pixy}
\end{eqnarray}
Substituting Eqs.~\eqref{eq_ansatz} and \eqref{eq_psi_comp} into Eq.~\eqref{eq_Dirac_matrix1} and using Eq.~\eqref{eq.Pixy}, we arrive at
\begin{eqnarray}
E  \begin{pmatrix}
e^{i(l-1) \phi} f_1 \\ e^{i l \phi} f_2 \\e^{i(l-1) \phi} f_3 \\ e^{i l \phi} f_4
\end{pmatrix} 
=
\hat{H} \begin{pmatrix}
e^{i(l-1) \phi} f_1 \\
e^{i l \phi} f_2 \\e^{i(l-1) \phi} f_3 \\ e^{i l \phi} f_4
\end{pmatrix}  
\label{eq_Dirac_matrix2}
\end{eqnarray}
where, $\hat{H}$ is given by,
\begin{eqnarray}
\hat{H} = \begin{pmatrix}
m-\mu -a B  &  0 &  \mu_5 +p_z & -i e^{- i \phi} \FB{{\frac{\partial}{\partial \rho} - \frac{i}{\rho} \frac{\partial}{\partial \phi}} - \frac{1}{2}eB \rho} \\ 
0 & m-\mu +a B & -i e^{ i \phi} \FB{{\frac{\partial}{\partial \rho} + \frac{i}{\rho} \frac{\partial}{\partial \phi}} + \frac{1}{2}eB \rho } & \mu_5 -p_z \\
\mu_5+p_z & -i e^{- i \phi} \FB{{\frac{\partial}{\partial \rho} - \frac{i}{\rho} \frac{\partial}{\partial \phi}} - \frac{1}{2}eB\rho  }  &-m-\mu +a B & 0 \\
-i e^{ i \phi} \FB{{\frac{\partial}{\partial \rho} + \frac{i}{\rho} \frac{\partial}{\partial \phi}} + \frac{1}{2}eB\rho  } & \mu_5 -p_z &0 & -m-\mu -a B 
\end{pmatrix}.  \nn
\end{eqnarray}
Writing down all the rows of the matrix in Eq.~\eqref{eq_Dirac_matrix2} separately, we arrive at the following four equations:
\begin{eqnarray}
\FB{E +\mu - m_1} f_1(\rho) &=& \FB{\mu_5 + p_z} f_3(\rho) - i\FB{\frac{d}{d \rho} + \frac{l}{\rho} - \xi \rho } f_4(\rho), \label{eq_coup1} \\
\FB{E +\mu - m_2} f_2(\rho) &=& \FB{\mu_5 - p_z} f_4(\rho) - i\FB{\frac{d}{d \rho} - \frac{l-1}{\rho} + \xi \rho } f_3(\rho), \label{eq_coup2}\\
\FB{E +\mu + m_1} f_3(\rho) &=& \FB{\mu_5 + p_z} f_1(\rho) - i\FB{\frac{d}{d \rho} + \frac{l}{\rho} - \xi \rho } f_2(\rho), \label{eq_coup3} \\
\FB{E +\mu + m_2} f_4(\rho) &=& \FB{\mu_5 - p_z} f_2(\rho) - i\FB{\frac{d}{d \rho} - \frac{l-1}{\rho} + \xi \rho } f_1(\rho) \label{eq_coup4}
\end{eqnarray}
where, $ m_1 = (m - a B)$, $ m_2 = (m+a B)$ and $ \xi = eB/2$. 

Now multiplying both sides of Eq.~\eqref{eq_coup1} by $ - i\FB{\frac{d}{d \rho} - \frac{l-1}{\rho} + \xi \rho } $, and, using Eqs.~\eqref{eq_coup2} and ~\eqref{eq_coup4}, we get after some simplifications
\begin{equation}
\FB{\derparttwo{}{\rho} + \frac{1}{\rho} \derpartone{}{\rho} + 2\xi (l-1) - \xi^2 \rho^2 -\frac{l^2}{\rho^2} + \mathfrak{B}_1^\prime} f_4(\rho) =  \mathfrak{B}_2^\prime f_2(\rho)  
\label{eq_diff_eq_r1}
\end{equation}
where, 
\begin{eqnarray}
\mathfrak{B}_1^\prime &=& \FB{E+\mu}^2  - \FB{E+\mu }\FB{m_1 - m_2 } - m_1 m_2 + \mu_5^2 - p_z^2 , \\
\mathfrak{B}_2^\prime &=& 2\FB{E+\mu}\mu_5 - \FB{m_1 + m_2 } \mu_5 + \FB{m_1 - m_2}p_z .
\end{eqnarray}
Similarly starting from Eq.~\eqref{eq_coup3} one arrives at
\begin{equation}
\FB{\derparttwo{}{\rho} + \frac{1}{\rho} \derpartone{}{\rho} + 2\xi (l-1) - \xi^2 \rho^2 -\frac{l^2}{\rho^2} + \mathfrak{D}_1^\prime} f_2(\rho) =  \mathfrak{D}_2^\prime f_4(\rho)  
\label{eq_diff_eq_r2}
\end{equation}
where
\begin{eqnarray}
\mathfrak{D}_1^\prime &=& \FB{E+\mu}^2  + \FB{E+\mu }\FB{m_1 - m_2 } - m_1 m_2 + \mu_5^2 - p_z^2 ,  \\
\mathfrak{D}_2^\prime &=& 2\FB{E+\mu}\mu_5 + \FB{m_1 + m_2 } \mu_5 - \FB{m_1 - m_2}p_z .
\end{eqnarray}
Introducing a dimensionless variable $ \lambda = \xi \rho^2 $, we have $ \derpartone{}{\rho} \to 2\sqrt{\lambda \xi} \derpartone{}{\lambda} $, so that, Eqs.~\eqref{eq_diff_eq_r1} and \eqref{eq_diff_eq_r2} become
\begin{eqnarray}
\TB{\lambda \derparttwo{}{\lambda } + \derpartone{}{\lambda } - \frac{l^2}{4\lambda} - \frac{\lambda}{4} -\frac{1}{2} \FB{l -1} + \mathfrak{B}_1}f_4(\lambda) &=& \mathfrak{B}_2 f_2(\lambda) ,\label{eq_diff_eqn_rho} \\
\TB{\lambda \derparttwo{}{\lambda } + \derpartone{}{\lambda } - \frac{l^2}{4\lambda} - \frac{\lambda}{4} -\frac{1}{2} \FB{l -1} + \mathfrak{D}_1}f_2(\lambda) &=& \mathfrak{D}_2 f_4(\lambda) \label{eq_diff_eqn_rho2}
\end{eqnarray}
where, $ \mathfrak{B}_i = \mathfrak{B}_i^\prime / 4\xi ,\mathfrak{D}_i = \mathfrak{D}_i^\prime / 4\xi $. The functions $ f_1 $ and $ f_3 $ obey similar kind of equations but they will not concern us here as  we are only interested  to obtain the energy eigenvalue. 
The above two differential equations have regular singularities at $ \lambda= 0  $. So we can solve them using Frobenius method in which we assume
\begin{equation}
f_4(\lambda ) = e^{-\lambda/2} \lambda^s \sum_{N = 0}^{\infty } c_N \lambda^N ~~~~,~~~~~
f_2(\lambda ) = e^{-\lambda/2} \lambda^s \sum_{N = 0}^{\infty }d_N \lambda^N.
\label{eq_series1}
\end{equation}
%
%
Substituting Eq.~\eqref{eq_series1} into Eqs.~\eqref{eq_diff_eqn_rho} and \eqref{eq_diff_eqn_rho2}and equating the coefficient of 
$ e^{-\lambda/2} \lambda^{s+N-1}$, we get after some simplifications the following recursion relations
\begin{eqnarray}
\SB{ \mathfrak{B}_1-N+\frac{1}{2}   - \frac{1}{2}(l-1) -s  }c_{N-1} + \SB{(N+ s)^2 -\frac{l^2}{4}  } c_N - \mathfrak{B}_2~ d_{N-1 }&=& 0 , \label{eq_series_coeff1} \\
\SB{ \mathfrak{D}_1-N+\frac{1}{2}   - \frac{1}{2}(l-1) -s  } d_{N-1} + \SB{(N+ s)^2 -\frac{l^2}{4}  } ~d_N - \mathfrak{D}_2 ~c_{N-1 }&=& 0 .\label{eq_series_coeff2} 
\end{eqnarray} 
If we take $ N = 0 $, we get $ 2s = \pm l $. But we will discard the solutions which diverge at $ \lambda = 0 $. 
Hence, considering $ 2s = l $, we get from Eqs.~\eqref{eq_series_coeff1} and \eqref{eq_series_coeff2} 
\begin{eqnarray}
\FB{ \mathfrak{B}_1-N+1-l  }c_{N-1} + N (N+l) c_N - \mathfrak{B}_2~ d_{N-1 }&=& 0,  \label{eq_series_coeff3}\\
\FB{ \mathfrak{D}_1-N+1-l  }d_{N-1} + N (N+l) d_N - \mathfrak{D}_2~ c_{N-1 }&=& 0 .\label{eq_series_coeff4} 
\end{eqnarray}
To obtain well behaved wave function, we assume that the series must terminate at some $ N = \Np $ 
(this ensures that, we get polynomial solution in $ \lambda $ and since we already have an $ e^{-\lambda/2} $ term, 
the solution must vanish as $ \lambda \to \infty $ ) so that, $c_{\Np + 1} = 0$ and $d_{\Np +1 } = 0$. 
Thus from Eqs.~\eqref{eq_series_coeff3} and \eqref{eq_series_coeff4}, we get
\begin{eqnarray}
\FB{ \mathfrak{B}_1-\Np-l  }c_{\Np}  - \mathfrak{B}_2 ~d_{\Np }&=& 0 , \label{eq_series_coeff5}\\
\FB{ \mathfrak{D}_1-\Np -l  }~d_{\Np}  - \mathfrak{D}_2` c_{\Np }&=& 0 . \label{eq_series_coeff6} 
\end{eqnarray}
For Eqs.~\eqref{eq_series_coeff5} and \eqref{eq_series_coeff6} to have non-trivial solutions, we must have
\begin{eqnarray}
\text{det} \begin{pmatrix}
\mathfrak{B}_1-(\Np+l) & - \mathfrak{B}_2 \\
- \mathfrak{D}_2 & \mathfrak{D}_1-(\Np +l) 
\end{pmatrix} = 0
\label{eq.det}
\end{eqnarray}
Simplification of Eq.~\eqref{eq.det} yields,
\begin{eqnarray}
\FB{E+\mu }^4 + \mathfrak{B} \FB{E+\mu}^2 + \mathfrak{C} = 0
\label{eq_Energy_solve1}
\end{eqnarray}
where 
\begin{eqnarray}
\mathfrak{B} &=& - 2\SB{p_z^2 + m^2 + \mu_5^2 + \fb{a B}^2 } - 8\xi (\Np + l) , \\
\mathfrak{C} &=& \SB{p_z^2 + m^2 -\fb{ \mu_5^2 + \fb{a B}^2} }^2 + 4\FB{ m\mu_5 + p_z a B  } \nn \\ 
&& +~ 8\xi (\Np + l ) \SB{p_z^2 + m^2 -\fb{ \mu_5^2 + \fb{a B}^2} } + 64 \xi^2 \fb{\Np+ l}^2 ,
\end{eqnarray}
and the discriminant is
\begin{equation}
(\mathfrak{B}^2 - 4\mathfrak{C}) = 16 \TB{ \SB{ m^2 + 4\xi (\Np + l) } \FB{a B}^2 +  \SB{ p_z^2 + 4\xi (\Np + l) } \mu_5^2  -  2 mp_z \mu_5 a B } .
\label{eq_discri}
\end{equation}
Hence, the solution of Eq.~\eqref{eq_Energy_solve1} is given by,
\begin{eqnarray}
\fb{E+ \mu}^2 =  \frac{-\mathfrak{B} \pm \sqrt{\mathfrak{B}^2- 4\mathfrak{C}}}{2} 
&=& p_z^2 + m^2 + \mu_5^2 + \FB{a B}^2  + 2 (\Np + l) eB \pm 2 \TB{  \SB{ m^2 + 2 (\Np + l)eB } \fb{a B}^2 \right. \nn \\ 
	&& \left. + \SB{ p_z^2 + 2 (\Np + l)eB } \mu_5^2  -  2 m p_z \mu_5 a B  }^{1/2}
\label{eq.E.1}
\end{eqnarray}
Following Ref.~\cite{OConnell:1968spc}, we replace $ \pm \to -s {~\rm sign}(e B) $ where $s$ is the helicity in the massless case~\cite{Fukushima:2008xe,Sheng:2017lfu}. 
This will ensure that we get back the correct result when the non-relativistic limit is taken. 
Also, we identify $(\Np+l) =n$ as the Landau level.
With these replacements, Eq.~\eqref{eq.E.1} becomes
\begin{eqnarray}
\fb{E+ \mu}^2 &=&  p_z^2 + m^2 + \mu_5^2 + \fb{a B}^2  + 2 n eB \nn \\ &&
- 2 s {~\rm sign} (eB) \TB{  \fb{ m^2 + 2 n eB } \fb{a B}^2 + \fb{ p_z^2 + 2 n eB } \mu_5^2  -  2 m p_z \mu_5 a B  }^{1/2}.
\end{eqnarray}
For ground state, we have $ n = 0 $ corresponding to the lowest Landau level (LLL), so that the energy eigenvalue becomes 
\begin{equation}
(E_0 +\mu)^2  =  (p_z - \mu_5)^2 + (m - \kappa \MB{e B})^2.
\label{Energy_EV_GS}
\end{equation}
It is to be noted that, for a positively (negatively) charged fermion, the ground sate contribution comes from spin down (up) state.
On the other hand, for higher Landau levels, $ n \ge 1 $ we get,
\begin{eqnarray}
\fb{E_{ns}+ \mu}^2 &=&  p_z^2 + m^2 + \mu_5^2 + \fb{\kappa e B}^2  + 2 n \MB{eB} \nn \\ && 
- 2 s {~\rm sign} (e B) \TB{  \fb{ m^2 + 2 n  \MB{eB} } \fb{\kappa e B}^2 + \fb{ p_z^2 + 2 n  \MB{eB} } \mu_5^2  -  2 m p_z \mu_5 \kappa e B  }^{1/2} .
\label{Energy_EV_ES}
\end{eqnarray}


\bibliography{Z-PNJL-mu5}

\begin{thebibliography}{113}%
\makeatletter
\providecommand \@ifxundefined [1]{%
 \@ifx{#1\undefined}
}%
\providecommand \@ifnum [1]{%
 \ifnum #1\expandafter \@firstoftwo
 \else \expandafter \@secondoftwo
 \fi
}%
\providecommand \@ifx [1]{%
 \ifx #1\expandafter \@firstoftwo
 \else \expandafter \@secondoftwo
 \fi
}%
\providecommand \natexlab [1]{#1}%
\providecommand \enquote  [1]{``#1''}%
\providecommand \bibnamefont  [1]{#1}%
\providecommand \bibfnamefont [1]{#1}%
\providecommand \citenamefont [1]{#1}%
\providecommand \href@noop [0]{\@secondoftwo}%
\providecommand \href [0]{\begingroup \@sanitize@url \@href}%
\providecommand \@href[1]{\@@startlink{#1}\@@href}%
\providecommand \@@href[1]{\endgroup#1\@@endlink}%
\providecommand \@sanitize@url [0]{\catcode `\\12\catcode `\$12\catcode
  `\&12\catcode `\#12\catcode `\^12\catcode `\_12\catcode `\%12\relax}%
\providecommand \@@startlink[1]{}%
\providecommand \@@endlink[0]{}%
\providecommand \url  [0]{\begingroup\@sanitize@url \@url }%
\providecommand \@url [1]{\endgroup\@href {#1}{\urlprefix }}%
\providecommand \urlprefix  [0]{URL }%
\providecommand \Eprint [0]{\href }%
\providecommand \doibase [0]{http://dx.doi.org/}%
\providecommand \selectlanguage [0]{\@gobble}%
\providecommand \bibinfo  [0]{\@secondoftwo}%
\providecommand \bibfield  [0]{\@secondoftwo}%
\providecommand \translation [1]{[#1]}%
\providecommand \BibitemOpen [0]{}%
\providecommand \bibitemStop [0]{}%
\providecommand \bibitemNoStop [0]{.\EOS\space}%
\providecommand \EOS [0]{\spacefactor3000\relax}%
\providecommand \BibitemShut  [1]{\csname bibitem#1\endcsname}%
\let\auto@bib@innerbib\@empty
\bibitem [{\citenamefont {Kharzeev}\ \emph
  {et~al.}(2013{\natexlab{a}})\citenamefont {Kharzeev}, \citenamefont
  {Landsteiner}, \citenamefont {Schmitt},\ and\ \citenamefont
  {Yee}}]{Kharzeev:2013jha}%
  \BibitemOpen
  \bibinfo {editor} {\bibfnamefont {D.}~\bibnamefont {Kharzeev}}, \bibinfo
  {editor} {\bibfnamefont {K.}~\bibnamefont {Landsteiner}}, \bibinfo {editor}
  {\bibfnamefont {A.}~\bibnamefont {Schmitt}}, \ and\ \bibinfo {editor}
  {\bibfnamefont {H.-U.}\ \bibnamefont {Yee}},\ eds.,\ \href {\doibase
  10.1007/978-3-642-37305-3} {\emph {\bibinfo {title} {{Strongly Interacting
  Matter in Magnetic Fields}}}},\ Vol.\ \bibinfo {volume} {871}\ (\bibinfo
  {year} {2013})\BibitemShut {NoStop}%
\bibitem [{\citenamefont {Miransky}\ and\ \citenamefont
  {Shovkovy}(2015)}]{Miransky:2015ava}%
  \BibitemOpen
  \bibfield  {author} {\bibinfo {author} {\bibfnamefont {V.~A.}\ \bibnamefont
  {Miransky}}\ and\ \bibinfo {author} {\bibfnamefont {I.~A.}\ \bibnamefont
  {Shovkovy}},\ }\href {\doibase 10.1016/j.physrep.2015.02.003} {\bibfield
  {journal} {\bibinfo  {journal} {Phys. Rept.}\ }\textbf {\bibinfo {volume}
  {576}},\ \bibinfo {pages} {1} (\bibinfo {year} {2015})},\ \Eprint
  {http://arxiv.org/abs/1503.00732} {arXiv:1503.00732 [hep-ph]} \BibitemShut
  {NoStop}%
\bibitem [{\citenamefont {Andersen}\ \emph {et~al.}(2016)\citenamefont
  {Andersen}, \citenamefont {Naylor},\ and\ \citenamefont
  {Tranberg}}]{Andersen:2014xxa}%
  \BibitemOpen
  \bibfield  {author} {\bibinfo {author} {\bibfnamefont {J.~O.}\ \bibnamefont
  {Andersen}}, \bibinfo {author} {\bibfnamefont {W.~R.}\ \bibnamefont
  {Naylor}}, \ and\ \bibinfo {author} {\bibfnamefont {A.}~\bibnamefont
  {Tranberg}},\ }\href {\doibase 10.1103/RevModPhys.88.025001} {\bibfield
  {journal} {\bibinfo  {journal} {Rev. Mod. Phys.}\ }\textbf {\bibinfo {volume}
  {88}},\ \bibinfo {pages} {025001} (\bibinfo {year} {2016})},\ \Eprint
  {http://arxiv.org/abs/1411.7176} {arXiv:1411.7176 [hep-ph]} \BibitemShut
  {NoStop}%
\bibitem [{\citenamefont {Kharzeev}\ \emph {et~al.}(2008)\citenamefont
  {Kharzeev}, \citenamefont {McLerran},\ and\ \citenamefont
  {Warringa}}]{Kharzeev:2007jp}%
  \BibitemOpen
  \bibfield  {author} {\bibinfo {author} {\bibfnamefont {D.~E.}\ \bibnamefont
  {Kharzeev}}, \bibinfo {author} {\bibfnamefont {L.~D.}\ \bibnamefont
  {McLerran}}, \ and\ \bibinfo {author} {\bibfnamefont {H.~J.}\ \bibnamefont
  {Warringa}},\ }\href {\doibase 10.1016/j.nuclphysa.2008.02.298} {\bibfield
  {journal} {\bibinfo  {journal} {Nucl. Phys.}\ }\textbf {\bibinfo {volume}
  {A803}},\ \bibinfo {pages} {227} (\bibinfo {year} {2008})},\ \Eprint
  {http://arxiv.org/abs/0711.0950} {arXiv:0711.0950 [hep-ph]} \BibitemShut
  {NoStop}%
\bibitem [{\citenamefont {Skokov}\ \emph {et~al.}(2009)\citenamefont {Skokov},
  \citenamefont {Illarionov},\ and\ \citenamefont {Toneev}}]{Skokov:2009qp}%
  \BibitemOpen
  \bibfield  {author} {\bibinfo {author} {\bibfnamefont {V.}~\bibnamefont
  {Skokov}}, \bibinfo {author} {\bibfnamefont {A.~{\relax Yu}.}\ \bibnamefont
  {Illarionov}}, \ and\ \bibinfo {author} {\bibfnamefont {V.}~\bibnamefont
  {Toneev}},\ }\href {\doibase 10.1142/S0217751X09047570} {\bibfield  {journal}
  {\bibinfo  {journal} {Int. J. Mod. Phys.}\ }\textbf {\bibinfo {volume}
  {A24}},\ \bibinfo {pages} {5925} (\bibinfo {year} {2009})},\ \Eprint
  {http://arxiv.org/abs/0907.1396} {arXiv:0907.1396 [nucl-th]} \BibitemShut
  {NoStop}%
\bibitem [{\citenamefont {Tuchin}(2013{\natexlab{a}})}]{Tuchin:2013apa}%
  \BibitemOpen
  \bibfield  {author} {\bibinfo {author} {\bibfnamefont {K.}~\bibnamefont
  {Tuchin}},\ }\href {\doibase 10.1103/PhysRevC.88.024911} {\bibfield
  {journal} {\bibinfo  {journal} {Phys. Rev. C}\ }\textbf {\bibinfo {volume}
  {88}},\ \bibinfo {pages} {024911} (\bibinfo {year} {2013}{\natexlab{a}})},\
  \Eprint {http://arxiv.org/abs/1305.5806} {arXiv:1305.5806 [hep-ph]}
  \BibitemShut {NoStop}%
\bibitem [{\citenamefont {Tuchin}(2016)}]{Tuchin:2015oka}%
  \BibitemOpen
  \bibfield  {author} {\bibinfo {author} {\bibfnamefont {K.}~\bibnamefont
  {Tuchin}},\ }\href {\doibase 10.1103/PhysRevC.93.014905} {\bibfield
  {journal} {\bibinfo  {journal} {Phys. Rev. C}\ }\textbf {\bibinfo {volume}
  {93}},\ \bibinfo {pages} {014905} (\bibinfo {year} {2016})},\ \Eprint
  {http://arxiv.org/abs/1508.06925} {arXiv:1508.06925 [hep-ph]} \BibitemShut
  {NoStop}%
\bibitem [{\citenamefont {Tuchin}(2013{\natexlab{b}})}]{Tuchin:2013ie}%
  \BibitemOpen
  \bibfield  {author} {\bibinfo {author} {\bibfnamefont {K.}~\bibnamefont
  {Tuchin}},\ }\href {\doibase 10.1155/2013/490495} {\bibfield  {journal}
  {\bibinfo  {journal} {Adv. High Energy Phys.}\ }\textbf {\bibinfo {volume}
  {2013}},\ \bibinfo {pages} {490495} (\bibinfo {year} {2013}{\natexlab{b}})},\
  \Eprint {http://arxiv.org/abs/1301.0099} {arXiv:1301.0099 [hep-ph]}
  \BibitemShut {NoStop}%
\bibitem [{\citenamefont {Gursoy}\ \emph {et~al.}(2014)\citenamefont {Gursoy},
  \citenamefont {Kharzeev},\ and\ \citenamefont {Rajagopal}}]{Gursoy:2014aka}%
  \BibitemOpen
  \bibfield  {author} {\bibinfo {author} {\bibfnamefont {U.}~\bibnamefont
  {Gursoy}}, \bibinfo {author} {\bibfnamefont {D.}~\bibnamefont {Kharzeev}}, \
  and\ \bibinfo {author} {\bibfnamefont {K.}~\bibnamefont {Rajagopal}},\ }\href
  {\doibase 10.1103/PhysRevC.89.054905} {\bibfield  {journal} {\bibinfo
  {journal} {Phys. Rev.}\ }\textbf {\bibinfo {volume} {C89}},\ \bibinfo {pages}
  {054905} (\bibinfo {year} {2014})},\ \Eprint {http://arxiv.org/abs/1401.3805}
  {arXiv:1401.3805 [hep-ph]} \BibitemShut {NoStop}%
\bibitem [{\citenamefont {Duncan}\ and\ \citenamefont
  {Thompson}(1992)}]{Duncan:1992hi}%
  \BibitemOpen
  \bibfield  {author} {\bibinfo {author} {\bibfnamefont {R.~C.}\ \bibnamefont
  {Duncan}}\ and\ \bibinfo {author} {\bibfnamefont {C.}~\bibnamefont
  {Thompson}},\ }\href {\doibase 10.1086/186413} {\bibfield  {journal}
  {\bibinfo  {journal} {Astrophys. J.}\ }\textbf {\bibinfo {volume} {392}},\
  \bibinfo {pages} {L9} (\bibinfo {year} {1992})}\BibitemShut {NoStop}%
\bibitem [{\citenamefont {Thompson}\ and\ \citenamefont
  {Duncan}(1993)}]{Thompson:1993hn}%
  \BibitemOpen
  \bibfield  {author} {\bibinfo {author} {\bibfnamefont {C.}~\bibnamefont
  {Thompson}}\ and\ \bibinfo {author} {\bibfnamefont {R.~C.}\ \bibnamefont
  {Duncan}},\ }\href {\doibase 10.1086/172580} {\bibfield  {journal} {\bibinfo
  {journal} {Astrophys. J.}\ }\textbf {\bibinfo {volume} {408}},\ \bibinfo
  {pages} {194} (\bibinfo {year} {1993})}\BibitemShut {NoStop}%
\bibitem [{\citenamefont {Vachaspati}(1991)}]{Vachaspati:1991nm}%
  \BibitemOpen
  \bibfield  {author} {\bibinfo {author} {\bibfnamefont {T.}~\bibnamefont
  {Vachaspati}},\ }\href {\doibase 10.1016/0370-2693(91)90051-Q} {\bibfield
  {journal} {\bibinfo  {journal} {Phys. Lett.}\ }\textbf {\bibinfo {volume}
  {B265}},\ \bibinfo {pages} {258} (\bibinfo {year} {1991})}\BibitemShut
  {NoStop}%
\bibitem [{\citenamefont {Campanelli}(2013)}]{Campanelli:2013mea}%
  \BibitemOpen
  \bibfield  {author} {\bibinfo {author} {\bibfnamefont {L.}~\bibnamefont
  {Campanelli}},\ }\href {\doibase 10.1103/PhysRevLett.111.061301} {\bibfield
  {journal} {\bibinfo  {journal} {Phys. Rev. Lett.}\ }\textbf {\bibinfo
  {volume} {111}},\ \bibinfo {pages} {061301} (\bibinfo {year} {2013})},\
  \Eprint {http://arxiv.org/abs/1304.6534} {arXiv:1304.6534 [astro-ph.CO]}
  \BibitemShut {NoStop}%
\bibitem [{\citenamefont {de~Forcrand}\ and\ \citenamefont
  {Philipsen}(2007)}]{deForcrand:2006pv}%
  \BibitemOpen
  \bibfield  {author} {\bibinfo {author} {\bibfnamefont {P.}~\bibnamefont
  {de~Forcrand}}\ and\ \bibinfo {author} {\bibfnamefont {O.}~\bibnamefont
  {Philipsen}},\ }\href {\doibase 10.1088/1126-6708/2007/01/077} {\bibfield
  {journal} {\bibinfo  {journal} {JHEP}\ }\textbf {\bibinfo {volume} {01}},\
  \bibinfo {pages} {077} (\bibinfo {year} {2007})},\ \Eprint
  {http://arxiv.org/abs/hep-lat/0607017} {arXiv:hep-lat/0607017} \BibitemShut
  {NoStop}%
\bibitem [{\citenamefont {Aoki}\ \emph {et~al.}(2006)\citenamefont {Aoki},
  \citenamefont {Fodor}, \citenamefont {Katz},\ and\ \citenamefont
  {Szabo}}]{Aoki:2006br}%
  \BibitemOpen
  \bibfield  {author} {\bibinfo {author} {\bibfnamefont {Y.}~\bibnamefont
  {Aoki}}, \bibinfo {author} {\bibfnamefont {Z.}~\bibnamefont {Fodor}},
  \bibinfo {author} {\bibfnamefont {S.~D.}\ \bibnamefont {Katz}}, \ and\
  \bibinfo {author} {\bibfnamefont {K.~K.}\ \bibnamefont {Szabo}},\ }\href
  {\doibase 10.1016/j.physletb.2006.10.021} {\bibfield  {journal} {\bibinfo
  {journal} {Phys. Lett. B}\ }\textbf {\bibinfo {volume} {643}},\ \bibinfo
  {pages} {46} (\bibinfo {year} {2006})},\ \Eprint
  {http://arxiv.org/abs/hep-lat/0609068} {arXiv:hep-lat/0609068} \BibitemShut
  {NoStop}%
\bibitem [{\citenamefont {Aoki}\ \emph {et~al.}(2009)\citenamefont {Aoki},
  \citenamefont {Borsanyi}, \citenamefont {Durr}, \citenamefont {Fodor},
  \citenamefont {Katz}, \citenamefont {Krieg},\ and\ \citenamefont
  {Szabo}}]{Aoki:2009sc}%
  \BibitemOpen
  \bibfield  {author} {\bibinfo {author} {\bibfnamefont {Y.}~\bibnamefont
  {Aoki}}, \bibinfo {author} {\bibfnamefont {S.}~\bibnamefont {Borsanyi}},
  \bibinfo {author} {\bibfnamefont {S.}~\bibnamefont {Durr}}, \bibinfo {author}
  {\bibfnamefont {Z.}~\bibnamefont {Fodor}}, \bibinfo {author} {\bibfnamefont
  {S.~D.}\ \bibnamefont {Katz}}, \bibinfo {author} {\bibfnamefont
  {S.}~\bibnamefont {Krieg}}, \ and\ \bibinfo {author} {\bibfnamefont {K.~K.}\
  \bibnamefont {Szabo}},\ }\href {\doibase 10.1088/1126-6708/2009/06/088}
  {\bibfield  {journal} {\bibinfo  {journal} {JHEP}\ }\textbf {\bibinfo
  {volume} {06}},\ \bibinfo {pages} {088} (\bibinfo {year} {2009})},\ \Eprint
  {http://arxiv.org/abs/0903.4155} {arXiv:0903.4155 [hep-lat]} \BibitemShut
  {NoStop}%
\bibitem [{\citenamefont {Bazavov}\ \emph {et~al.}(2009)\citenamefont {Bazavov}
  \emph {et~al.}}]{Bazavov:2009zn}%
  \BibitemOpen
  \bibfield  {author} {\bibinfo {author} {\bibfnamefont {A.}~\bibnamefont
  {Bazavov}} \emph {et~al.},\ }\href {\doibase 10.1103/PhysRevD.80.014504}
  {\bibfield  {journal} {\bibinfo  {journal} {Phys. Rev. D}\ }\textbf {\bibinfo
  {volume} {80}},\ \bibinfo {pages} {014504} (\bibinfo {year} {2009})},\
  \Eprint {http://arxiv.org/abs/0903.4379} {arXiv:0903.4379 [hep-lat]}
  \BibitemShut {NoStop}%
\bibitem [{\citenamefont {Cheng}\ \emph {et~al.}(2008)\citenamefont {Cheng}
  \emph {et~al.}}]{Cheng:2007jq}%
  \BibitemOpen
  \bibfield  {author} {\bibinfo {author} {\bibfnamefont {M.}~\bibnamefont
  {Cheng}} \emph {et~al.},\ }\href {\doibase 10.1103/PhysRevD.77.014511}
  {\bibfield  {journal} {\bibinfo  {journal} {Phys. Rev. D}\ }\textbf {\bibinfo
  {volume} {77}},\ \bibinfo {pages} {014511} (\bibinfo {year} {2008})},\
  \Eprint {http://arxiv.org/abs/0710.0354} {arXiv:0710.0354 [hep-lat]}
  \BibitemShut {NoStop}%
\bibitem [{\citenamefont {Muroya}\ \emph {et~al.}(2003)\citenamefont {Muroya},
  \citenamefont {Nakamura}, \citenamefont {Nonaka},\ and\ \citenamefont
  {Takaishi}}]{Muroya:2003qs}%
  \BibitemOpen
  \bibfield  {author} {\bibinfo {author} {\bibfnamefont {S.}~\bibnamefont
  {Muroya}}, \bibinfo {author} {\bibfnamefont {A.}~\bibnamefont {Nakamura}},
  \bibinfo {author} {\bibfnamefont {C.}~\bibnamefont {Nonaka}}, \ and\ \bibinfo
  {author} {\bibfnamefont {T.}~\bibnamefont {Takaishi}},\ }\href {\doibase
  10.1143/PTP.110.615} {\bibfield  {journal} {\bibinfo  {journal} {Prog. Theor.
  Phys.}\ }\textbf {\bibinfo {volume} {110}},\ \bibinfo {pages} {615} (\bibinfo
  {year} {2003})},\ \Eprint {http://arxiv.org/abs/hep-lat/0306031}
  {arXiv:hep-lat/0306031} \BibitemShut {NoStop}%
\bibitem [{\citenamefont {Splittorff}\ and\ \citenamefont
  {Verbaarschot}(2007)}]{Splittorff:2006fu}%
  \BibitemOpen
  \bibfield  {author} {\bibinfo {author} {\bibfnamefont {K.}~\bibnamefont
  {Splittorff}}\ and\ \bibinfo {author} {\bibfnamefont {J.~J.~M.}\ \bibnamefont
  {Verbaarschot}},\ }\href {\doibase 10.1103/PhysRevLett.98.031601} {\bibfield
  {journal} {\bibinfo  {journal} {Phys. Rev. Lett.}\ }\textbf {\bibinfo
  {volume} {98}},\ \bibinfo {pages} {031601} (\bibinfo {year} {2007})},\
  \Eprint {http://arxiv.org/abs/hep-lat/0609076} {arXiv:hep-lat/0609076}
  \BibitemShut {NoStop}%
\bibitem [{\citenamefont {Fukushima}\ and\ \citenamefont
  {Hidaka}(2007)}]{Fukushima:2006uv}%
  \BibitemOpen
  \bibfield  {author} {\bibinfo {author} {\bibfnamefont {K.}~\bibnamefont
  {Fukushima}}\ and\ \bibinfo {author} {\bibfnamefont {Y.}~\bibnamefont
  {Hidaka}},\ }\href {\doibase 10.1103/PhysRevD.75.036002} {\bibfield
  {journal} {\bibinfo  {journal} {Phys. Rev. D}\ }\textbf {\bibinfo {volume}
  {75}},\ \bibinfo {pages} {036002} (\bibinfo {year} {2007})},\ \Eprint
  {http://arxiv.org/abs/hep-ph/0610323} {arXiv:hep-ph/0610323} \BibitemShut
  {NoStop}%
\bibitem [{\citenamefont {Bazavov}\ \emph {et~al.}(2017)\citenamefont {Bazavov}
  \emph {et~al.}}]{Bazavov:2017dus}%
  \BibitemOpen
  \bibfield  {author} {\bibinfo {author} {\bibfnamefont {A.}~\bibnamefont
  {Bazavov}} \emph {et~al.},\ }\href {\doibase 10.1103/PhysRevD.95.054504}
  {\bibfield  {journal} {\bibinfo  {journal} {Phys. Rev. D}\ }\textbf {\bibinfo
  {volume} {95}},\ \bibinfo {pages} {054504} (\bibinfo {year} {2017})},\
  \Eprint {http://arxiv.org/abs/1701.04325} {arXiv:1701.04325 [hep-lat]}
  \BibitemShut {NoStop}%
\bibitem [{\citenamefont {Sharma}(2019)}]{Sharma:2019wiv}%
  \BibitemOpen
  \bibfield  {author} {\bibinfo {author} {\bibfnamefont {S.}~\bibnamefont
  {Sharma}},\ }\href {\doibase 10.22323/1.334.0009} {\bibfield  {journal}
  {\bibinfo  {journal} {PoS}\ }\textbf {\bibinfo {volume} {LATTICE2018}},\
  \bibinfo {pages} {009} (\bibinfo {year} {2019})},\ \Eprint
  {http://arxiv.org/abs/1901.07190} {arXiv:1901.07190 [hep-lat]} \BibitemShut
  {NoStop}%
\bibitem [{\citenamefont {Nambu}\ and\ \citenamefont
  {Jona-Lasinio}(1961{\natexlab{a}})}]{Nambu1}%
  \BibitemOpen
  \bibfield  {author} {\bibinfo {author} {\bibfnamefont {Y.}~\bibnamefont
  {Nambu}}\ and\ \bibinfo {author} {\bibfnamefont {G.}~\bibnamefont
  {Jona-Lasinio}},\ }\href {\doibase 10.1103/PhysRev.124.246} {\bibfield
  {journal} {\bibinfo  {journal} {Phys. Rev.}\ }\textbf {\bibinfo {volume}
  {124}},\ \bibinfo {pages} {246} (\bibinfo {year} {1961}{\natexlab{a}})},\
  \bibinfo {note} {[,141(1961)]}\BibitemShut {NoStop}%
\bibitem [{\citenamefont {Nambu}\ and\ \citenamefont
  {Jona-Lasinio}(1961{\natexlab{b}})}]{Nambu2}%
  \BibitemOpen
  \bibfield  {author} {\bibinfo {author} {\bibfnamefont {Y.}~\bibnamefont
  {Nambu}}\ and\ \bibinfo {author} {\bibfnamefont {G.}~\bibnamefont
  {Jona-Lasinio}},\ }\href {\doibase 10.1103/PhysRev.122.345} {\bibfield
  {journal} {\bibinfo  {journal} {Phys. Rev.}\ }\textbf {\bibinfo {volume}
  {122}},\ \bibinfo {pages} {345} (\bibinfo {year} {1961}{\natexlab{b}})},\
  \bibinfo {note} {[,127(1961)]}\BibitemShut {NoStop}%
\bibitem [{\citenamefont {Klevansky}(1992)}]{Klevansky}%
  \BibitemOpen
  \bibfield  {author} {\bibinfo {author} {\bibfnamefont {S.~P.}\ \bibnamefont
  {Klevansky}},\ }\href {\doibase 10.1103/RevModPhys.64.649} {\bibfield
  {journal} {\bibinfo  {journal} {Rev. Mod. Phys.}\ }\textbf {\bibinfo {volume}
  {64}},\ \bibinfo {pages} {649} (\bibinfo {year} {1992})}\BibitemShut
  {NoStop}%
\bibitem [{\citenamefont {Vogl}\ and\ \citenamefont {Weise}(1991)}]{Vogl}%
  \BibitemOpen
  \bibfield  {author} {\bibinfo {author} {\bibfnamefont {U.}~\bibnamefont
  {Vogl}}\ and\ \bibinfo {author} {\bibfnamefont {W.}~\bibnamefont {Weise}},\
  }\href {\doibase 10.1016/0146-6410(91)90005-9} {\bibfield  {journal}
  {\bibinfo  {journal} {Prog. Part. Nucl. Phys.}\ }\textbf {\bibinfo {volume}
  {27}},\ \bibinfo {pages} {195} (\bibinfo {year} {1991})}\BibitemShut
  {NoStop}%
\bibitem [{\citenamefont {Buballa}(2005)}]{Buballa}%
  \BibitemOpen
  \bibfield  {author} {\bibinfo {author} {\bibfnamefont {M.}~\bibnamefont
  {Buballa}},\ }\href {\doibase 10.1016/j.physrep.2004.11.004} {\bibfield
  {journal} {\bibinfo  {journal} {Phys. Rept.}\ }\textbf {\bibinfo {volume}
  {407}},\ \bibinfo {pages} {205} (\bibinfo {year} {2005})},\ \Eprint
  {http://arxiv.org/abs/hep-ph/0402234} {arXiv:hep-ph/0402234 [hep-ph]}
  \BibitemShut {NoStop}%
\bibitem [{\citenamefont {Volkov}\ and\ \citenamefont
  {Radzhabov}(2006)}]{Volkov:2005kw}%
  \BibitemOpen
  \bibfield  {author} {\bibinfo {author} {\bibfnamefont {M.~K.}\ \bibnamefont
  {Volkov}}\ and\ \bibinfo {author} {\bibfnamefont {A.~E.}\ \bibnamefont
  {Radzhabov}},\ }\href {\doibase 10.1070/PU2006v049n06ABEH005905} {\bibfield
  {journal} {\bibinfo  {journal} {Phys. Usp.}\ }\textbf {\bibinfo {volume}
  {49}},\ \bibinfo {pages} {551} (\bibinfo {year} {2006})},\ \Eprint
  {http://arxiv.org/abs/hep-ph/0508263} {arXiv:hep-ph/0508263} \BibitemShut
  {NoStop}%
\bibitem [{\citenamefont {McLerran}\ and\ \citenamefont
  {Svetitsky}(1981)}]{McLerran:1981pb}%
  \BibitemOpen
  \bibfield  {author} {\bibinfo {author} {\bibfnamefont {L.~D.}\ \bibnamefont
  {McLerran}}\ and\ \bibinfo {author} {\bibfnamefont {B.}~\bibnamefont
  {Svetitsky}},\ }\href {\doibase 10.1103/PhysRevD.24.450} {\bibfield
  {journal} {\bibinfo  {journal} {Phys. Rev. D}\ }\textbf {\bibinfo {volume}
  {24}},\ \bibinfo {pages} {450} (\bibinfo {year} {1981})}\BibitemShut
  {NoStop}%
\bibitem [{\citenamefont {Ratti}\ \emph {et~al.}(2006)\citenamefont {Ratti},
  \citenamefont {Thaler},\ and\ \citenamefont {Weise}}]{Ratti:2005jh}%
  \BibitemOpen
  \bibfield  {author} {\bibinfo {author} {\bibfnamefont {C.}~\bibnamefont
  {Ratti}}, \bibinfo {author} {\bibfnamefont {M.~A.}\ \bibnamefont {Thaler}}, \
  and\ \bibinfo {author} {\bibfnamefont {W.}~\bibnamefont {Weise}},\ }\href
  {\doibase 10.1103/PhysRevD.73.014019} {\bibfield  {journal} {\bibinfo
  {journal} {Phys. Rev. D}\ }\textbf {\bibinfo {volume} {73}},\ \bibinfo
  {pages} {014019} (\bibinfo {year} {2006})},\ \Eprint
  {http://arxiv.org/abs/hep-ph/0506234} {arXiv:hep-ph/0506234} \BibitemShut
  {NoStop}%
\bibitem [{\citenamefont {Ratti}\ \emph {et~al.}(2007)\citenamefont {Ratti},
  \citenamefont {Roessner}, \citenamefont {Thaler},\ and\ \citenamefont
  {Weise}}]{Ratti:2006wg}%
  \BibitemOpen
  \bibfield  {author} {\bibinfo {author} {\bibfnamefont {C.}~\bibnamefont
  {Ratti}}, \bibinfo {author} {\bibfnamefont {S.}~\bibnamefont {Roessner}},
  \bibinfo {author} {\bibfnamefont {M.}~\bibnamefont {Thaler}}, \ and\ \bibinfo
  {author} {\bibfnamefont {W.}~\bibnamefont {Weise}},\ }\href {\doibase
  10.1140/epjc/s10052-006-0065-x} {\bibfield  {journal} {\bibinfo  {journal}
  {Eur. Phys. J. C}\ }\textbf {\bibinfo {volume} {49}},\ \bibinfo {pages} {213}
  (\bibinfo {year} {2007})},\ \Eprint {http://arxiv.org/abs/hep-ph/0609218}
  {arXiv:hep-ph/0609218} \BibitemShut {NoStop}%
\bibitem [{\citenamefont {Shifman}(1989)}]{Shifman:1988zk}%
  \BibitemOpen
  \bibfield  {author} {\bibinfo {author} {\bibfnamefont {M.~A.}\ \bibnamefont
  {Shifman}},\ }\href {\doibase 10.1016/0370-1573(91)90020-M} {\bibfield
  {journal} {\bibinfo  {journal} {Sov. Phys. Usp.}\ }\textbf {\bibinfo {volume}
  {32}},\ \bibinfo {pages} {289} (\bibinfo {year} {1989})}\BibitemShut
  {NoStop}%
\bibitem [{\citenamefont {Belavin}\ \emph {et~al.}(1975)\citenamefont
  {Belavin}, \citenamefont {Polyakov}, \citenamefont {Schwartz},\ and\
  \citenamefont {Tyupkin}}]{Belavin:1975fg}%
  \BibitemOpen
  \bibfield  {author} {\bibinfo {author} {\bibfnamefont {A.~A.}\ \bibnamefont
  {Belavin}}, \bibinfo {author} {\bibfnamefont {A.~M.}\ \bibnamefont
  {Polyakov}}, \bibinfo {author} {\bibfnamefont {A.~S.}\ \bibnamefont
  {Schwartz}}, \ and\ \bibinfo {author} {\bibfnamefont {Y.~S.}\ \bibnamefont
  {Tyupkin}},\ }\href {\doibase 10.1016/0370-2693(75)90163-X} {\bibfield
  {journal} {\bibinfo  {journal} {Phys. Lett. B}\ }\textbf {\bibinfo {volume}
  {59}},\ \bibinfo {pages} {85} (\bibinfo {year} {1975})}\BibitemShut {NoStop}%
\bibitem [{\citenamefont {'t~Hooft}(1976{\natexlab{a}})}]{tHooft:1976rip}%
  \BibitemOpen
  \bibfield  {author} {\bibinfo {author} {\bibfnamefont {G.}~\bibnamefont
  {'t~Hooft}},\ }\href {\doibase 10.1103/PhysRevLett.37.8} {\bibfield
  {journal} {\bibinfo  {journal} {Phys. Rev. Lett.}\ }\textbf {\bibinfo
  {volume} {37}},\ \bibinfo {pages} {8} (\bibinfo {year}
  {1976}{\natexlab{a}})}\BibitemShut {NoStop}%
\bibitem [{\citenamefont {'t~Hooft}(1976{\natexlab{b}})}]{tHooft:1976snw}%
  \BibitemOpen
  \bibfield  {author} {\bibinfo {author} {\bibfnamefont {G.}~\bibnamefont
  {'t~Hooft}},\ }\href {\doibase 10.1103/PhysRevD.14.3432} {\bibfield
  {journal} {\bibinfo  {journal} {Phys. Rev. D}\ }\textbf {\bibinfo {volume}
  {14}},\ \bibinfo {pages} {3432} (\bibinfo {year} {1976}{\natexlab{b}})},\
  \bibinfo {note} {[Erratum: Phys.Rev.D 18, 2199 (1978)]}\BibitemShut {NoStop}%
\bibitem [{\citenamefont {Sch\"afer}\ and\ \citenamefont
  {Shuryak}(1998)}]{Schafer:1996wv}%
  \BibitemOpen
  \bibfield  {author} {\bibinfo {author} {\bibfnamefont {T.}~\bibnamefont
  {Sch\"afer}}\ and\ \bibinfo {author} {\bibfnamefont {E.~V.}\ \bibnamefont
  {Shuryak}},\ }\href {\doibase 10.1103/RevModPhys.70.323} {\bibfield
  {journal} {\bibinfo  {journal} {Rev. Mod. Phys.}\ }\textbf {\bibinfo {volume}
  {70}},\ \bibinfo {pages} {323} (\bibinfo {year} {1998})},\ \Eprint
  {http://arxiv.org/abs/hep-ph/9610451} {arXiv:hep-ph/9610451} \BibitemShut
  {NoStop}%
\bibitem [{\citenamefont {Manton}(1983)}]{Manton:1983nd}%
  \BibitemOpen
  \bibfield  {author} {\bibinfo {author} {\bibfnamefont {N.~S.}\ \bibnamefont
  {Manton}},\ }\href {\doibase 10.1103/PhysRevD.28.2019} {\bibfield  {journal}
  {\bibinfo  {journal} {Phys. Rev. D}\ }\textbf {\bibinfo {volume} {28}},\
  \bibinfo {pages} {2019} (\bibinfo {year} {1983})}\BibitemShut {NoStop}%
\bibitem [{\citenamefont {Klinkhamer}\ and\ \citenamefont
  {Manton}(1984)}]{Klinkhamer:1984di}%
  \BibitemOpen
  \bibfield  {author} {\bibinfo {author} {\bibfnamefont {F.~R.}\ \bibnamefont
  {Klinkhamer}}\ and\ \bibinfo {author} {\bibfnamefont {N.~S.}\ \bibnamefont
  {Manton}},\ }\href {\doibase 10.1103/PhysRevD.30.2212} {\bibfield  {journal}
  {\bibinfo  {journal} {Phys. Rev. D}\ }\textbf {\bibinfo {volume} {30}},\
  \bibinfo {pages} {2212} (\bibinfo {year} {1984})}\BibitemShut {NoStop}%
\bibitem [{\citenamefont {Kuzmin}\ \emph {et~al.}(1985)\citenamefont {Kuzmin},
  \citenamefont {Rubakov},\ and\ \citenamefont {Shaposhnikov}}]{Kuzmin:1985mm}%
  \BibitemOpen
  \bibfield  {author} {\bibinfo {author} {\bibfnamefont {V.~A.}\ \bibnamefont
  {Kuzmin}}, \bibinfo {author} {\bibfnamefont {V.~A.}\ \bibnamefont {Rubakov}},
  \ and\ \bibinfo {author} {\bibfnamefont {M.~E.}\ \bibnamefont
  {Shaposhnikov}},\ }\href {\doibase 10.1016/0370-2693(85)91028-7} {\bibfield
  {journal} {\bibinfo  {journal} {Phys. Lett. B}\ }\textbf {\bibinfo {volume}
  {155}},\ \bibinfo {pages} {36} (\bibinfo {year} {1985})}\BibitemShut
  {NoStop}%
\bibitem [{\citenamefont {Arnold}\ and\ \citenamefont
  {McLerran}(1987)}]{Arnold:1987mh}%
  \BibitemOpen
  \bibfield  {author} {\bibinfo {author} {\bibfnamefont {P.~B.}\ \bibnamefont
  {Arnold}}\ and\ \bibinfo {author} {\bibfnamefont {L.~D.}\ \bibnamefont
  {McLerran}},\ }\href {\doibase 10.1103/PhysRevD.36.581} {\bibfield  {journal}
  {\bibinfo  {journal} {Phys. Rev. D}\ }\textbf {\bibinfo {volume} {36}},\
  \bibinfo {pages} {581} (\bibinfo {year} {1987})}\BibitemShut {NoStop}%
\bibitem [{\citenamefont {Khlebnikov}\ and\ \citenamefont
  {Shaposhnikov}(1988)}]{Khlebnikov:1988sr}%
  \BibitemOpen
  \bibfield  {author} {\bibinfo {author} {\bibfnamefont {S.~Y.}\ \bibnamefont
  {Khlebnikov}}\ and\ \bibinfo {author} {\bibfnamefont {M.~E.}\ \bibnamefont
  {Shaposhnikov}},\ }\href {\doibase 10.1016/0550-3213(88)90133-2} {\bibfield
  {journal} {\bibinfo  {journal} {Nucl. Phys. B}\ }\textbf {\bibinfo {volume}
  {308}},\ \bibinfo {pages} {885} (\bibinfo {year} {1988})}\BibitemShut
  {NoStop}%
\bibitem [{\citenamefont {Arnold}\ and\ \citenamefont
  {McLerran}(1988)}]{Arnold:1987zg}%
  \BibitemOpen
  \bibfield  {author} {\bibinfo {author} {\bibfnamefont {P.~B.}\ \bibnamefont
  {Arnold}}\ and\ \bibinfo {author} {\bibfnamefont {L.~D.}\ \bibnamefont
  {McLerran}},\ }\href {\doibase 10.1103/PhysRevD.37.1020} {\bibfield
  {journal} {\bibinfo  {journal} {Phys. Rev. D}\ }\textbf {\bibinfo {volume}
  {37}},\ \bibinfo {pages} {1020} (\bibinfo {year} {1988})}\BibitemShut
  {NoStop}%
\bibitem [{\citenamefont {McLerran}\ \emph {et~al.}(1991)\citenamefont
  {McLerran}, \citenamefont {Mottola},\ and\ \citenamefont
  {Shaposhnikov}}]{McLerran:1990de}%
  \BibitemOpen
  \bibfield  {author} {\bibinfo {author} {\bibfnamefont {L.~D.}\ \bibnamefont
  {McLerran}}, \bibinfo {author} {\bibfnamefont {E.}~\bibnamefont {Mottola}}, \
  and\ \bibinfo {author} {\bibfnamefont {M.~E.}\ \bibnamefont {Shaposhnikov}},\
  }\href {\doibase 10.1103/PhysRevD.43.2027} {\bibfield  {journal} {\bibinfo
  {journal} {Phys. Rev. D}\ }\textbf {\bibinfo {volume} {43}},\ \bibinfo
  {pages} {2027} (\bibinfo {year} {1991})}\BibitemShut {NoStop}%
\bibitem [{\citenamefont {Moore}\ and\ \citenamefont
  {Tassler}(2011)}]{Moore:2010jd}%
  \BibitemOpen
  \bibfield  {author} {\bibinfo {author} {\bibfnamefont {G.~D.}\ \bibnamefont
  {Moore}}\ and\ \bibinfo {author} {\bibfnamefont {M.}~\bibnamefont
  {Tassler}},\ }\href {\doibase 10.1007/JHEP02(2011)105} {\bibfield  {journal}
  {\bibinfo  {journal} {JHEP}\ }\textbf {\bibinfo {volume} {02}},\ \bibinfo
  {pages} {105} (\bibinfo {year} {2011})},\ \Eprint
  {http://arxiv.org/abs/1011.1167} {arXiv:1011.1167 [hep-ph]} \BibitemShut
  {NoStop}%
\bibitem [{\citenamefont {Fukushima}\ \emph {et~al.}(2008)\citenamefont
  {Fukushima}, \citenamefont {Kharzeev},\ and\ \citenamefont
  {Warringa}}]{Fukushima:2008xe}%
  \BibitemOpen
  \bibfield  {author} {\bibinfo {author} {\bibfnamefont {K.}~\bibnamefont
  {Fukushima}}, \bibinfo {author} {\bibfnamefont {D.~E.}\ \bibnamefont
  {Kharzeev}}, \ and\ \bibinfo {author} {\bibfnamefont {H.~J.}\ \bibnamefont
  {Warringa}},\ }\href {\doibase 10.1103/PhysRevD.78.074033} {\bibfield
  {journal} {\bibinfo  {journal} {Phys. Rev.}\ }\textbf {\bibinfo {volume}
  {D78}},\ \bibinfo {pages} {074033} (\bibinfo {year} {2008})},\ \Eprint
  {http://arxiv.org/abs/0808.3382} {arXiv:0808.3382 [hep-ph]} \BibitemShut
  {NoStop}%
\bibitem [{\citenamefont {Kharzeev}\ and\ \citenamefont
  {Warringa}(2009)}]{Kharzeev:2009pj}%
  \BibitemOpen
  \bibfield  {author} {\bibinfo {author} {\bibfnamefont {D.~E.}\ \bibnamefont
  {Kharzeev}}\ and\ \bibinfo {author} {\bibfnamefont {H.~J.}\ \bibnamefont
  {Warringa}},\ }\href {\doibase 10.1103/PhysRevD.80.034028} {\bibfield
  {journal} {\bibinfo  {journal} {Phys. Rev.}\ }\textbf {\bibinfo {volume}
  {D80}},\ \bibinfo {pages} {034028} (\bibinfo {year} {2009})},\ \Eprint
  {http://arxiv.org/abs/0907.5007} {arXiv:0907.5007 [hep-ph]} \BibitemShut
  {NoStop}%
\bibitem [{\citenamefont {Bali}\ \emph {et~al.}(2012)\citenamefont {Bali},
  \citenamefont {Bruckmann}, \citenamefont {Endrodi}, \citenamefont {Fodor},
  \citenamefont {Katz}, \citenamefont {Krieg}, \citenamefont {Schafer},\ and\
  \citenamefont {Szabo}}]{Bali:2011qj}%
  \BibitemOpen
  \bibfield  {author} {\bibinfo {author} {\bibfnamefont {G.~S.}\ \bibnamefont
  {Bali}}, \bibinfo {author} {\bibfnamefont {F.}~\bibnamefont {Bruckmann}},
  \bibinfo {author} {\bibfnamefont {G.}~\bibnamefont {Endrodi}}, \bibinfo
  {author} {\bibfnamefont {Z.}~\bibnamefont {Fodor}}, \bibinfo {author}
  {\bibfnamefont {S.~D.}\ \bibnamefont {Katz}}, \bibinfo {author}
  {\bibfnamefont {S.}~\bibnamefont {Krieg}}, \bibinfo {author} {\bibfnamefont
  {A.}~\bibnamefont {Schafer}}, \ and\ \bibinfo {author} {\bibfnamefont
  {K.~K.}\ \bibnamefont {Szabo}},\ }\href {\doibase 10.1007/JHEP02(2012)044}
  {\bibfield  {journal} {\bibinfo  {journal} {JHEP}\ }\textbf {\bibinfo
  {volume} {02}},\ \bibinfo {pages} {044} (\bibinfo {year} {2012})},\ \Eprint
  {http://arxiv.org/abs/1111.4956} {arXiv:1111.4956 [hep-lat]} \BibitemShut
  {NoStop}%
\bibitem [{\citenamefont {Vilenkin}(1979)}]{Vilenkin:1979ui}%
  \BibitemOpen
  \bibfield  {author} {\bibinfo {author} {\bibfnamefont {A.}~\bibnamefont
  {Vilenkin}},\ }\href {\doibase 10.1103/PhysRevD.20.1807} {\bibfield
  {journal} {\bibinfo  {journal} {Phys. Rev. D}\ }\textbf {\bibinfo {volume}
  {20}},\ \bibinfo {pages} {1807} (\bibinfo {year} {1979})}\BibitemShut
  {NoStop}%
\bibitem [{\citenamefont {Vilenkin}(1980)}]{Vilenkin:1980fu}%
  \BibitemOpen
  \bibfield  {author} {\bibinfo {author} {\bibfnamefont {A.}~\bibnamefont
  {Vilenkin}},\ }\href {\doibase 10.1103/PhysRevD.22.3080} {\bibfield
  {journal} {\bibinfo  {journal} {Phys. Rev. D}\ }\textbf {\bibinfo {volume}
  {22}},\ \bibinfo {pages} {3080} (\bibinfo {year} {1980})}\BibitemShut
  {NoStop}%
\bibitem [{\citenamefont {Son}\ and\ \citenamefont
  {Surowka}(2009)}]{Son:2009tf}%
  \BibitemOpen
  \bibfield  {author} {\bibinfo {author} {\bibfnamefont {D.~T.}\ \bibnamefont
  {Son}}\ and\ \bibinfo {author} {\bibfnamefont {P.}~\bibnamefont {Surowka}},\
  }\href {\doibase 10.1103/PhysRevLett.103.191601} {\bibfield  {journal}
  {\bibinfo  {journal} {Phys. Rev. Lett.}\ }\textbf {\bibinfo {volume} {103}},\
  \bibinfo {pages} {191601} (\bibinfo {year} {2009})},\ \Eprint
  {http://arxiv.org/abs/0906.5044} {arXiv:0906.5044 [hep-th]} \BibitemShut
  {NoStop}%
\bibitem [{\citenamefont {Akamatsu}\ and\ \citenamefont
  {Yamamoto}(2013)}]{Akamatsu:2013pjd}%
  \BibitemOpen
  \bibfield  {author} {\bibinfo {author} {\bibfnamefont {Y.}~\bibnamefont
  {Akamatsu}}\ and\ \bibinfo {author} {\bibfnamefont {N.}~\bibnamefont
  {Yamamoto}},\ }\href {\doibase 10.1103/PhysRevLett.111.052002} {\bibfield
  {journal} {\bibinfo  {journal} {Phys. Rev. Lett.}\ }\textbf {\bibinfo
  {volume} {111}},\ \bibinfo {pages} {052002} (\bibinfo {year} {2013})},\
  \Eprint {http://arxiv.org/abs/1302.2125} {arXiv:1302.2125 [nucl-th]}
  \BibitemShut {NoStop}%
\bibitem [{\citenamefont {Carignano}\ and\ \citenamefont
  {Manuel}(2019)}]{Carignano:2018thu}%
  \BibitemOpen
  \bibfield  {author} {\bibinfo {author} {\bibfnamefont {S.}~\bibnamefont
  {Carignano}}\ and\ \bibinfo {author} {\bibfnamefont {C.}~\bibnamefont
  {Manuel}},\ }\href {\doibase 10.1103/PhysRevD.99.096022} {\bibfield
  {journal} {\bibinfo  {journal} {Phys. Rev. D}\ }\textbf {\bibinfo {volume}
  {99}},\ \bibinfo {pages} {096022} (\bibinfo {year} {2019})},\ \Eprint
  {http://arxiv.org/abs/1811.06394} {arXiv:1811.06394 [hep-ph]} \BibitemShut
  {NoStop}%
\bibitem [{\citenamefont {Carignano}\ and\ \citenamefont
  {Manuel}(2021)}]{Carignano:2021mrn}%
  \BibitemOpen
  \bibfield  {author} {\bibinfo {author} {\bibfnamefont {S.}~\bibnamefont
  {Carignano}}\ and\ \bibinfo {author} {\bibfnamefont {C.}~\bibnamefont
  {Manuel}},\ }\href {\doibase 10.1103/PhysRevD.103.116002} {\bibfield
  {journal} {\bibinfo  {journal} {Phys. Rev. D}\ }\textbf {\bibinfo {volume}
  {103}},\ \bibinfo {pages} {116002} (\bibinfo {year} {2021})},\ \Eprint
  {http://arxiv.org/abs/2103.02491} {arXiv:2103.02491 [hep-ph]} \BibitemShut
  {NoStop}%
\bibitem [{\citenamefont {Carignano}\ and\ \citenamefont
  {Buballa}(2020)}]{Carignano:2019ivp}%
  \BibitemOpen
  \bibfield  {author} {\bibinfo {author} {\bibfnamefont {S.}~\bibnamefont
  {Carignano}}\ and\ \bibinfo {author} {\bibfnamefont {M.}~\bibnamefont
  {Buballa}},\ }\href {\doibase 10.1103/PhysRevD.101.014026} {\bibfield
  {journal} {\bibinfo  {journal} {Phys. Rev. D}\ }\textbf {\bibinfo {volume}
  {101}},\ \bibinfo {pages} {014026} (\bibinfo {year} {2020})},\ \Eprint
  {http://arxiv.org/abs/1910.03604} {arXiv:1910.03604 [hep-ph]} \BibitemShut
  {NoStop}%
\bibitem [{\citenamefont {Kharzeev}(2014)}]{Kharzeev:2013ffa}%
  \BibitemOpen
  \bibfield  {author} {\bibinfo {author} {\bibfnamefont {D.~E.}\ \bibnamefont
  {Kharzeev}},\ }\href {\doibase 10.1016/j.ppnp.2014.01.002} {\bibfield
  {journal} {\bibinfo  {journal} {Prog. Part. Nucl. Phys.}\ }\textbf {\bibinfo
  {volume} {75}},\ \bibinfo {pages} {133} (\bibinfo {year} {2014})},\ \Eprint
  {http://arxiv.org/abs/1312.3348} {arXiv:1312.3348 [hep-ph]} \BibitemShut
  {NoStop}%
\bibitem [{\citenamefont {Kharzeev}\ \emph {et~al.}(2016)\citenamefont
  {Kharzeev}, \citenamefont {Liao}, \citenamefont {Voloshin},\ and\
  \citenamefont {Wang}}]{Kharzeev:2015znc}%
  \BibitemOpen
  \bibfield  {author} {\bibinfo {author} {\bibfnamefont {D.~E.}\ \bibnamefont
  {Kharzeev}}, \bibinfo {author} {\bibfnamefont {J.}~\bibnamefont {Liao}},
  \bibinfo {author} {\bibfnamefont {S.~A.}\ \bibnamefont {Voloshin}}, \ and\
  \bibinfo {author} {\bibfnamefont {G.}~\bibnamefont {Wang}},\ }\href {\doibase
  10.1016/j.ppnp.2016.01.001} {\bibfield  {journal} {\bibinfo  {journal} {Prog.
  Part. Nucl. Phys.}\ }\textbf {\bibinfo {volume} {88}},\ \bibinfo {pages} {1}
  (\bibinfo {year} {2016})},\ \Eprint {http://arxiv.org/abs/1511.04050}
  {arXiv:1511.04050 [hep-ph]} \BibitemShut {NoStop}%
\bibitem [{\citenamefont {Huang}(2016)}]{Huang:2015oca}%
  \BibitemOpen
  \bibfield  {author} {\bibinfo {author} {\bibfnamefont {X.-G.}\ \bibnamefont
  {Huang}},\ }\href {\doibase 10.1088/0034-4885/79/7/076302} {\bibfield
  {journal} {\bibinfo  {journal} {Rept. Prog. Phys.}\ }\textbf {\bibinfo
  {volume} {79}},\ \bibinfo {pages} {076302} (\bibinfo {year} {2016})},\
  \Eprint {http://arxiv.org/abs/1509.04073} {arXiv:1509.04073 [nucl-th]}
  \BibitemShut {NoStop}%
\bibitem [{\citenamefont {Landsteiner}(2016)}]{Landsteiner:2016led}%
  \BibitemOpen
  \bibfield  {author} {\bibinfo {author} {\bibfnamefont {K.}~\bibnamefont
  {Landsteiner}},\ }\href {\doibase 10.5506/APhysPolB.47.2617} {\bibfield
  {journal} {\bibinfo  {journal} {Acta Phys. Polon. B}\ }\textbf {\bibinfo
  {volume} {47}},\ \bibinfo {pages} {2617} (\bibinfo {year} {2016})},\ \Eprint
  {http://arxiv.org/abs/1610.04413} {arXiv:1610.04413 [hep-th]} \BibitemShut
  {NoStop}%
\bibitem [{\citenamefont {Gorbar}\ \emph {et~al.}(2018)\citenamefont {Gorbar},
  \citenamefont {Miransky}, \citenamefont {Shovkovy},\ and\ \citenamefont
  {Sukhachov}}]{Gorbar:2017lnp}%
  \BibitemOpen
  \bibfield  {author} {\bibinfo {author} {\bibfnamefont {E.~V.}\ \bibnamefont
  {Gorbar}}, \bibinfo {author} {\bibfnamefont {V.~A.}\ \bibnamefont
  {Miransky}}, \bibinfo {author} {\bibfnamefont {I.~A.}\ \bibnamefont
  {Shovkovy}}, \ and\ \bibinfo {author} {\bibfnamefont {P.~O.}\ \bibnamefont
  {Sukhachov}},\ }\href {\doibase 10.1063/1.5037551} {\bibfield  {journal}
  {\bibinfo  {journal} {Low Temp. Phys.}\ }\textbf {\bibinfo {volume} {44}},\
  \bibinfo {pages} {487} (\bibinfo {year} {2018})},\ \Eprint
  {http://arxiv.org/abs/1712.08947} {arXiv:1712.08947 [cond-mat.mes-hall]}
  \BibitemShut {NoStop}%
\bibitem [{\citenamefont {Joyce}\ and\ \citenamefont
  {Shaposhnikov}(1997)}]{Joyce:1997uy}%
  \BibitemOpen
  \bibfield  {author} {\bibinfo {author} {\bibfnamefont {M.}~\bibnamefont
  {Joyce}}\ and\ \bibinfo {author} {\bibfnamefont {M.~E.}\ \bibnamefont
  {Shaposhnikov}},\ }\href {\doibase 10.1103/PhysRevLett.79.1193} {\bibfield
  {journal} {\bibinfo  {journal} {Phys. Rev. Lett.}\ }\textbf {\bibinfo
  {volume} {79}},\ \bibinfo {pages} {1193} (\bibinfo {year} {1997})},\ \Eprint
  {http://arxiv.org/abs/astro-ph/9703005} {arXiv:astro-ph/9703005} \BibitemShut
  {NoStop}%
\bibitem [{\citenamefont {Tashiro}\ \emph {et~al.}(2012)\citenamefont
  {Tashiro}, \citenamefont {Vachaspati},\ and\ \citenamefont
  {Vilenkin}}]{Tashiro:2012mf}%
  \BibitemOpen
  \bibfield  {author} {\bibinfo {author} {\bibfnamefont {H.}~\bibnamefont
  {Tashiro}}, \bibinfo {author} {\bibfnamefont {T.}~\bibnamefont {Vachaspati}},
  \ and\ \bibinfo {author} {\bibfnamefont {A.}~\bibnamefont {Vilenkin}},\
  }\href {\doibase 10.1103/PhysRevD.86.105033} {\bibfield  {journal} {\bibinfo
  {journal} {Phys. Rev. D}\ }\textbf {\bibinfo {volume} {86}},\ \bibinfo
  {pages} {105033} (\bibinfo {year} {2012})},\ \Eprint
  {http://arxiv.org/abs/1206.5549} {arXiv:1206.5549 [astro-ph.CO]} \BibitemShut
  {NoStop}%
\bibitem [{\citenamefont {Ruggieri}\ and\ \citenamefont
  {Peng}(2016)}]{Ruggieri:2016lrn}%
  \BibitemOpen
  \bibfield  {author} {\bibinfo {author} {\bibfnamefont {M.}~\bibnamefont
  {Ruggieri}}\ and\ \bibinfo {author} {\bibfnamefont {G.~X.}\ \bibnamefont
  {Peng}},\ }\href {\doibase 10.1103/PhysRevD.93.094021} {\bibfield  {journal}
  {\bibinfo  {journal} {Phys. Rev. D}\ }\textbf {\bibinfo {volume} {93}},\
  \bibinfo {pages} {094021} (\bibinfo {year} {2016})},\ \Eprint
  {http://arxiv.org/abs/1602.08994} {arXiv:1602.08994 [hep-ph]} \BibitemShut
  {NoStop}%
\bibitem [{\citenamefont {Ruggieri}\ \emph {et~al.}(2016)\citenamefont
  {Ruggieri}, \citenamefont {Peng},\ and\ \citenamefont
  {Chernodub}}]{Ruggieri:2016asg}%
  \BibitemOpen
  \bibfield  {author} {\bibinfo {author} {\bibfnamefont {M.}~\bibnamefont
  {Ruggieri}}, \bibinfo {author} {\bibfnamefont {G.~X.}\ \bibnamefont {Peng}},
  \ and\ \bibinfo {author} {\bibfnamefont {M.}~\bibnamefont {Chernodub}},\
  }\href {\doibase 10.1103/PhysRevD.94.054011} {\bibfield  {journal} {\bibinfo
  {journal} {Phys. Rev. D}\ }\textbf {\bibinfo {volume} {94}},\ \bibinfo
  {pages} {054011} (\bibinfo {year} {2016})},\ \Eprint
  {http://arxiv.org/abs/1606.03287} {arXiv:1606.03287 [hep-ph]} \BibitemShut
  {NoStop}%
\bibitem [{\citenamefont {Ruggieri}\ \emph {et~al.}(2020)\citenamefont
  {Ruggieri}, \citenamefont {Chernodub},\ and\ \citenamefont
  {Lu}}]{Ruggieri:2020qtq}%
  \BibitemOpen
  \bibfield  {author} {\bibinfo {author} {\bibfnamefont {M.}~\bibnamefont
  {Ruggieri}}, \bibinfo {author} {\bibfnamefont {M.~N.}\ \bibnamefont
  {Chernodub}}, \ and\ \bibinfo {author} {\bibfnamefont {Z.-Y.}\ \bibnamefont
  {Lu}},\ }\href {\doibase 10.1103/PhysRevD.102.014031} {\bibfield  {journal}
  {\bibinfo  {journal} {Phys. Rev. D}\ }\textbf {\bibinfo {volume} {102}},\
  \bibinfo {pages} {014031} (\bibinfo {year} {2020})},\ \Eprint
  {http://arxiv.org/abs/2004.09393} {arXiv:2004.09393 [hep-ph]} \BibitemShut
  {NoStop}%
\bibitem [{\citenamefont {Fukushima}\ \emph {et~al.}(2010)\citenamefont
  {Fukushima}, \citenamefont {Ruggieri},\ and\ \citenamefont
  {Gatto}}]{Fukushima:2010fe}%
  \BibitemOpen
  \bibfield  {author} {\bibinfo {author} {\bibfnamefont {K.}~\bibnamefont
  {Fukushima}}, \bibinfo {author} {\bibfnamefont {M.}~\bibnamefont {Ruggieri}},
  \ and\ \bibinfo {author} {\bibfnamefont {R.}~\bibnamefont {Gatto}},\ }\href
  {\doibase 10.1103/PhysRevD.81.114031} {\bibfield  {journal} {\bibinfo
  {journal} {Phys. Rev. D}\ }\textbf {\bibinfo {volume} {81}},\ \bibinfo
  {pages} {114031} (\bibinfo {year} {2010})},\ \Eprint
  {http://arxiv.org/abs/1003.0047} {arXiv:1003.0047 [hep-ph]} \BibitemShut
  {NoStop}%
\bibitem [{\citenamefont {Yu}\ \emph {et~al.}(2015)\citenamefont {Yu},
  \citenamefont {Van~Doorsselaere},\ and\ \citenamefont {Huang}}]{Yu:2014xoa}%
  \BibitemOpen
  \bibfield  {author} {\bibinfo {author} {\bibfnamefont {L.}~\bibnamefont
  {Yu}}, \bibinfo {author} {\bibfnamefont {J.}~\bibnamefont
  {Van~Doorsselaere}}, \ and\ \bibinfo {author} {\bibfnamefont
  {M.}~\bibnamefont {Huang}},\ }\href {\doibase 10.1103/PhysRevD.91.074011}
  {\bibfield  {journal} {\bibinfo  {journal} {Phys. Rev. D}\ }\textbf {\bibinfo
  {volume} {91}},\ \bibinfo {pages} {074011} (\bibinfo {year} {2015})},\
  \Eprint {http://arxiv.org/abs/1411.7552} {arXiv:1411.7552 [hep-ph]}
  \BibitemShut {NoStop}%
\bibitem [{\citenamefont {Yu}\ \emph {et~al.}(2016)\citenamefont {Yu},
  \citenamefont {Liu},\ and\ \citenamefont {Huang}}]{Yu:2015hym}%
  \BibitemOpen
  \bibfield  {author} {\bibinfo {author} {\bibfnamefont {L.}~\bibnamefont
  {Yu}}, \bibinfo {author} {\bibfnamefont {H.}~\bibnamefont {Liu}}, \ and\
  \bibinfo {author} {\bibfnamefont {M.}~\bibnamefont {Huang}},\ }\href
  {\doibase 10.1103/PhysRevD.94.014026} {\bibfield  {journal} {\bibinfo
  {journal} {Phys. Rev. D}\ }\textbf {\bibinfo {volume} {94}},\ \bibinfo
  {pages} {014026} (\bibinfo {year} {2016})},\ \Eprint
  {http://arxiv.org/abs/1511.03073} {arXiv:1511.03073 [hep-ph]} \BibitemShut
  {NoStop}%
\bibitem [{\citenamefont {Chao}\ \emph {et~al.}(2013)\citenamefont {Chao},
  \citenamefont {Chu},\ and\ \citenamefont {Huang}}]{Chao:2013qpa}%
  \BibitemOpen
  \bibfield  {author} {\bibinfo {author} {\bibfnamefont {J.}~\bibnamefont
  {Chao}}, \bibinfo {author} {\bibfnamefont {P.}~\bibnamefont {Chu}}, \ and\
  \bibinfo {author} {\bibfnamefont {M.}~\bibnamefont {Huang}},\ }\href
  {\doibase 10.1103/PhysRevD.88.054009} {\bibfield  {journal} {\bibinfo
  {journal} {Phys. Rev. D}\ }\textbf {\bibinfo {volume} {88}},\ \bibinfo
  {pages} {054009} (\bibinfo {year} {2013})},\ \Eprint
  {http://arxiv.org/abs/1305.1100} {arXiv:1305.1100 [hep-ph]} \BibitemShut
  {NoStop}%
\bibitem [{\citenamefont {Gusynin}\ \emph {et~al.}(1994)\citenamefont
  {Gusynin}, \citenamefont {Miransky},\ and\ \citenamefont
  {Shovkovy}}]{Gusynin:1994re}%
  \BibitemOpen
  \bibfield  {author} {\bibinfo {author} {\bibfnamefont {V.~P.}\ \bibnamefont
  {Gusynin}}, \bibinfo {author} {\bibfnamefont {V.~A.}\ \bibnamefont
  {Miransky}}, \ and\ \bibinfo {author} {\bibfnamefont {I.~A.}\ \bibnamefont
  {Shovkovy}},\ }\href {\doibase 10.1103/PhysRevLett.76.1005,
  10.1103/PhysRevLett.73.3499} {\bibfield  {journal} {\bibinfo  {journal}
  {Phys. Rev. Lett.}\ }\textbf {\bibinfo {volume} {73}},\ \bibinfo {pages}
  {3499} (\bibinfo {year} {1994})},\ \bibinfo {note} {[Erratum: Phys. Rev.
  Lett.76,1005(1996)]},\ \Eprint {http://arxiv.org/abs/hep-ph/9405262}
  {arXiv:hep-ph/9405262 [hep-ph]} \BibitemShut {NoStop}%
\bibitem [{\citenamefont {Gusynin}\ \emph {et~al.}(1996)\citenamefont
  {Gusynin}, \citenamefont {Miransky},\ and\ \citenamefont
  {Shovkovy}}]{Gusynin:1995nb}%
  \BibitemOpen
  \bibfield  {author} {\bibinfo {author} {\bibfnamefont {V.~P.}\ \bibnamefont
  {Gusynin}}, \bibinfo {author} {\bibfnamefont {V.~A.}\ \bibnamefont
  {Miransky}}, \ and\ \bibinfo {author} {\bibfnamefont {I.~A.}\ \bibnamefont
  {Shovkovy}},\ }\href {\doibase 10.1016/0550-3213(96)00021-1} {\bibfield
  {journal} {\bibinfo  {journal} {Nucl. Phys.}\ }\textbf {\bibinfo {volume}
  {B462}},\ \bibinfo {pages} {249} (\bibinfo {year} {1996})},\ \Eprint
  {http://arxiv.org/abs/hep-ph/9509320} {arXiv:hep-ph/9509320 [hep-ph]}
  \BibitemShut {NoStop}%
\bibitem [{\citenamefont {Fayazbakhsh}\ \emph {et~al.}(2012)\citenamefont
  {Fayazbakhsh}, \citenamefont {Sadeghian},\ and\ \citenamefont
  {Sadooghi}}]{Fayazbakhsh:2012vr}%
  \BibitemOpen
  \bibfield  {author} {\bibinfo {author} {\bibfnamefont {S.}~\bibnamefont
  {Fayazbakhsh}}, \bibinfo {author} {\bibfnamefont {S.}~\bibnamefont
  {Sadeghian}}, \ and\ \bibinfo {author} {\bibfnamefont {N.}~\bibnamefont
  {Sadooghi}},\ }\href {\doibase 10.1103/PhysRevD.86.085042} {\bibfield
  {journal} {\bibinfo  {journal} {Phys. Rev. D}\ }\textbf {\bibinfo {volume}
  {86}},\ \bibinfo {pages} {085042} (\bibinfo {year} {2012})},\ \Eprint
  {http://arxiv.org/abs/1206.6051} {arXiv:1206.6051 [hep-ph]} \BibitemShut
  {NoStop}%
\bibitem [{\citenamefont {Mao}(2016)}]{Mao:2016fha}%
  \BibitemOpen
  \bibfield  {author} {\bibinfo {author} {\bibfnamefont {S.}~\bibnamefont
  {Mao}},\ }\href {\doibase 10.1016/j.physletb.2016.05.018} {\bibfield
  {journal} {\bibinfo  {journal} {Phys. Lett. B}\ }\textbf {\bibinfo {volume}
  {758}},\ \bibinfo {pages} {195} (\bibinfo {year} {2016})},\ \Eprint
  {http://arxiv.org/abs/1602.06503} {arXiv:1602.06503 [hep-ph]} \BibitemShut
  {NoStop}%
\bibitem [{\citenamefont {Fayazbakhsh}\ and\ \citenamefont
  {Sadooghi}(2014)}]{Fayazbakhsh:2014mca}%
  \BibitemOpen
  \bibfield  {author} {\bibinfo {author} {\bibfnamefont {S.}~\bibnamefont
  {Fayazbakhsh}}\ and\ \bibinfo {author} {\bibfnamefont {N.}~\bibnamefont
  {Sadooghi}},\ }\href {\doibase 10.1103/PhysRevD.90.105030} {\bibfield
  {journal} {\bibinfo  {journal} {Phys. Rev. D}\ }\textbf {\bibinfo {volume}
  {90}},\ \bibinfo {pages} {105030} (\bibinfo {year} {2014})},\ \Eprint
  {http://arxiv.org/abs/1408.5457} {arXiv:1408.5457 [hep-ph]} \BibitemShut
  {NoStop}%
\bibitem [{\citenamefont {Farias}\ \emph {et~al.}(2021)\citenamefont {Farias},
  \citenamefont {Tavares}, \citenamefont {Nunes},\ and\ \citenamefont
  {Avancini}}]{Farias:2021fci}%
  \BibitemOpen
  \bibfield  {author} {\bibinfo {author} {\bibfnamefont {R.~L.~S.}\
  \bibnamefont {Farias}}, \bibinfo {author} {\bibfnamefont {W.~R.}\
  \bibnamefont {Tavares}}, \bibinfo {author} {\bibfnamefont {R.~M.}\
  \bibnamefont {Nunes}}, \ and\ \bibinfo {author} {\bibfnamefont {S.~S.}\
  \bibnamefont {Avancini}},\ }\href@noop {} {\  (\bibinfo {year} {2021})},\
  \Eprint {http://arxiv.org/abs/2109.11112} {arXiv:2109.11112 [hep-ph]}
  \BibitemShut {NoStop}%
\bibitem [{\citenamefont {Chaudhuri}\ \emph {et~al.}(2019)\citenamefont
  {Chaudhuri}, \citenamefont {Ghosh}, \citenamefont {Sarkar},\ and\
  \citenamefont {Roy}}]{Chaudhuri:2019lbw}%
  \BibitemOpen
  \bibfield  {author} {\bibinfo {author} {\bibfnamefont {N.}~\bibnamefont
  {Chaudhuri}}, \bibinfo {author} {\bibfnamefont {S.}~\bibnamefont {Ghosh}},
  \bibinfo {author} {\bibfnamefont {S.}~\bibnamefont {Sarkar}}, \ and\ \bibinfo
  {author} {\bibfnamefont {P.}~\bibnamefont {Roy}},\ }\href {\doibase
  10.1103/PhysRevD.99.116025} {\bibfield  {journal} {\bibinfo  {journal} {Phys.
  Rev.}\ }\textbf {\bibinfo {volume} {D99}},\ \bibinfo {pages} {116025}
  (\bibinfo {year} {2019})},\ \Eprint {http://arxiv.org/abs/1907.03990}
  {arXiv:1907.03990 [nucl-th]} \BibitemShut {NoStop}%
\bibitem [{\citenamefont {Chaudhuri}\ \emph {et~al.}(2020)\citenamefont
  {Chaudhuri}, \citenamefont {Ghosh}, \citenamefont {Sarkar},\ and\
  \citenamefont {Roy}}]{Chaudhuri:2020lga}%
  \BibitemOpen
  \bibfield  {author} {\bibinfo {author} {\bibfnamefont {N.}~\bibnamefont
  {Chaudhuri}}, \bibinfo {author} {\bibfnamefont {S.}~\bibnamefont {Ghosh}},
  \bibinfo {author} {\bibfnamefont {S.}~\bibnamefont {Sarkar}}, \ and\ \bibinfo
  {author} {\bibfnamefont {P.}~\bibnamefont {Roy}},\ }\href {\doibase
  10.1140/epja/s10050-020-00222-9} {\bibfield  {journal} {\bibinfo  {journal}
  {Eur. Phys. J. A}\ }\textbf {\bibinfo {volume} {56}},\ \bibinfo {pages} {213}
  (\bibinfo {year} {2020})},\ \Eprint {http://arxiv.org/abs/2003.05692}
  {arXiv:2003.05692 [nucl-th]} \BibitemShut {NoStop}%
\bibitem [{\citenamefont {Ghosh}\ \emph {et~al.}(2021)\citenamefont {Ghosh},
  \citenamefont {Chaudhuri}, \citenamefont {Roy},\ and\ \citenamefont
  {Sarkar}}]{Ghosh:2021dlo}%
  \BibitemOpen
  \bibfield  {author} {\bibinfo {author} {\bibfnamefont {S.}~\bibnamefont
  {Ghosh}}, \bibinfo {author} {\bibfnamefont {N.}~\bibnamefont {Chaudhuri}},
  \bibinfo {author} {\bibfnamefont {P.}~\bibnamefont {Roy}}, \ and\ \bibinfo
  {author} {\bibfnamefont {S.}~\bibnamefont {Sarkar}},\ }\href {\doibase
  10.1103/PhysRevD.103.116008} {\bibfield  {journal} {\bibinfo  {journal}
  {Phys. Rev. D}\ }\textbf {\bibinfo {volume} {103}},\ \bibinfo {pages}
  {116008} (\bibinfo {year} {2021})},\ \Eprint
  {http://arxiv.org/abs/2104.14112} {arXiv:2104.14112 [hep-ph]} \BibitemShut
  {NoStop}%
\bibitem [{\citenamefont {Ghosh}\ \emph
  {et~al.}(2020{\natexlab{a}})\citenamefont {Ghosh}, \citenamefont {Chaudhuri},
  \citenamefont {Sarkar},\ and\ \citenamefont {Roy}}]{Ghosh:2020xwp}%
  \BibitemOpen
  \bibfield  {author} {\bibinfo {author} {\bibfnamefont {S.}~\bibnamefont
  {Ghosh}}, \bibinfo {author} {\bibfnamefont {N.}~\bibnamefont {Chaudhuri}},
  \bibinfo {author} {\bibfnamefont {S.}~\bibnamefont {Sarkar}}, \ and\ \bibinfo
  {author} {\bibfnamefont {P.}~\bibnamefont {Roy}},\ }\href {\doibase
  10.1103/PhysRevD.101.096002} {\bibfield  {journal} {\bibinfo  {journal}
  {Phys. Rev. D}\ }\textbf {\bibinfo {volume} {101}},\ \bibinfo {pages}
  {096002} (\bibinfo {year} {2020}{\natexlab{a}})},\ \Eprint
  {http://arxiv.org/abs/2004.09203} {arXiv:2004.09203 [nucl-th]} \BibitemShut
  {NoStop}%
\bibitem [{\citenamefont {Chaudhuri}\ \emph {et~al.}(2021)\citenamefont
  {Chaudhuri}, \citenamefont {Ghosh}, \citenamefont {Sarkar},\ and\
  \citenamefont {Roy}}]{Chaudhuri:2021skc}%
  \BibitemOpen
  \bibfield  {author} {\bibinfo {author} {\bibfnamefont {N.}~\bibnamefont
  {Chaudhuri}}, \bibinfo {author} {\bibfnamefont {S.}~\bibnamefont {Ghosh}},
  \bibinfo {author} {\bibfnamefont {S.}~\bibnamefont {Sarkar}}, \ and\ \bibinfo
  {author} {\bibfnamefont {P.}~\bibnamefont {Roy}},\ }\href {\doibase
  10.1103/PhysRevD.103.096021} {\bibfield  {journal} {\bibinfo  {journal}
  {Phys. Rev. D}\ }\textbf {\bibinfo {volume} {103}},\ \bibinfo {pages}
  {096021} (\bibinfo {year} {2021})},\ \Eprint
  {http://arxiv.org/abs/2104.11425} {arXiv:2104.11425 [hep-ph]} \BibitemShut
  {NoStop}%
\bibitem [{\citenamefont {O'Connell}(1968)}]{OConnell:1968spc}%
  \BibitemOpen
  \bibfield  {author} {\bibinfo {author} {\bibfnamefont {R.~F.}\ \bibnamefont
  {O'Connell}},\ }\href {\doibase 10.1103/PhysRev.176.1433} {\bibfield
  {journal} {\bibinfo  {journal} {Phys. Rev.}\ }\textbf {\bibinfo {volume}
  {176}},\ \bibinfo {pages} {1433} (\bibinfo {year} {1968})}\BibitemShut
  {NoStop}%
\bibitem [{\citenamefont {Sheng}\ \emph {et~al.}(2018)\citenamefont {Sheng},
  \citenamefont {Rischke}, \citenamefont {Vasak},\ and\ \citenamefont
  {Wang}}]{Sheng:2017lfu}%
  \BibitemOpen
  \bibfield  {author} {\bibinfo {author} {\bibfnamefont {X.-l.}\ \bibnamefont
  {Sheng}}, \bibinfo {author} {\bibfnamefont {D.~H.}\ \bibnamefont {Rischke}},
  \bibinfo {author} {\bibfnamefont {D.}~\bibnamefont {Vasak}}, \ and\ \bibinfo
  {author} {\bibfnamefont {Q.}~\bibnamefont {Wang}},\ }\href {\doibase
  10.1140/epja/i2018-12414-9} {\bibfield  {journal} {\bibinfo  {journal} {Eur.
  Phys. J. A}\ }\textbf {\bibinfo {volume} {54}},\ \bibinfo {pages} {21}
  (\bibinfo {year} {2018})},\ \Eprint {http://arxiv.org/abs/1707.01388}
  {arXiv:1707.01388 [hep-ph]} \BibitemShut {NoStop}%
\bibitem [{\citenamefont {Peskin}\ and\ \citenamefont
  {Schroeder}(1995)}]{Peskin:1995ev}%
  \BibitemOpen
  \bibfield  {author} {\bibinfo {author} {\bibfnamefont {M.~E.}\ \bibnamefont
  {Peskin}}\ and\ \bibinfo {author} {\bibfnamefont {D.~V.}\ \bibnamefont
  {Schroeder}},\ }\href@noop {} {\emph {\bibinfo {title} {{An Introduction to
  quantum field theory}}}}\ (\bibinfo  {publisher} {Addison-Wesley},\ \bibinfo
  {address} {Reading, USA},\ \bibinfo {year} {1995})\BibitemShut {NoStop}%
\bibitem [{\citenamefont {Xu}\ \emph {et~al.}(2021)\citenamefont {Xu},
  \citenamefont {Chao},\ and\ \citenamefont {Huang}}]{Xu:2020yag}%
  \BibitemOpen
  \bibfield  {author} {\bibinfo {author} {\bibfnamefont {K.}~\bibnamefont
  {Xu}}, \bibinfo {author} {\bibfnamefont {J.}~\bibnamefont {Chao}}, \ and\
  \bibinfo {author} {\bibfnamefont {M.}~\bibnamefont {Huang}},\ }\href
  {\doibase 10.1103/PhysRevD.103.076015} {\bibfield  {journal} {\bibinfo
  {journal} {Phys. Rev. D}\ }\textbf {\bibinfo {volume} {103}},\ \bibinfo
  {pages} {076015} (\bibinfo {year} {2021})},\ \Eprint
  {http://arxiv.org/abs/2007.13122} {arXiv:2007.13122 [hep-ph]} \BibitemShut
  {NoStop}%
\bibitem [{\citenamefont {Roessner}\ \emph {et~al.}(2007)\citenamefont
  {Roessner}, \citenamefont {Ratti},\ and\ \citenamefont
  {Weise}}]{Roessner:2006xn}%
  \BibitemOpen
  \bibfield  {author} {\bibinfo {author} {\bibfnamefont {S.}~\bibnamefont
  {Roessner}}, \bibinfo {author} {\bibfnamefont {C.}~\bibnamefont {Ratti}}, \
  and\ \bibinfo {author} {\bibfnamefont {W.}~\bibnamefont {Weise}},\ }\href
  {\doibase 10.1103/PhysRevD.75.034007} {\bibfield  {journal} {\bibinfo
  {journal} {Phys. Rev. D}\ }\textbf {\bibinfo {volume} {75}},\ \bibinfo
  {pages} {034007} (\bibinfo {year} {2007})},\ \Eprint
  {http://arxiv.org/abs/hep-ph/0609281} {arXiv:hep-ph/0609281} \BibitemShut
  {NoStop}%
\bibitem [{\citenamefont {Bicudo}\ \emph {et~al.}(1999)\citenamefont {Bicudo},
  \citenamefont {Ribeiro},\ and\ \citenamefont {Fernandes}}]{Bicudo:1998qb}%
  \BibitemOpen
  \bibfield  {author} {\bibinfo {author} {\bibfnamefont {P.~J.}\ \bibnamefont
  {Bicudo}}, \bibinfo {author} {\bibfnamefont {J.~F.}\ \bibnamefont {Ribeiro}},
  \ and\ \bibinfo {author} {\bibfnamefont {R.}~\bibnamefont {Fernandes}},\
  }\href {\doibase 10.1103/PhysRevC.59.1107} {\bibfield  {journal} {\bibinfo
  {journal} {Phys. Rev. C}\ }\textbf {\bibinfo {volume} {59}},\ \bibinfo
  {pages} {1107} (\bibinfo {year} {1999})},\ \Eprint
  {http://arxiv.org/abs/hep-ph/9806243} {arXiv:hep-ph/9806243} \BibitemShut
  {NoStop}%
\bibitem [{\citenamefont {Shovkovy}(2013)}]{Shovkovy:2012zn}%
  \BibitemOpen
  \bibfield  {author} {\bibinfo {author} {\bibfnamefont {I.~A.}\ \bibnamefont
  {Shovkovy}},\ }\href {\doibase 10.1007/978-3-642-37305-3_2} {\bibfield
  {journal} {\bibinfo  {journal} {Lect. Notes Phys.}\ }\textbf {\bibinfo
  {volume} {871}},\ \bibinfo {pages} {13} (\bibinfo {year} {2013})},\ \Eprint
  {http://arxiv.org/abs/1207.5081} {arXiv:1207.5081 [hep-ph]} \BibitemShut
  {NoStop}%
\bibitem [{\citenamefont {Gusynin}\ \emph {et~al.}(1999)\citenamefont
  {Gusynin}, \citenamefont {Miransky},\ and\ \citenamefont
  {Shovkovy}}]{Gusynin:1999pq}%
  \BibitemOpen
  \bibfield  {author} {\bibinfo {author} {\bibfnamefont {V.~P.}\ \bibnamefont
  {Gusynin}}, \bibinfo {author} {\bibfnamefont {V.~A.}\ \bibnamefont
  {Miransky}}, \ and\ \bibinfo {author} {\bibfnamefont {I.~A.}\ \bibnamefont
  {Shovkovy}},\ }\href {\doibase 10.1016/S0550-3213(99)00573-8} {\bibfield
  {journal} {\bibinfo  {journal} {Nucl. Phys.}\ }\textbf {\bibinfo {volume}
  {B563}},\ \bibinfo {pages} {361} (\bibinfo {year} {1999})},\ \Eprint
  {http://arxiv.org/abs/hep-ph/9908320} {arXiv:hep-ph/9908320 [hep-ph]}
  \BibitemShut {NoStop}%
\bibitem [{\citenamefont {Ghosh}\ \emph
  {et~al.}(2020{\natexlab{b}})\citenamefont {Ghosh}, \citenamefont {Mukherjee},
  \citenamefont {Chaudhuri}, \citenamefont {Roy},\ and\ \citenamefont
  {Sarkar}}]{Ghosh:2020qvg}%
  \BibitemOpen
  \bibfield  {author} {\bibinfo {author} {\bibfnamefont {S.}~\bibnamefont
  {Ghosh}}, \bibinfo {author} {\bibfnamefont {A.}~\bibnamefont {Mukherjee}},
  \bibinfo {author} {\bibfnamefont {N.}~\bibnamefont {Chaudhuri}}, \bibinfo
  {author} {\bibfnamefont {P.}~\bibnamefont {Roy}}, \ and\ \bibinfo {author}
  {\bibfnamefont {S.}~\bibnamefont {Sarkar}},\ }\href {\doibase
  10.1103/PhysRevD.101.056023} {\bibfield  {journal} {\bibinfo  {journal}
  {Phys. Rev. D}\ }\textbf {\bibinfo {volume} {101}},\ \bibinfo {pages}
  {056023} (\bibinfo {year} {2020}{\natexlab{b}})},\ \Eprint
  {http://arxiv.org/abs/2003.02024} {arXiv:2003.02024 [hep-ph]} \BibitemShut
  {NoStop}%
\bibitem [{\citenamefont {Kharzeev}\ \emph
  {et~al.}(2013{\natexlab{b}})\citenamefont {Kharzeev}, \citenamefont
  {Landsteiner}, \citenamefont {Schmitt},\ and\ \citenamefont
  {Yee}}]{Kharzeev:2012ph}%
  \BibitemOpen
  \bibfield  {author} {\bibinfo {author} {\bibfnamefont {D.~E.}\ \bibnamefont
  {Kharzeev}}, \bibinfo {author} {\bibfnamefont {K.}~\bibnamefont
  {Landsteiner}}, \bibinfo {author} {\bibfnamefont {A.}~\bibnamefont
  {Schmitt}}, \ and\ \bibinfo {author} {\bibfnamefont {H.-U.}\ \bibnamefont
  {Yee}},\ }\href {\doibase 10.1007/978-3-642-37305-3_1} {\bibfield  {journal}
  {\bibinfo  {journal} {Lect. Notes Phys.}\ }\textbf {\bibinfo {volume}
  {871}},\ \bibinfo {pages} {1} (\bibinfo {year} {2013}{\natexlab{b}})},\
  \Eprint {http://arxiv.org/abs/1211.6245} {arXiv:1211.6245 [hep-ph]}
  \BibitemShut {NoStop}%
\bibitem [{\citenamefont {Avancini}\ \emph {et~al.}(2019)\citenamefont
  {Avancini}, \citenamefont {Farias},\ and\ \citenamefont
  {Tavares}}]{Avancini:2018svs}%
  \BibitemOpen
  \bibfield  {author} {\bibinfo {author} {\bibfnamefont {S.~S.}\ \bibnamefont
  {Avancini}}, \bibinfo {author} {\bibfnamefont {R.~L.}\ \bibnamefont
  {Farias}}, \ and\ \bibinfo {author} {\bibfnamefont {W.~R.}\ \bibnamefont
  {Tavares}},\ }\href {\doibase 10.1103/PhysRevD.99.056009} {\bibfield
  {journal} {\bibinfo  {journal} {Phys. Rev. D}\ }\textbf {\bibinfo {volume}
  {99}},\ \bibinfo {pages} {056009} (\bibinfo {year} {2019})},\ \Eprint
  {http://arxiv.org/abs/1812.00945} {arXiv:1812.00945 [hep-ph]} \BibitemShut
  {NoStop}%
\bibitem [{\citenamefont {Gatto}\ and\ \citenamefont
  {Ruggieri}(2010)}]{Gatto:2010qs}%
  \BibitemOpen
  \bibfield  {author} {\bibinfo {author} {\bibfnamefont {R.}~\bibnamefont
  {Gatto}}\ and\ \bibinfo {author} {\bibfnamefont {M.}~\bibnamefont
  {Ruggieri}},\ }\href {\doibase 10.1103/PhysRevD.82.054027} {\bibfield
  {journal} {\bibinfo  {journal} {Phys. Rev. D}\ }\textbf {\bibinfo {volume}
  {82}},\ \bibinfo {pages} {054027} (\bibinfo {year} {2010})},\ \Eprint
  {http://arxiv.org/abs/1007.0790} {arXiv:1007.0790 [hep-ph]} \BibitemShut
  {NoStop}%
\bibitem [{\citenamefont {Landau}\ and\ \citenamefont
  {Lifshitz}(1980)}]{Landau:1980mil}%
  \BibitemOpen
  \bibfield  {author} {\bibinfo {author} {\bibfnamefont {L.~D.}\ \bibnamefont
  {Landau}}\ and\ \bibinfo {author} {\bibfnamefont {E.~M.}\ \bibnamefont
  {Lifshitz}},\ }\href@noop {} {\emph {\bibinfo {title} {{Statistical Physics,
  Part 1}}}},\ \bibinfo {series} {Course of Theoretical Physics}, Vol.~\bibinfo
  {volume} {5}\ (\bibinfo  {publisher} {Butterworth-Heinemann},\ \bibinfo
  {address} {Oxford},\ \bibinfo {year} {1980})\BibitemShut {NoStop}%
\bibitem [{\citenamefont {Fayazbakhsh}\ and\ \citenamefont
  {Sadooghi}(2011)}]{Fayazbakhsh:2010bh}%
  \BibitemOpen
  \bibfield  {author} {\bibinfo {author} {\bibfnamefont {S.}~\bibnamefont
  {Fayazbakhsh}}\ and\ \bibinfo {author} {\bibfnamefont {N.}~\bibnamefont
  {Sadooghi}},\ }\href {\doibase 10.1103/PhysRevD.83.025026} {\bibfield
  {journal} {\bibinfo  {journal} {Phys. Rev.}\ }\textbf {\bibinfo {volume}
  {D83}},\ \bibinfo {pages} {025026} (\bibinfo {year} {2011})},\ \Eprint
  {http://arxiv.org/abs/1009.6125} {arXiv:1009.6125 [hep-ph]} \BibitemShut
  {NoStop}%
\bibitem [{\citenamefont {Fayazbakhsh}\ and\ \citenamefont
  {Sadooghi}(2010)}]{Fayazbakhsh:2010gc}%
  \BibitemOpen
  \bibfield  {author} {\bibinfo {author} {\bibfnamefont {S.}~\bibnamefont
  {Fayazbakhsh}}\ and\ \bibinfo {author} {\bibfnamefont {N.}~\bibnamefont
  {Sadooghi}},\ }\href {\doibase 10.1103/PhysRevD.82.045010} {\bibfield
  {journal} {\bibinfo  {journal} {Phys. Rev.}\ }\textbf {\bibinfo {volume}
  {D82}},\ \bibinfo {pages} {045010} (\bibinfo {year} {2010})},\ \Eprint
  {http://arxiv.org/abs/1005.5022} {arXiv:1005.5022 [hep-ph]} \BibitemShut
  {NoStop}%
\bibitem [{\citenamefont {Ebert}\ and\ \citenamefont
  {Vshivtsev}(1998)}]{Ebert:1998gx}%
  \BibitemOpen
  \bibfield  {author} {\bibinfo {author} {\bibfnamefont {D.}~\bibnamefont
  {Ebert}}\ and\ \bibinfo {author} {\bibfnamefont {A.~S.}\ \bibnamefont
  {Vshivtsev}},\ }\href@noop {} {\  (\bibinfo {year} {1998})},\ \Eprint
  {http://arxiv.org/abs/hep-ph/9806421} {arXiv:hep-ph/9806421 [hep-ph]}
  \BibitemShut {NoStop}%
\bibitem [{\citenamefont {Inagaki}\ \emph {et~al.}(2004)\citenamefont
  {Inagaki}, \citenamefont {Kimura},\ and\ \citenamefont
  {Murata}}]{Inagaki:2004ih}%
  \BibitemOpen
  \bibfield  {author} {\bibinfo {author} {\bibfnamefont {T.}~\bibnamefont
  {Inagaki}}, \bibinfo {author} {\bibfnamefont {D.}~\bibnamefont {Kimura}}, \
  and\ \bibinfo {author} {\bibfnamefont {T.}~\bibnamefont {Murata}},\
  }\bibfield  {booktitle} {\emph {\bibinfo {booktitle} {{Finite density QCD.
  Proceedings, International Workshop, Nara, Japan, July 10-12, 2003}}},\
  }\href {\doibase 10.1143/PTPS.153.321} {\bibfield  {journal} {\bibinfo
  {journal} {Prog. Theor. Phys. Suppl.}\ }\textbf {\bibinfo {volume} {153}},\
  \bibinfo {pages} {321} (\bibinfo {year} {2004})},\ \Eprint
  {http://arxiv.org/abs/hep-ph/0404219} {arXiv:hep-ph/0404219 [hep-ph]}
  \BibitemShut {NoStop}%
\bibitem [{\citenamefont {Noronha}\ and\ \citenamefont
  {Shovkovy}(2007)}]{Noronha:2007wg}%
  \BibitemOpen
  \bibfield  {author} {\bibinfo {author} {\bibfnamefont {J.~L.}\ \bibnamefont
  {Noronha}}\ and\ \bibinfo {author} {\bibfnamefont {I.~A.}\ \bibnamefont
  {Shovkovy}},\ }\href {\doibase 10.1103/PhysRevD.76.105030,
  10.1103/PhysRevD.86.049901} {\bibfield  {journal} {\bibinfo  {journal} {Phys.
  Rev.}\ }\textbf {\bibinfo {volume} {D76}},\ \bibinfo {pages} {105030}
  (\bibinfo {year} {2007})},\ \bibinfo {note} {[Erratum: Phys.
  Rev.D86,049901(2012)]},\ \Eprint {http://arxiv.org/abs/0708.0307}
  {arXiv:0708.0307 [hep-ph]} \BibitemShut {NoStop}%
\bibitem [{\citenamefont {Fukushima}\ and\ \citenamefont
  {Warringa}(2008)}]{Fukushima:2007fc}%
  \BibitemOpen
  \bibfield  {author} {\bibinfo {author} {\bibfnamefont {K.}~\bibnamefont
  {Fukushima}}\ and\ \bibinfo {author} {\bibfnamefont {H.~J.}\ \bibnamefont
  {Warringa}},\ }\href {\doibase 10.1103/PhysRevLett.100.032007} {\bibfield
  {journal} {\bibinfo  {journal} {Phys. Rev. Lett.}\ }\textbf {\bibinfo
  {volume} {100}},\ \bibinfo {pages} {032007} (\bibinfo {year} {2008})},\
  \Eprint {http://arxiv.org/abs/0707.3785} {arXiv:0707.3785 [hep-ph]}
  \BibitemShut {NoStop}%
\bibitem [{\citenamefont {Aguirre}(2020)}]{Aguirre:2020tiy}%
  \BibitemOpen
  \bibfield  {author} {\bibinfo {author} {\bibfnamefont {R.}~\bibnamefont
  {Aguirre}},\ }\href {\doibase 10.1103/PhysRevD.102.096025} {\bibfield
  {journal} {\bibinfo  {journal} {Phys. Rev. D}\ }\textbf {\bibinfo {volume}
  {102}},\ \bibinfo {pages} {096025} (\bibinfo {year} {2020})},\ \Eprint
  {http://arxiv.org/abs/2009.01828} {arXiv:2009.01828 [hep-ph]} \BibitemShut
  {NoStop}%
\bibitem [{\citenamefont {Mei}\ and\ \citenamefont {Mao}(2020)}]{Mei:2020jzn}%
  \BibitemOpen
  \bibfield  {author} {\bibinfo {author} {\bibfnamefont {J.}~\bibnamefont
  {Mei}}\ and\ \bibinfo {author} {\bibfnamefont {S.}~\bibnamefont {Mao}},\
  }\href@noop {} {\  (\bibinfo {year} {2020})},\ \Eprint
  {http://arxiv.org/abs/2008.12123} {arXiv:2008.12123 [hep-ph]} \BibitemShut
  {NoStop}%
\bibitem [{\citenamefont {Mukherjee}\ \emph {et~al.}(2018)\citenamefont
  {Mukherjee}, \citenamefont {Ghosh}, \citenamefont {Mandal}, \citenamefont
  {Sarkar},\ and\ \citenamefont {Roy}}]{Mukherjee:2018ebw}%
  \BibitemOpen
  \bibfield  {author} {\bibinfo {author} {\bibfnamefont {A.}~\bibnamefont
  {Mukherjee}}, \bibinfo {author} {\bibfnamefont {S.}~\bibnamefont {Ghosh}},
  \bibinfo {author} {\bibfnamefont {M.}~\bibnamefont {Mandal}}, \bibinfo
  {author} {\bibfnamefont {S.}~\bibnamefont {Sarkar}}, \ and\ \bibinfo {author}
  {\bibfnamefont {P.}~\bibnamefont {Roy}},\ }\href {\doibase
  10.1103/PhysRevD.98.056024} {\bibfield  {journal} {\bibinfo  {journal} {Phys.
  Rev. D}\ }\textbf {\bibinfo {volume} {98}},\ \bibinfo {pages} {056024}
  (\bibinfo {year} {2018})},\ \Eprint {http://arxiv.org/abs/1809.07028}
  {arXiv:1809.07028 [hep-ph]} \BibitemShut {NoStop}%
\bibitem [{\citenamefont {Yu}\ \emph {et~al.}(2014)\citenamefont {Yu},
  \citenamefont {Liu},\ and\ \citenamefont {Huang}}]{Yu:2014sla}%
  \BibitemOpen
  \bibfield  {author} {\bibinfo {author} {\bibfnamefont {L.}~\bibnamefont
  {Yu}}, \bibinfo {author} {\bibfnamefont {H.}~\bibnamefont {Liu}}, \ and\
  \bibinfo {author} {\bibfnamefont {M.}~\bibnamefont {Huang}},\ }\href
  {\doibase 10.1103/PhysRevD.90.074009} {\bibfield  {journal} {\bibinfo
  {journal} {Phys. Rev. D}\ }\textbf {\bibinfo {volume} {90}},\ \bibinfo
  {pages} {074009} (\bibinfo {year} {2014})},\ \Eprint
  {http://arxiv.org/abs/1404.6969} {arXiv:1404.6969 [hep-ph]} \BibitemShut
  {NoStop}%
\bibitem [{\citenamefont {Astrakhantsev}\ \emph {et~al.}(2021)\citenamefont
  {Astrakhantsev}, \citenamefont {Braguta}, \citenamefont {Kotov},
  \citenamefont {Kuznedelev},\ and\ \citenamefont
  {Nikolaev}}]{Astrakhantsev:2019wnp}%
  \BibitemOpen
  \bibfield  {author} {\bibinfo {author} {\bibfnamefont {N.~Y.}\ \bibnamefont
  {Astrakhantsev}}, \bibinfo {author} {\bibfnamefont {V.~V.}\ \bibnamefont
  {Braguta}}, \bibinfo {author} {\bibfnamefont {A.~Y.}\ \bibnamefont {Kotov}},
  \bibinfo {author} {\bibfnamefont {D.~D.}\ \bibnamefont {Kuznedelev}}, \ and\
  \bibinfo {author} {\bibfnamefont {A.~A.}\ \bibnamefont {Nikolaev}},\ }\href
  {\doibase 10.1140/epja/s10050-020-00326-2} {\bibfield  {journal} {\bibinfo
  {journal} {Eur. Phys. J. A}\ }\textbf {\bibinfo {volume} {57}},\ \bibinfo
  {pages} {15} (\bibinfo {year} {2021})},\ \Eprint
  {http://arxiv.org/abs/1902.09325} {arXiv:1902.09325 [hep-lat]} \BibitemShut
  {NoStop}%
\bibitem [{\citenamefont {Braguta}\ \emph {et~al.}(2016)\citenamefont
  {Braguta}, \citenamefont {Ilgenfritz}, \citenamefont {Kotov}, \citenamefont
  {Petersson},\ and\ \citenamefont {Skinderev}}]{Braguta:2015owi}%
  \BibitemOpen
  \bibfield  {author} {\bibinfo {author} {\bibfnamefont {V.~V.}\ \bibnamefont
  {Braguta}}, \bibinfo {author} {\bibfnamefont {E.~M.}\ \bibnamefont
  {Ilgenfritz}}, \bibinfo {author} {\bibfnamefont {A.~Y.}\ \bibnamefont
  {Kotov}}, \bibinfo {author} {\bibfnamefont {B.}~\bibnamefont {Petersson}}, \
  and\ \bibinfo {author} {\bibfnamefont {S.~A.}\ \bibnamefont {Skinderev}},\
  }\href {\doibase 10.1103/PhysRevD.93.034509} {\bibfield  {journal} {\bibinfo
  {journal} {Phys. Rev. D}\ }\textbf {\bibinfo {volume} {93}},\ \bibinfo
  {pages} {034509} (\bibinfo {year} {2016})},\ \Eprint
  {http://arxiv.org/abs/1512.05873} {arXiv:1512.05873 [hep-lat]} \BibitemShut
  {NoStop}%
\bibitem [{\citenamefont {Braguta}\ \emph {et~al.}(2015)\citenamefont
  {Braguta}, \citenamefont {Goy}, \citenamefont {Ilgenfritz}, \citenamefont
  {Kotov}, \citenamefont {Molochkov}, \citenamefont {Muller-Preussker},\ and\
  \citenamefont {Petersson}}]{Braguta:2015zta}%
  \BibitemOpen
  \bibfield  {author} {\bibinfo {author} {\bibfnamefont {V.~V.}\ \bibnamefont
  {Braguta}}, \bibinfo {author} {\bibfnamefont {V.~A.}\ \bibnamefont {Goy}},
  \bibinfo {author} {\bibfnamefont {E.~M.}\ \bibnamefont {Ilgenfritz}},
  \bibinfo {author} {\bibfnamefont {A.~Y.}\ \bibnamefont {Kotov}}, \bibinfo
  {author} {\bibfnamefont {A.~V.}\ \bibnamefont {Molochkov}}, \bibinfo {author}
  {\bibfnamefont {M.}~\bibnamefont {Muller-Preussker}}, \ and\ \bibinfo
  {author} {\bibfnamefont {B.}~\bibnamefont {Petersson}},\ }\href {\doibase
  10.1007/JHEP06(2015)094} {\bibfield  {journal} {\bibinfo  {journal} {JHEP}\
  }\textbf {\bibinfo {volume} {06}},\ \bibinfo {pages} {094} (\bibinfo {year}
  {2015})},\ \Eprint {http://arxiv.org/abs/1503.06670} {arXiv:1503.06670
  [hep-lat]} \BibitemShut {NoStop}%
\bibitem [{\citenamefont {Xu}\ \emph {et~al.}(2015)\citenamefont {Xu},
  \citenamefont {Cui}, \citenamefont {Wang}, \citenamefont {Shi}, \citenamefont
  {Yang},\ and\ \citenamefont {Zong}}]{Xu:2015vna}%
  \BibitemOpen
  \bibfield  {author} {\bibinfo {author} {\bibfnamefont {S.-S.}\ \bibnamefont
  {Xu}}, \bibinfo {author} {\bibfnamefont {Z.-F.}\ \bibnamefont {Cui}},
  \bibinfo {author} {\bibfnamefont {B.}~\bibnamefont {Wang}}, \bibinfo {author}
  {\bibfnamefont {Y.-M.}\ \bibnamefont {Shi}}, \bibinfo {author} {\bibfnamefont
  {Y.-C.}\ \bibnamefont {Yang}}, \ and\ \bibinfo {author} {\bibfnamefont
  {H.-S.}\ \bibnamefont {Zong}},\ }\href {\doibase 10.1103/PhysRevD.91.056003}
  {\bibfield  {journal} {\bibinfo  {journal} {Phys. Rev. D}\ }\textbf {\bibinfo
  {volume} {91}},\ \bibinfo {pages} {056003} (\bibinfo {year} {2015})},\
  \Eprint {http://arxiv.org/abs/1505.00316} {arXiv:1505.00316 [hep-ph]}
  \BibitemShut {NoStop}%
\bibitem [{\citenamefont {Wang}\ \emph {et~al.}(2015)\citenamefont {Wang},
  \citenamefont {Wang}, \citenamefont {Cui},\ and\ \citenamefont
  {Zong}}]{Wang:2015tia}%
  \BibitemOpen
  \bibfield  {author} {\bibinfo {author} {\bibfnamefont {B.}~\bibnamefont
  {Wang}}, \bibinfo {author} {\bibfnamefont {Y.-L.}\ \bibnamefont {Wang}},
  \bibinfo {author} {\bibfnamefont {Z.-F.}\ \bibnamefont {Cui}}, \ and\
  \bibinfo {author} {\bibfnamefont {H.-S.}\ \bibnamefont {Zong}},\ }\href
  {\doibase 10.1103/PhysRevD.91.034017} {\bibfield  {journal} {\bibinfo
  {journal} {Phys. Rev. D}\ }\textbf {\bibinfo {volume} {91}},\ \bibinfo
  {pages} {034017} (\bibinfo {year} {2015})}\BibitemShut {NoStop}%
\bibitem [{\citenamefont {Cui}\ \emph {et~al.}(2016)\citenamefont {Cui},
  \citenamefont {Cloet}, \citenamefont {Lu}, \citenamefont {Roberts},
  \citenamefont {Schmidt}, \citenamefont {Xu},\ and\ \citenamefont
  {Zong}}]{Cui:2016zqp}%
  \BibitemOpen
  \bibfield  {author} {\bibinfo {author} {\bibfnamefont {Z.-F.}\ \bibnamefont
  {Cui}}, \bibinfo {author} {\bibfnamefont {I.~C.}\ \bibnamefont {Cloet}},
  \bibinfo {author} {\bibfnamefont {Y.}~\bibnamefont {Lu}}, \bibinfo {author}
  {\bibfnamefont {C.~D.}\ \bibnamefont {Roberts}}, \bibinfo {author}
  {\bibfnamefont {S.~M.}\ \bibnamefont {Schmidt}}, \bibinfo {author}
  {\bibfnamefont {S.-S.}\ \bibnamefont {Xu}}, \ and\ \bibinfo {author}
  {\bibfnamefont {H.-S.}\ \bibnamefont {Zong}},\ }\href {\doibase
  10.1103/PhysRevD.94.071503} {\bibfield  {journal} {\bibinfo  {journal} {Phys.
  Rev. D}\ }\textbf {\bibinfo {volume} {94}},\ \bibinfo {pages} {071503}
  (\bibinfo {year} {2016})},\ \Eprint {http://arxiv.org/abs/1604.08454}
  {arXiv:1604.08454 [nucl-th]} \BibitemShut {NoStop}%
\bibitem [{\citenamefont {Shi}\ \emph {et~al.}(2020)\citenamefont {Shi},
  \citenamefont {He}, \citenamefont {Jia}, \citenamefont {Wang}, \citenamefont
  {Xu},\ and\ \citenamefont {Zong}}]{Shi:2020uyb}%
  \BibitemOpen
  \bibfield  {author} {\bibinfo {author} {\bibfnamefont {C.}~\bibnamefont
  {Shi}}, \bibinfo {author} {\bibfnamefont {X.-T.}\ \bibnamefont {He}},
  \bibinfo {author} {\bibfnamefont {W.-B.}\ \bibnamefont {Jia}}, \bibinfo
  {author} {\bibfnamefont {Q.-W.}\ \bibnamefont {Wang}}, \bibinfo {author}
  {\bibfnamefont {S.-S.}\ \bibnamefont {Xu}}, \ and\ \bibinfo {author}
  {\bibfnamefont {H.-S.}\ \bibnamefont {Zong}},\ }\href {\doibase
  10.1007/JHEP06(2020)122} {\bibfield  {journal} {\bibinfo  {journal} {JHEP}\
  }\textbf {\bibinfo {volume} {06}},\ \bibinfo {pages} {122} (\bibinfo {year}
  {2020})},\ \Eprint {http://arxiv.org/abs/2004.09918} {arXiv:2004.09918
  [hep-ph]} \BibitemShut {NoStop}%
\bibitem [{\citenamefont {Chernodub}\ and\ \citenamefont
  {Nedelin}(2011)}]{Chernodub:2011fr}%
  \BibitemOpen
  \bibfield  {author} {\bibinfo {author} {\bibfnamefont {M.~N.}\ \bibnamefont
  {Chernodub}}\ and\ \bibinfo {author} {\bibfnamefont {A.~S.}\ \bibnamefont
  {Nedelin}},\ }\href {\doibase 10.1103/PhysRevD.83.105008} {\bibfield
  {journal} {\bibinfo  {journal} {Phys. Rev. D}\ }\textbf {\bibinfo {volume}
  {83}},\ \bibinfo {pages} {105008} (\bibinfo {year} {2011})},\ \Eprint
  {http://arxiv.org/abs/1102.0188} {arXiv:1102.0188 [hep-ph]} \BibitemShut
  {NoStop}%
\bibitem [{\citenamefont {Braguta}\ and\ \citenamefont
  {Kotov}(2016)}]{Braguta:2016aov}%
  \BibitemOpen
  \bibfield  {author} {\bibinfo {author} {\bibfnamefont {V.~V.}\ \bibnamefont
  {Braguta}}\ and\ \bibinfo {author} {\bibfnamefont {A.~Y.}\ \bibnamefont
  {Kotov}},\ }\href {\doibase 10.1103/PhysRevD.93.105025} {\bibfield  {journal}
  {\bibinfo  {journal} {Phys. Rev. D}\ }\textbf {\bibinfo {volume} {93}},\
  \bibinfo {pages} {105025} (\bibinfo {year} {2016})},\ \Eprint
  {http://arxiv.org/abs/1601.04957} {arXiv:1601.04957 [hep-th]} \BibitemShut
  {NoStop}%
\bibitem [{\citenamefont {Farias}\ \emph {et~al.}(2016)\citenamefont {Farias},
  \citenamefont {Duarte}, \citenamefont {Krein},\ and\ \citenamefont
  {Ramos}}]{Farias:2016let}%
  \BibitemOpen
  \bibfield  {author} {\bibinfo {author} {\bibfnamefont {R.~L.~S.}\
  \bibnamefont {Farias}}, \bibinfo {author} {\bibfnamefont {D.~C.}\
  \bibnamefont {Duarte}}, \bibinfo {author} {\bibfnamefont {G.~a.}\
  \bibnamefont {Krein}}, \ and\ \bibinfo {author} {\bibfnamefont {R.~O.}\
  \bibnamefont {Ramos}},\ }\href {\doibase 10.1103/PhysRevD.94.074011}
  {\bibfield  {journal} {\bibinfo  {journal} {Phys. Rev. D}\ }\textbf {\bibinfo
  {volume} {94}},\ \bibinfo {pages} {074011} (\bibinfo {year} {2016})},\
  \Eprint {http://arxiv.org/abs/1604.04518} {arXiv:1604.04518 [hep-ph]}
  \BibitemShut {NoStop}%
\end{thebibliography}%

\end{document}